\documentclass[nonatbib]{elsarticle}
\makeatletter
\let\c@author\relax
\makeatother
\usepackage{lineno,hyperref} 
\usepackage{amsmath} 
\modulolinenumbers[5] 
\usepackage{xcolor} 
\usepackage[labelformat=simple]{subcaption} 

\usepackage{chemformula} 
\usepackage{tikz} 
\usetikzlibrary{shapes.geometric, arrows, patterns} 
\usepackage{circuitikz} 
\usepackage{todonotes}
\usepackage{booktabs} 
\usepackage{threeparttable} 
\usepackage{adjustbox} 
\usepackage{siunitx} 
\usepackage{tabularx} 
\usepackage{placeins} 
\usepackage{caption} 
\usepackage{parskip} 
\journal{Journal}

\usepackage[
backend=biber,
style=nature,
doi=false,isbn=false,url=false,eprint=false
]{biblatex}
\begin{document}

\begin{frontmatter}

\title{Reservoir Computing with Colloidal Mixtures of ZnO and Proteinoids}


\author[a]{Raphael Fortulan}
\author[a]{Noushin Raeisi Kheirabadi}
\author[a]{Panagiotis Mougkogiannis}
\author[c,a]{Alessandro Chiolerio}
\author[a]{Andrew Adamatzky}

\address[a]{Unconventional Computing Laboratory, UWE, Bristol, UK}

\address[c]{Bioinspired Soft Robotics, Istituto Italiano di Tecnologia, Via Morego 30, 16165 Genova, Italy}

\begin{abstract}
Liquid computers use incompressible fluids for computational processes. Here we present experimental laboratory prototypes of liquid computers using colloids composed of zinc oxide (ZnO) nanoparticles and microspheres containing thermal proteins (proteinoids). The choice of proteinoids is based on their distinctive neuron-like electrical behaviour and their similarity to protocells. In addition, ZnO nanoparticles are chosen for their non-trivial electrical properties. Our research demonstrates the successful extraction of 2-, 4- and 8-bit logic functions in ZnO proteinoid colloids. Our analysis shows that each material has a distinct set of logic functions, and that the complexity of the expressions is directly related to each material present in a mixture. These findings provide a basis for the development of future hybris liquid devices capable of general-purpose computing.
\end{abstract}


\end{frontmatter}


\section{Introduction}

Liquid-based computers use incompressible fluids (i.e., substances featuring a zero divergence of the flow velocity) as a medium for driving or hosting computational processes. Examples of liquid computers include hydraulic algebraic machines~\cite{emch1901two,gibb1914}, hydraulic integrators~\cite{moore1936hydrocal,luk1939hydraulic}, fluid mappers~\cite{moore1949fields, fuerstman2003solving}, fluidic logic~\cite{hobbs1963fluid,peter1965and}, reaction-diffusion computers~\cite{adamatzky2005reaction}, fluid maze solver~ \cite{fuerstman2003solving}, liquid marbles logic~\cite{draper2017liquid}. 

Despite their potentially non-linear and computationally rich behaviour, colloids have not been previously considered unconventional computing devices. Earlier, the concept of liquid cybernetic systems was developed \cite{chiolerioLiquidCyberneticSystems2020}, introducing colloidal autonomous soft holonomic processors with {autolographic} features \cite{chiolerioSmartFluidSystems2017}. Recently, the computational capabilities of \ch{Fe3O4} ferrofluid were demonstrated, showcasing its ability to recognise digits using an $8\times 8$ pixel-grid dataset. This ferrofluid system is programmable and readable via electrical signalling \cite{crepaldiExperimentalDemonstrationInMemory2023}. This advancement signals a departure from conventional notions, revealing the untapped potential of colloids in the realm of unconventional computing and offering new avenues for exploration and innovation.

In this work, we explore the capabilities of colloidal mixtures of ZnO and proteinoid. ZnO, a non-toxic biocompatible~\cite{kielbikBiodegradationZnOEu2017}, being a direct band gap semiconductor with a large exciton binding energy of 60 meV~\cite{tsukazakiRepeatedTemperatureModulation2005}, is considered a capable semiconductor to be used in numerous applications such as solar cells~\cite{keisEfficientPhotoelectrochemicalSolar2002,wibowoZnONanostructuredMaterials2020,a.mohammedSurfaceTreatmentZnO2021}, photocatalysis~\cite{fengPhotocatalyticPhenolDegradation2020}, catalysts~\cite{kellyZnOActiveSelective2019}, and LEDs~\cite{qianStableEfficientQuantumdot2011}. Its non-symmetric hexagonal wurtzite structure~\cite{leeAtomicscaleOriginPiezoelectricity2015} enables excellent piezoelectric~\cite{kumarEnergyHarvestingBased2012}, 
pyroelectric properties~\cite{yangPyroelectricNanogeneratorsHarvesting2012} and memristance~\cite{laurenti2017}. Our group has also conducted controlled experiments to reveal how ZnO colloids can act as electrical-analogue neurons, demonstrating synaptic-like learning~\cite{kheirabadi2023neuromorphic,kheirabadi2023learning}, and Pavlovian reflexes~\cite{kheirabadi2022pavlovian}.

Proteinoids, or thermal proteins, on the other hand, are a significant area of investigation in the field of origin of life research~\cite{202311.0064}. These molecules, resembling proteins, have the ability to spontaneously emerge under simulated prebiotic conditions using amino acids. Examining proteinoids offers a valuable understanding of the potential origins of the initial rudimentary biomolecules on early Earth, thereby establishing the foundation for the development of life.  

Recent studies have demonstrated that when proteinoids are assembled in an aqueous solution, they can display intriguing computational capabilities~\cite{mougkogiannis2023light,mougkogiannis2023electrical} and indicate that proteinoids may have possessed the ability to do basic information processing before the emergence of the first living species, which is a significant step towards the development of life. Exploring the mechanisms by which proteinoids can store and manipulate information may uncover novel biomolecular computing models that can be applied in bio-inspired engineering~\cite{adamatzky2021towards},\cite{mougkogiannis2023transfer},~\cite{mougkogiannis2023logical},~\cite{mougkogiannis2023low}. Colloidal proteinoids, similar to neural networks, demonstrate the ability to process inputs in a manner that resembles biological computing. Exploiting the computational capacity of colloidal systems through the development of customised proteinoids could pave the way for novel and unconventional computing devices~\cite{adamatzky2014unconventional}. Additional examination of proteinoids will provide insights into prebiotic chemistry, biomolecular computing, and potentially a novel cohort of bio-inspired computational systems~\cite{fei2007bio}.

In order to formally assess the computational abilities of colloids beyond previous studies, we tested the implementation of Boolean logic in the colloidal material. We applied binary strings by electrically stimulating the liquid, recorded the resulting electrical output responses, classified them, and, finally, obtained the logical expressions. This approach is inspired by reservoir computing~\cite{daleSubstrateindependentFrameworkCharacterize2019} and in-materia computing~\cite{daleReservoirComputingModel2017,roberts2023logical} techniques that characterise the properties of computational substrates. The exact approach, however, is based on simplified, bare-bone techniques of extracting Boolean functions, developed in computational experiments with acting computing devices~\cite{adamatzky2019computing} and experimental laboratory implementation of fungal Boolean circuits~\cite{roberts2023mining}.

\section{Experimental details}
\subsection{Synthesis}

\textbf{Preparation of ZnO dispersion} Sodium dodecyl sulphate (SDS, Merck) was added to deionised water and stirred to create a homogeneous surfactant solution with a concentration of 0.22 wt.\%. Under stirring, 2 ml of the solution and 1 ml of NaOH (Reagent grade, Merck, $\geq$98\%) were added to 7 ml of dimethyl sulphoxide (DMSO, Pharmaceutical grade 99.9\%, Fisher Scientific). Then, 3 mg of ZnO nanoparticles (+99\%, 10-25 nm, Plasmachem) were added to the mixture while constant stirring, with a resulting dispersion concentration of 0.3 mg/ml. The resulting suspension was placed in an ultrasonic bath (DK Sonic Ultrasonic, \SI{40}{\kilo\hertz}) for 30 min. After ultrasonication, stirring was repeated for a few hours to ensure a homogeneous dispersion of ZnO.

\textbf{Preparation of proteinoids} Proteinoids were synthesised by mixing glycine (99.5\%, Sigma-Aldrich), alanine (99.5\%, Sigma-Aldrich), aspartate (99.5\%, Sigma-Aldrich), and glutamate (99.5\%, Sigma-Aldrich) in a reaction flask. This combination was thermally polymerised at \SI{433}{\kelvin}--\SI{453}{\kelvin} where these elevated temperatures activated condensation processes between amino acid amine and carboxyl moieties. These condensation processes, which form peptide bonds, polymerise polypeptide chains of different lengths into proteinoids. Polymerisation was stopped by cooling the reaction vessel to room temperature. Proteinoids precipitated from solution after the cooled reaction mixture was dissolved in \SI{353}{\kelvin} aqueous medium. Lyophilisation freeze-dries the precipitate to isolate the proteinoids from the aqueous phase. Prior to lyophilisation, the solid was washed to remove impurities and to recover the proteinoids. This multi-stage thermochemical process produces proteinoids, short protein-like polymer chains, from amino acid precursor monomers.

\textbf{Preparation of mixture} The mixture of ZnO and proteinoid was created by mixing half and half of each component, volumetric wise.

\subsection{Materials characterisation}
\textbf{Suspensions characterisation} The morphology of the suspensions was determined on a FEI Quanta 650 scanning electron microscope (SEM) at 2 kV. Room temperature absorbance of the samples was measured using a Perkin Elmer Lambda XLS ultraviolet-visible spectrometer. $I–V$ characteristics of the samples were analysed using a Keithley 2400 Source Measure Unit. Impedance and conductivity measurements were performed using a precision B\&K Precision 891 Benchtop LCR Meter at frequencies varying from 1 to \SI{300}{\kilo\hertz}. All electrical measurements were conducted using platinum/iridium electrodes with \SI{10}{\micro\metre} of diameter.

\begin{figure}[!tbp]
\centering
    \begin{subfigure}[c]{0.3\textwidth}
        \centering
        \caption{}
        \includegraphics[width=\textwidth, keepaspectratio]{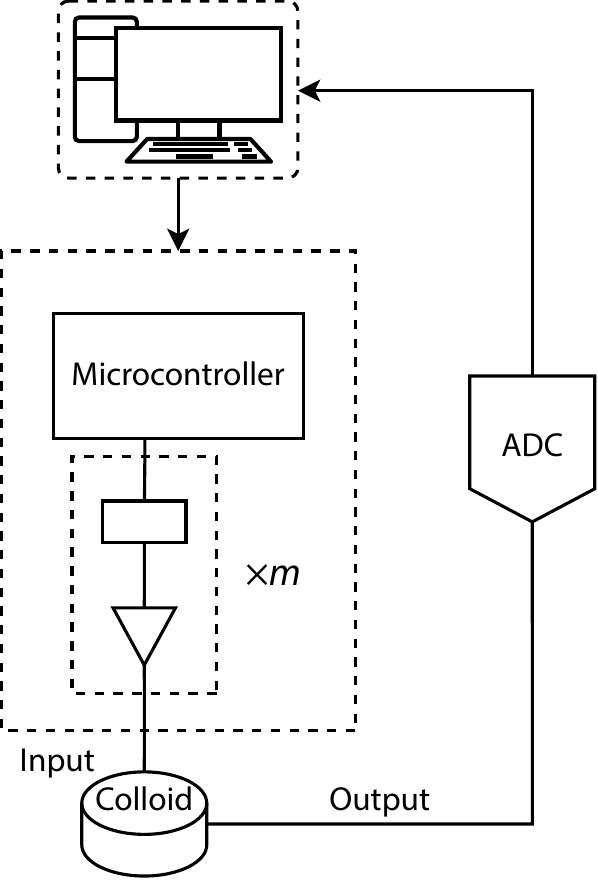}
        \label{fig:extraction_scheme}
    \end{subfigure}
    \hfill
    \begin{subfigure}[c]{0.3\textwidth}
        \centering
        \caption{}
        \includegraphics[width=\textwidth, keepaspectratio]{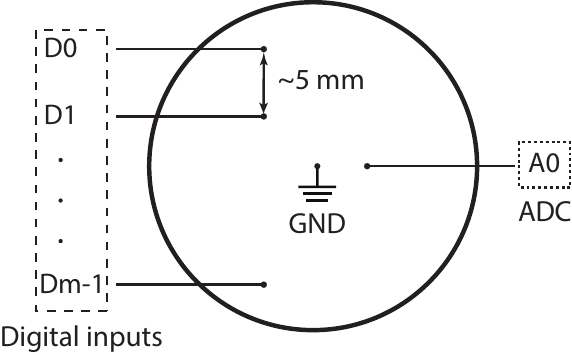}
        \label{fig:probs_placement}
    \end{subfigure}
    \hfill
    \begin{subfigure}[c]{0.375\textwidth}
        \centering
        \caption{}
        \includegraphics[width=\textwidth]{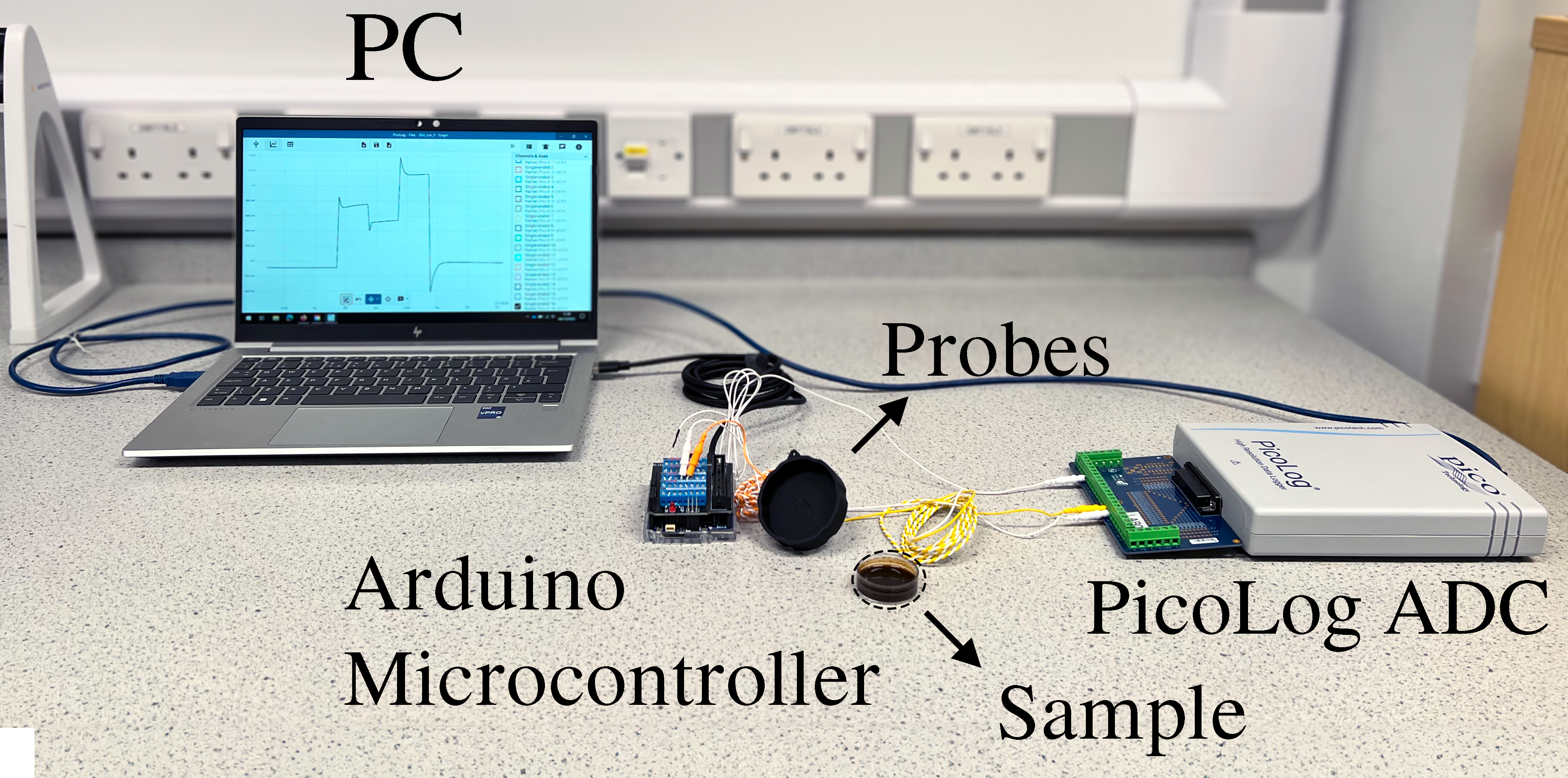}
        \label{fig:experimental_setup}
    \end{subfigure}
    \caption{(a) Logic circuit extraction experimental scheme, where ADC refers to the analogue-to-digital converter (b) probe placement in the samples, and (c) picture of the experimental setup for 2-bit extraction.}
    \label{fig:hardware}
\end{figure}

\textbf{Logical circuits extraction} 
The experimental approach to extracting logical functions is based on the theoretical methodology proposed in \cite{adamatzky2019computing,adamatzky2021towards}.
The logic gates were extracted using an in-house device constructed around a microcontroller board (Arduino Uno, Arduino), as illustrated in Fig.~\ref{fig:hardware}(a). The strings were encoded in line with the unipolar return-to-zero logic, wherein a logical 0 (false) was encoded as \SI{0}{V} and a logical 1 (true) as \SI{5}{V}. A series of 2, 4, 8, and 16 platinum/iridium electrodes, each with a diameter of \SI{10}{\micro\metre}, were positioned at a distance of approximately \SI{5}{\milli\metre} from one another in order to extract logic gate circuits of 2-, 4-, and 8-bit, respectively (as seen in Fig.~\ref{fig:hardware}(b)). Two additional platinum/iridium electrodes, also with a diameter of \SI{10}{\micro\metre} and separated by approximately \SI{5}{\milli\metre}, were placed in parallel to measure the output potential. The output electrodes were connected to a 24-bit analogue-to-digital converter (ADC-24, PICO Technology).

A series of binary strings ($b = \lbrace 0,1\rbrace ^m$, where $m$= 2, 4, 8, and 16) were applied to the colloidal mixture. In all binary strings, each bit was changed every \SI{15}{\second}. The output voltage was sampled at a frequency of \SI{1}{\hertz}. Fig.~\ref{fig:hardware}

\section{Results and discussion}
\subsection{Structural and electronic properties of the colloidal mixture}
\begin{figure}[!tbp]
    \centering
    \includegraphics[width=0.5\textwidth, keepaspectratio]{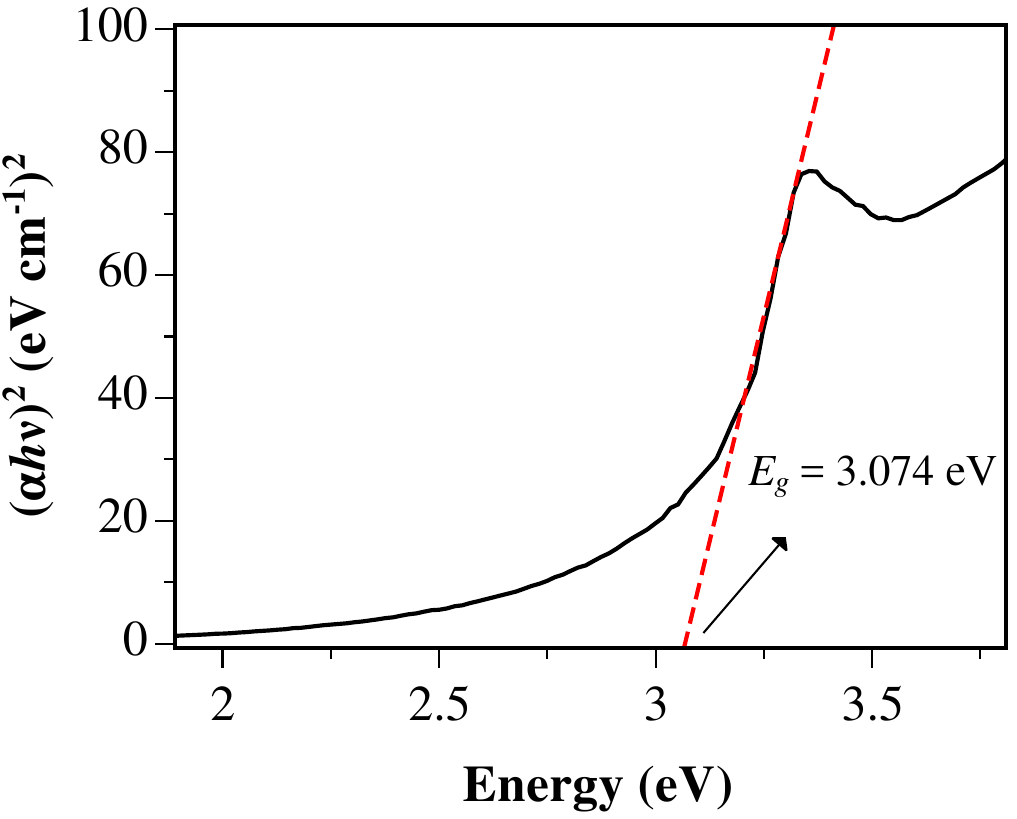}
    \caption{Tauc plot of ZnO.}
    \label{fig:tauc_plot}
\end{figure}

UV-vis spectroscopy was used to study the optical properties of the ZnO semiconductor. The Tauc method~\cite{makulaHowCorrectlyDetermine2018} was employed to calculate the optical bandgap, which is based on the following equation
\begin{equation}
    (\alpha h\nu)^{1/n} = B (h\nu-E_g),
    \label{eq:kubelka-munk}
\end{equation}
where $h$ is the Planck constant, $\nu$ is the photon frequency, $E_g$ is the bandgap, and $B$ is a constant. The value of $n$ is determined by the nature of the carrier transition, being equal to 1/2 or 2 for direct and indirect bandgaps, respectively. Since ZnO is a direct bandgap semiconductor~\cite{huangRoomTemperatureUltravioletNanowire2001, reynoldsValencebandOrderingZnO1999}, $n = 1/2$ was used. The optical bandgap is then estimated from a linear fit of $(\alpha h\nu)^{1/n}$ against the photon energy ($h\nu$) near the absorption edge. Fig.~\ref{fig:tauc_plot} shows the reflectance spectrum of ZnO mapped according to Eq.~\ref{eq:kubelka-munk} plotted against the photon energy. The linear fit of the data indicates a bandgap of $\sim$\SI{3.074}{\electronvolt}, which is in agreement with the reported values of 3.1 -- 3.37 eV~\cite{srikantOpticalBandGap1998, saenz-trevizoOpticalBandGap2016, davisBandGapEngineered2019}.

\begin{figure}[!tbp]
    \centering
    \includegraphics[width=0.5\textwidth, keepaspectratio]{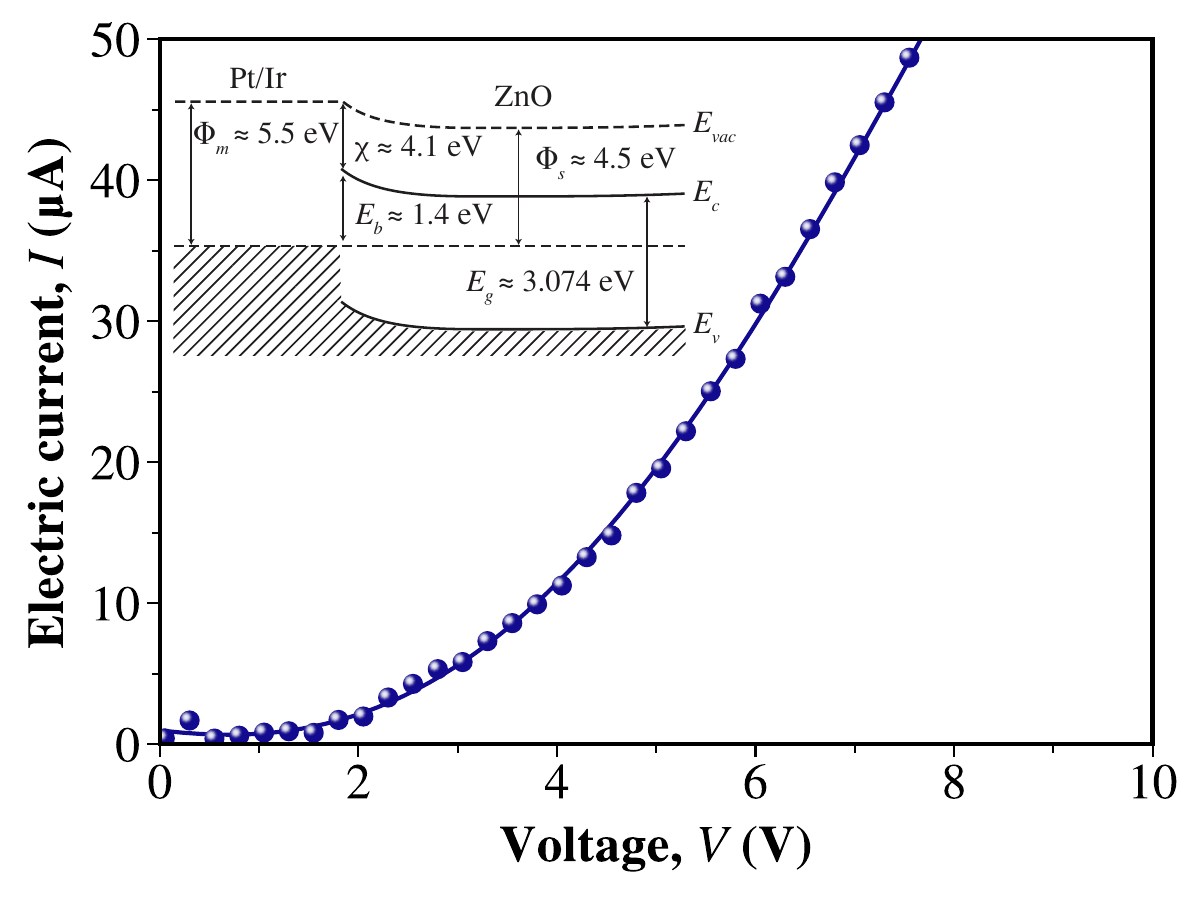}
    \caption{I quadrant portion of the $I-V$ curve recorded from the colloidal solution. The inset represents the proposed band structure of the Pt/Ir contact with the ZnO nanoparticles.}
    \label{fig:iv-curve}
\end{figure}

The $I-V$ characteristics of the colloidal suspension are shown in Fig.~\ref{fig:iv-curve}. The exponential-like shape of the $I-V$ curve demonstrates the occurrence of a Schottky barrier between the Pt/Ir probes and the ZnO nanoparticles~\cite{tungPhysicsChemistrySchottky2014}. In the inset in Fig.~\ref{fig:iv-curve}, the proposed band structure of the metal/semiconductor contact is drawn. The values for the work functions of the Pt/Ir alloy ($\Phi_m\approx \SI{5.5}{\electronvolt}$~\cite{reddyPoly4ethylenedioxypyrroleAu2015}) and ZnO ($\Phi_s\approx $~\cite{sundaramWorkFunctionDetermination1997}), and electron affinity of ZnO ($\chi\approx \SI{4.1}{\electronvolt}$~\cite{hussainElectronAffinityBandgap2019}) were obtained from literature, while the bandgap was experimentally measured as described above. The barrier height of $E_b = \SI{1.4}{\electronvolt}$ was estimate using Schottky–Mott rule ($E_b\approx \Phi_m- \chi$)~\cite{parkSchottkyMottRule2021}. This nonlinearity brought by the metal/semiconductor junction is favourable for mining more complex logic gates from the material~\cite{tadokoroHighlySensitiveImplementation2021, kiaNonlinearDynamicsBased2015}.

\begin{figure}[h]
    \centering
    \begin{subfigure}[c]{0.45\textwidth}
        \centering
        \caption{}
        \includegraphics[width=1\textwidth, keepaspectratio]{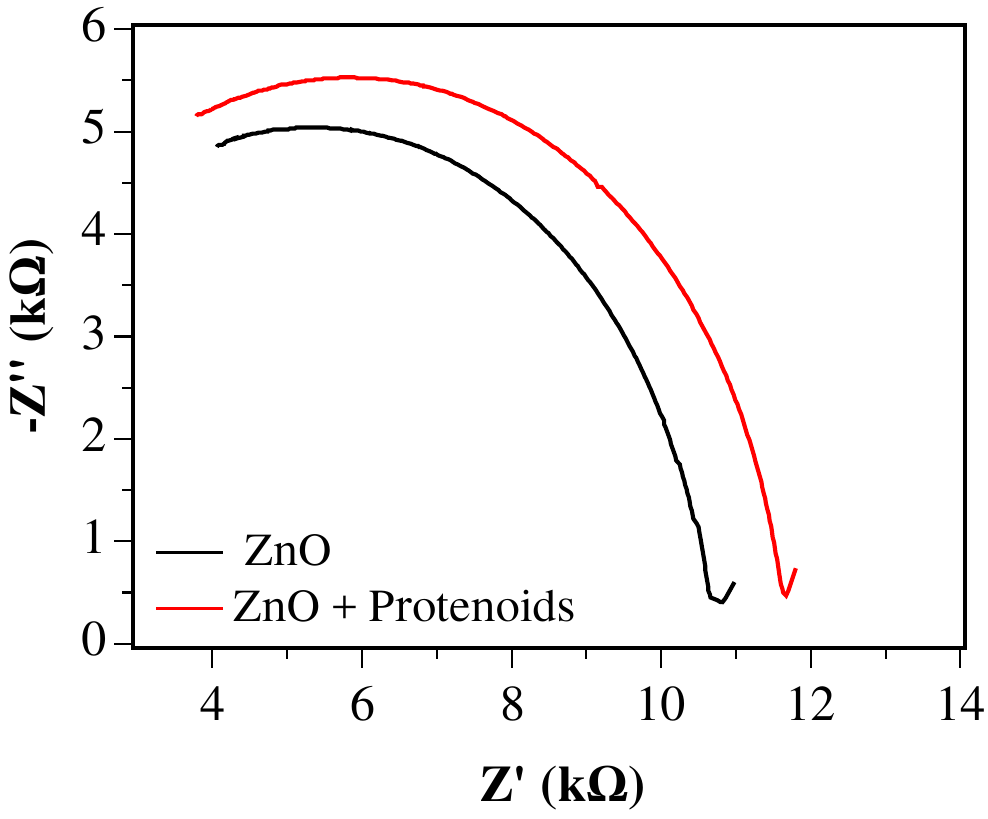}
        \label{fig:cole1}
    \end{subfigure}
    \hfill
    \begin{subfigure}[c]{0.45\textwidth}
        \centering
        \includegraphics[width=1\textwidth, keepaspectratio]{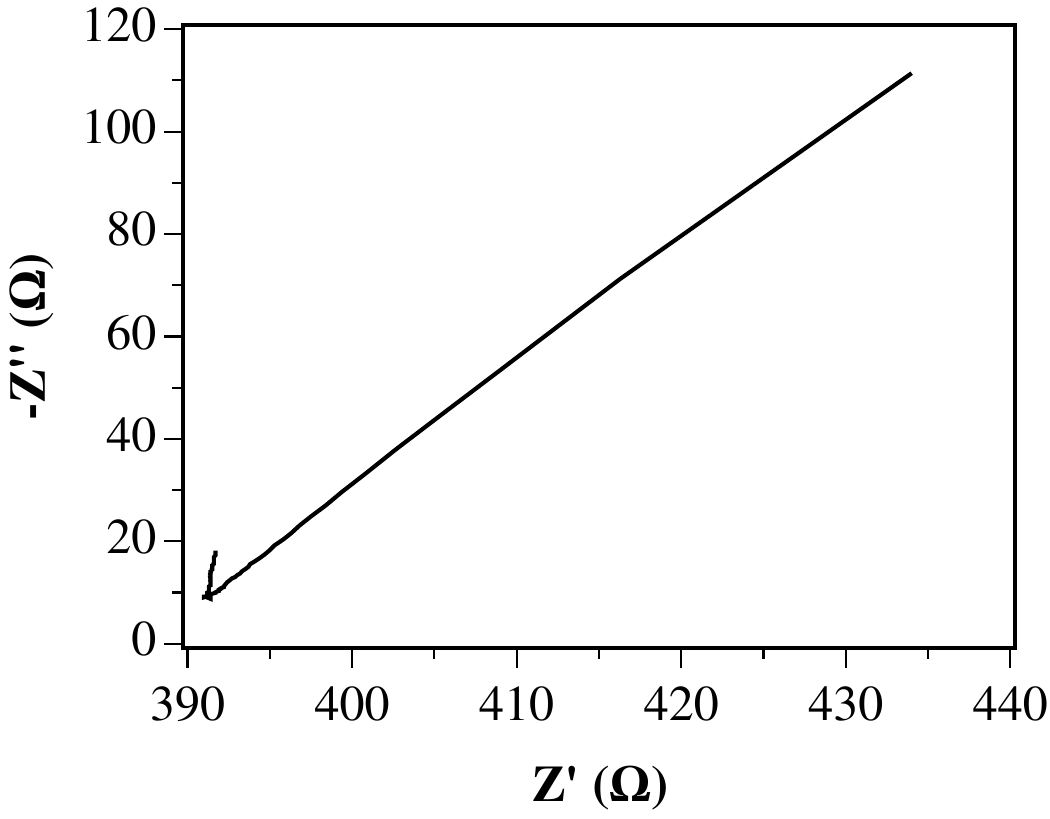} 
        \caption{}
        \label{fig:cole2}
    \end{subfigure}
    \caption{Cole-Cole plots for (a) ZnO colloidal suspension and ZnO + proteinoids and (b) proteinoids.}
    \label{fig:cole-cole}
\end{figure}

\begin{figure}[!h]
    \centering
    \includegraphics[width=0.5\textwidth, keepaspectratio]{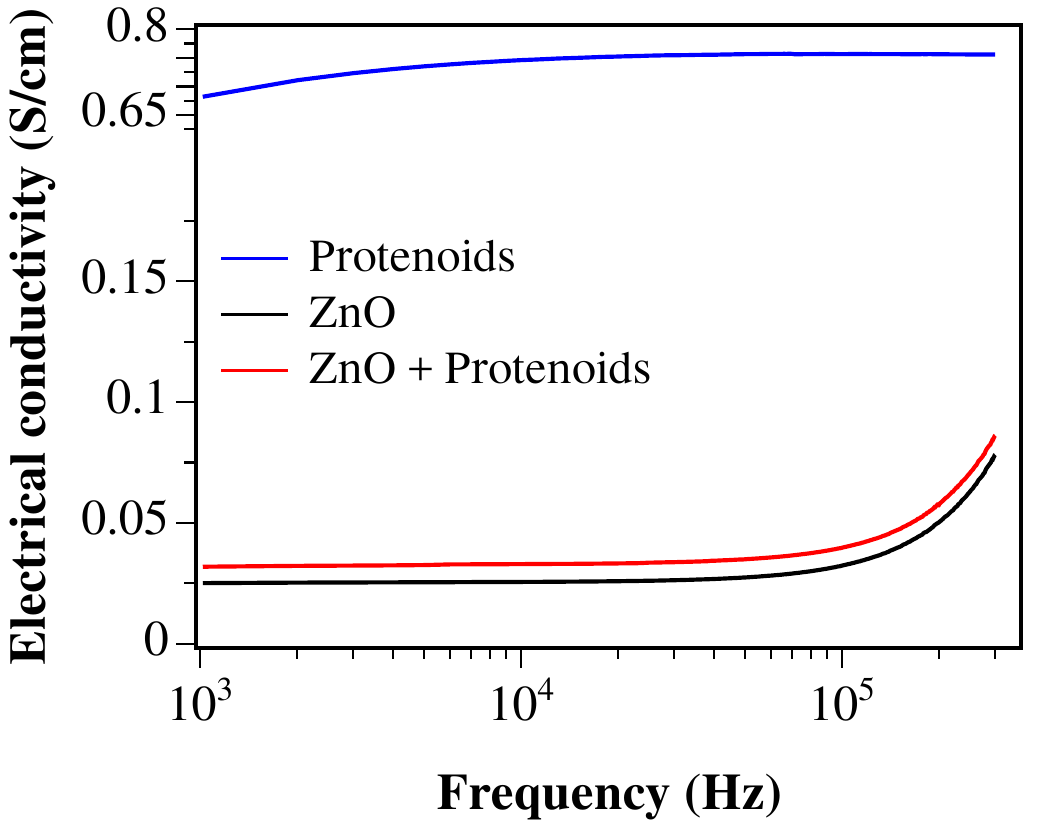}
    \caption{Frequency dependency of the electrical conductivity of proteinoids, ZnO colloidal suspension, and mixture of ZnO and proteinoids.}
    \label{fig:sigma}
\end{figure}

\FloatBarrier
Cole-Cole plots obtained from impedance spectroscopy revealed an ionic conductivity mechanism for the sample containing only proteinoid~\cite{mohamedInvestigationTransportMechanism2022, mohamadisaIonicConductivityConduction2014} (see Fig.~\ref{fig:cole1}), while the sample containing ZnO showed a depressed semicircle with an spike at lower frequencies possible due to the polarisation of the electrodes~\cite{wibowoZnONanostructuredMaterials2020} (see Fig.~\ref{fig:cole2}). The frequency dependence of the electrical conductivity ($\sigma = L/(R\cdot A)$, where $R$ is the resistance, $L$ is the length between the electrodes, and $A$ is the electrode area) for all samples is shown in Fig~\ref{fig:sigma}. The inclusion of proteinoid in the ZnO colloidal suspension increases the bulk electrical conductivity, which is in agreement with the inclusion of a more conductive phase in the material. The electrical conductivity of the proteinoid exhibits a very weak frequency dependence, in contrast with the samples containing ZnO. In terms of this work, this disparity is beneficial as it shows that the different materials change the frequency response and, as it is shown later, their response to electric stimuli in binary strings.

\begin{figure}[!tbp]
    \centering
    \begin{subfigure}[c]{0.32\textwidth}
        \centering
        \caption{}
        \includegraphics[width=\textwidth, keepaspectratio]{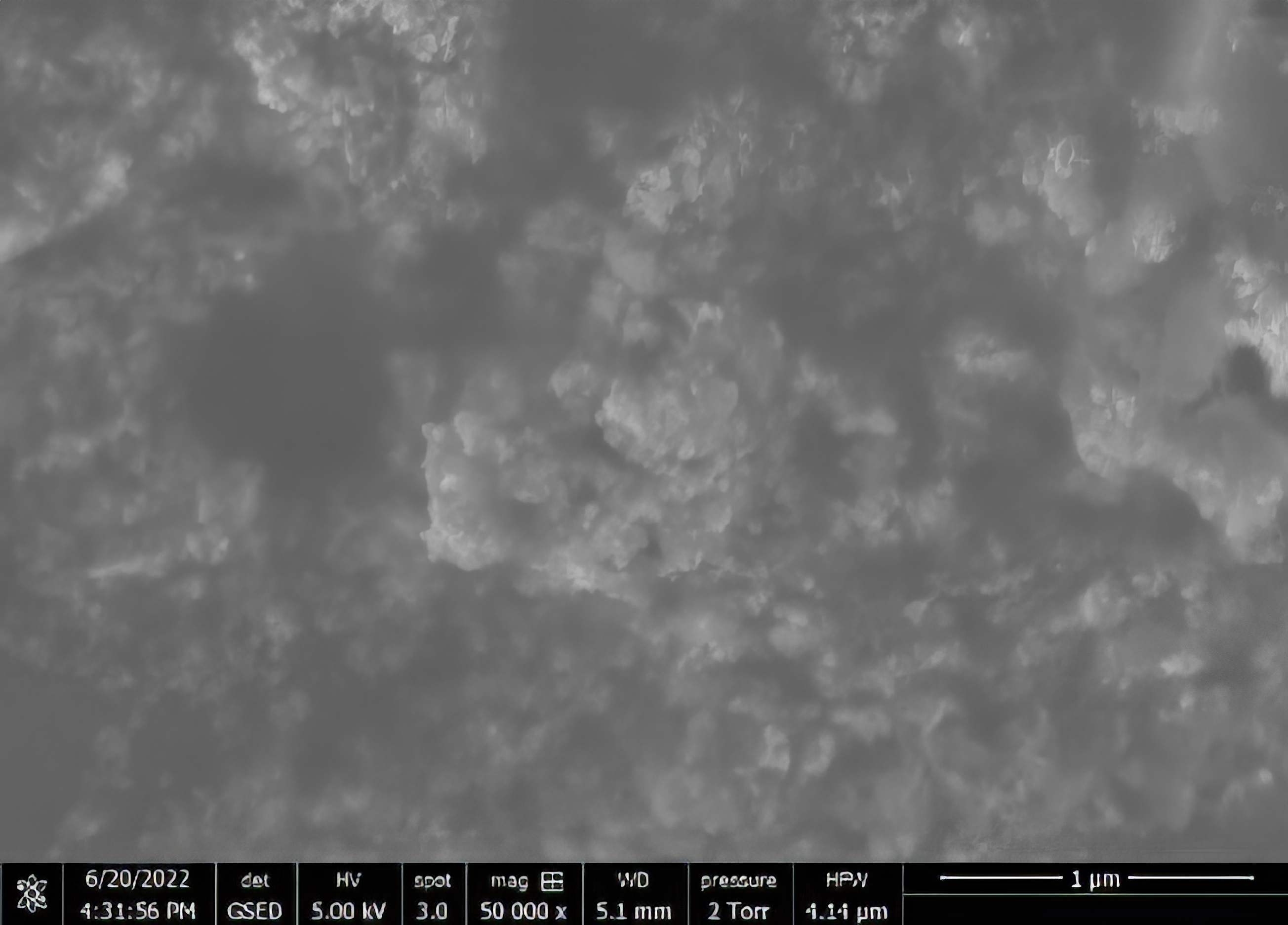}
        \label{fig:sem_image_1}
    \end{subfigure}
    \hfill
    \begin{subfigure}[c]{0.32\textwidth}
         \centering
         \caption{}
         \includegraphics[width=\textwidth, keepaspectratio]{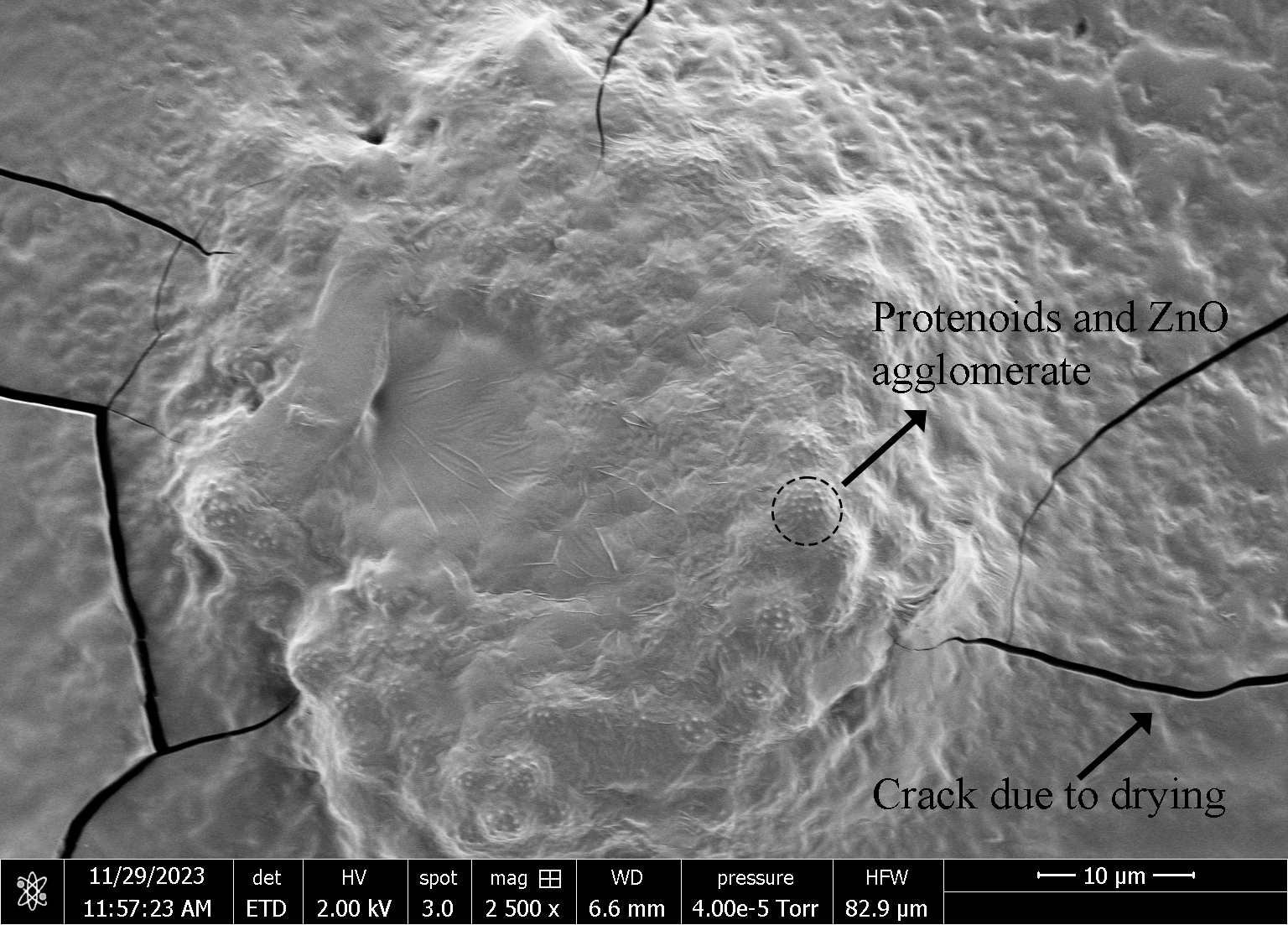}
         \label{fig:sem_image_2}
    \end{subfigure}
    \hfill
    \begin{subfigure}[c]{0.32\textwidth}
        \centering
        \caption{}
        \includegraphics[width=1\textwidth, keepaspectratio]{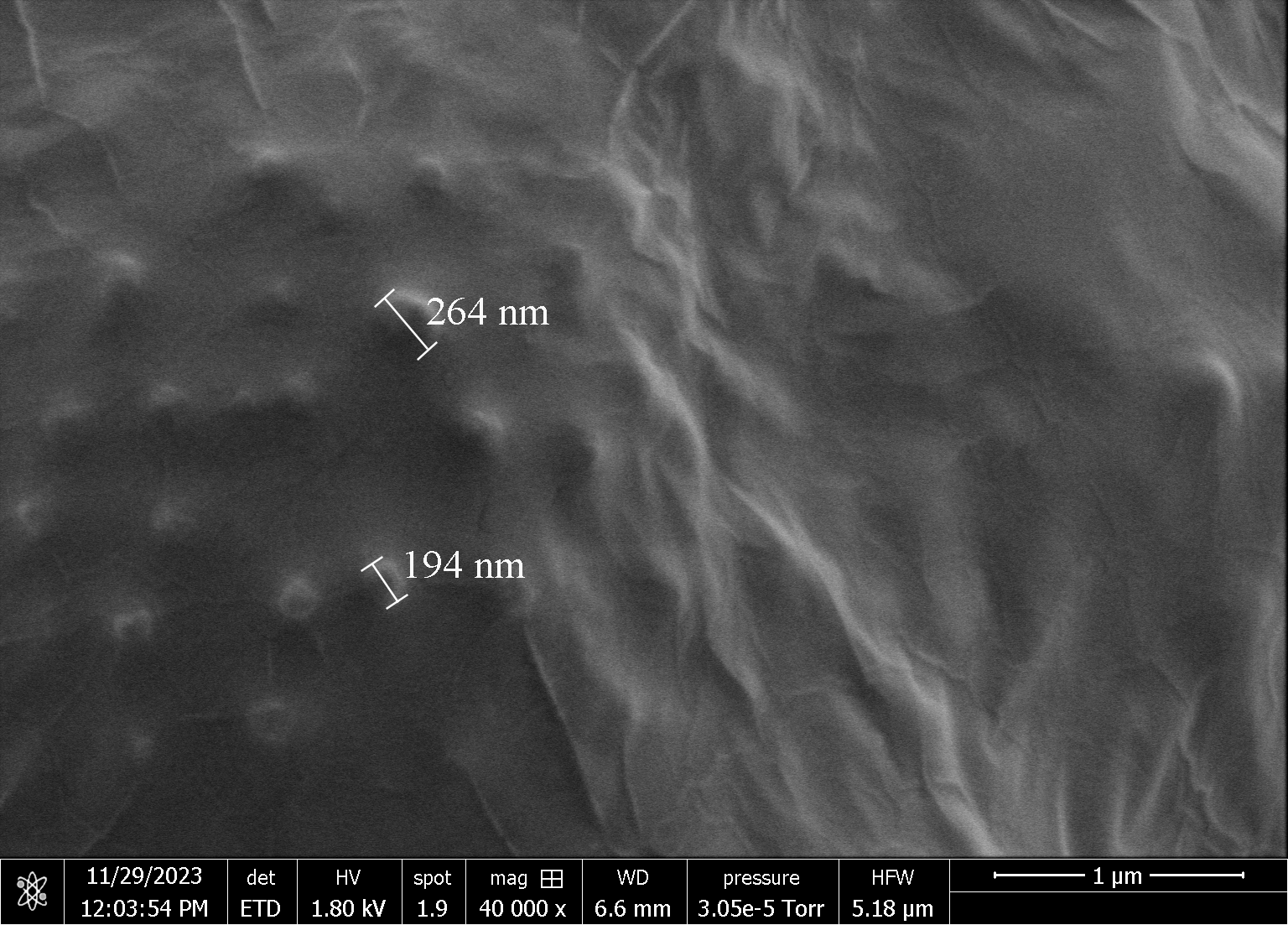}
        \label{fig:sem_image_5}
    \end{subfigure}
    \hfill
    \begin{subfigure}[c]{0.32\textwidth}
        \centering
        \caption{}
        \includegraphics[width=\textwidth, keepaspectratio]{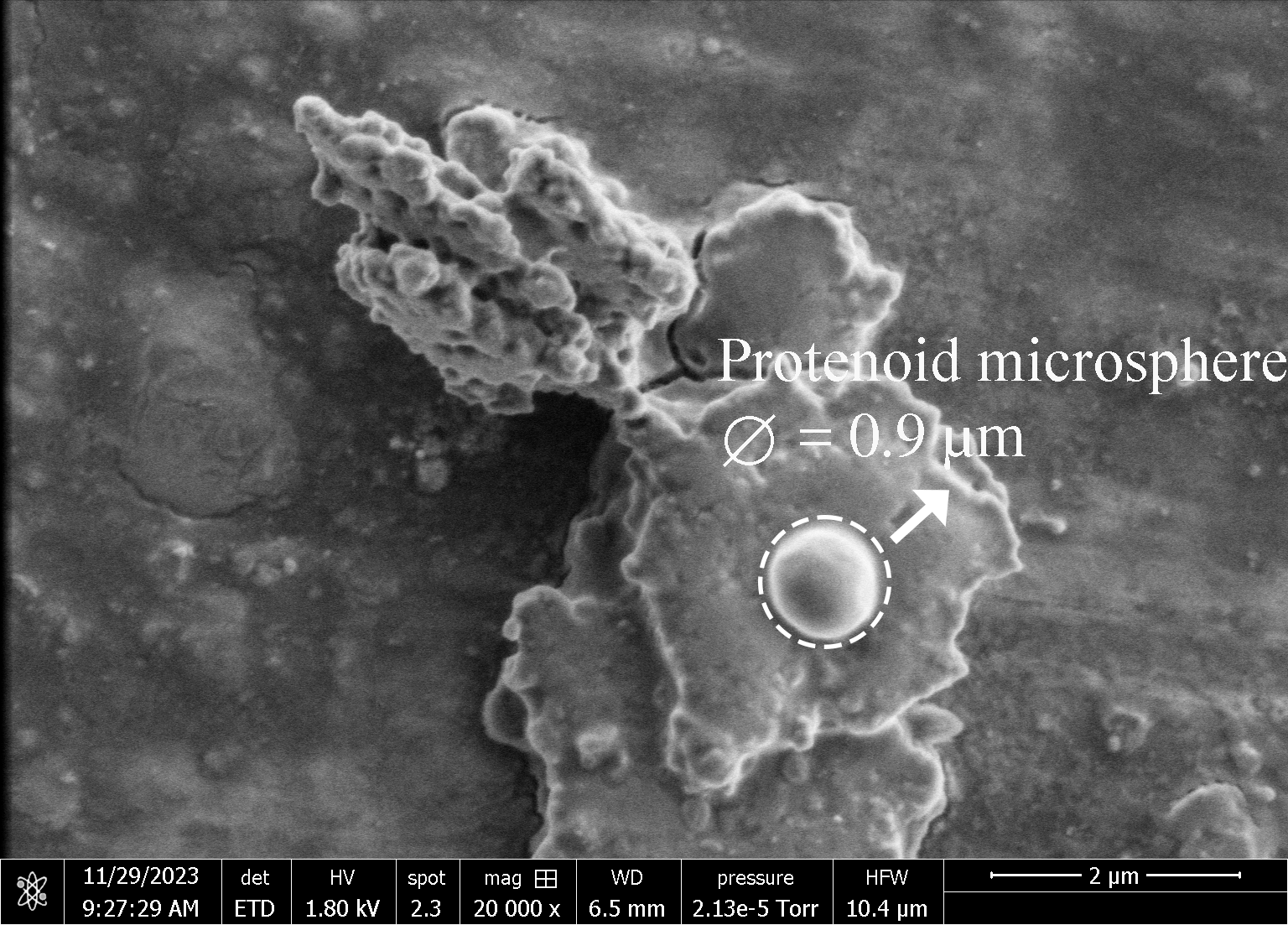}
        \label{fig:sem_image_3}
    \end{subfigure}
        \hfill
    \begin{subfigure}[c]{0.32\textwidth}
        \centering
        \caption{}
        \includegraphics[width=\textwidth, keepaspectratio]{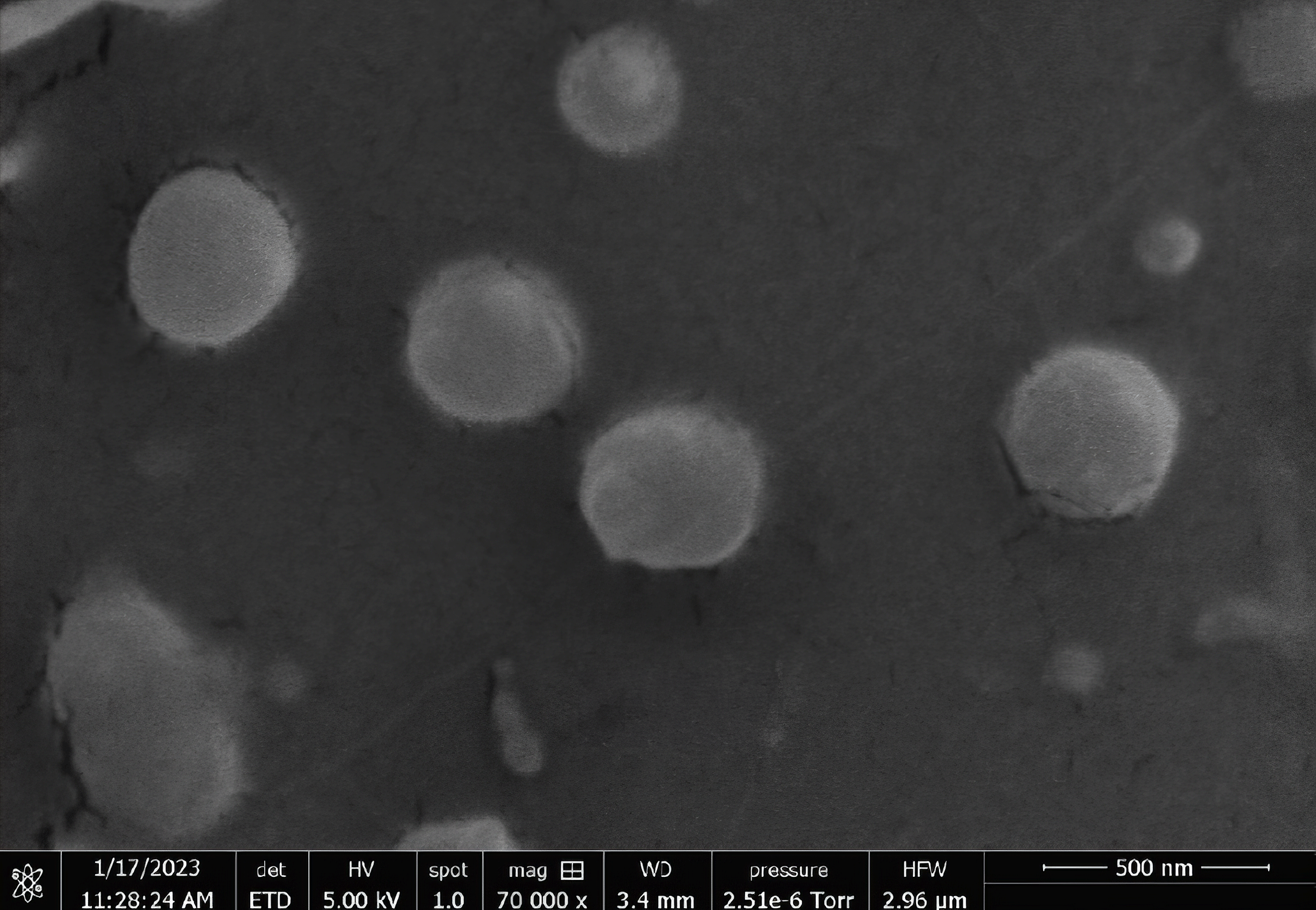}
        \label{fig:sem_image_4}
    \end{subfigure}
        \caption{Scanning electron microscopy images for (a) ZnO colloidal suspension, (b) mixture of ZnO + proteinoids, (c) zoomed at region rich in ZnO, (d) proteinoids microspheres in ZnO + protenoids, and (e) proteinoids.}
        \label{fig:sem_images}
\end{figure}

The samples were drop-casted onto a Cu substrate of $\sim\SI{200}{\micro\metre}$ at room temperature and left to dry for around 4 days to fabricate a thin layer of the mixture for SEM imaging. SEM images reveal the agglomeration of ZnO nanoparticles during sample preparation (see Fig.~\ref{fig:sem_image_1}. It is likely that surface tension causes the nanoparticles to realign as the solvent evaporates, resulting in most ZnO particles appearing layered. In the sample consisting of a mixture of ZnO and proteinoids, large proteinoid particles were observed to clump together with spots of agglomerated ZnO (Figs.~\ref{fig:sem_image_2} and~\ref{fig:sem_image_5}) and proteinoid microspheres~\ref{fig:sem_image_3}. The cracks depicted in Fig.~\ref{fig:sem_image_2} were a natural result of the drying process. Dispersed microspheres protenoids were observed for the sample containing just proteinoids (Fig.~\ref{fig:sem_image_4}).

\begin{figure}[!tbp]
    \centering
    \includegraphics[width=.5\textwidth, keepaspectratio]{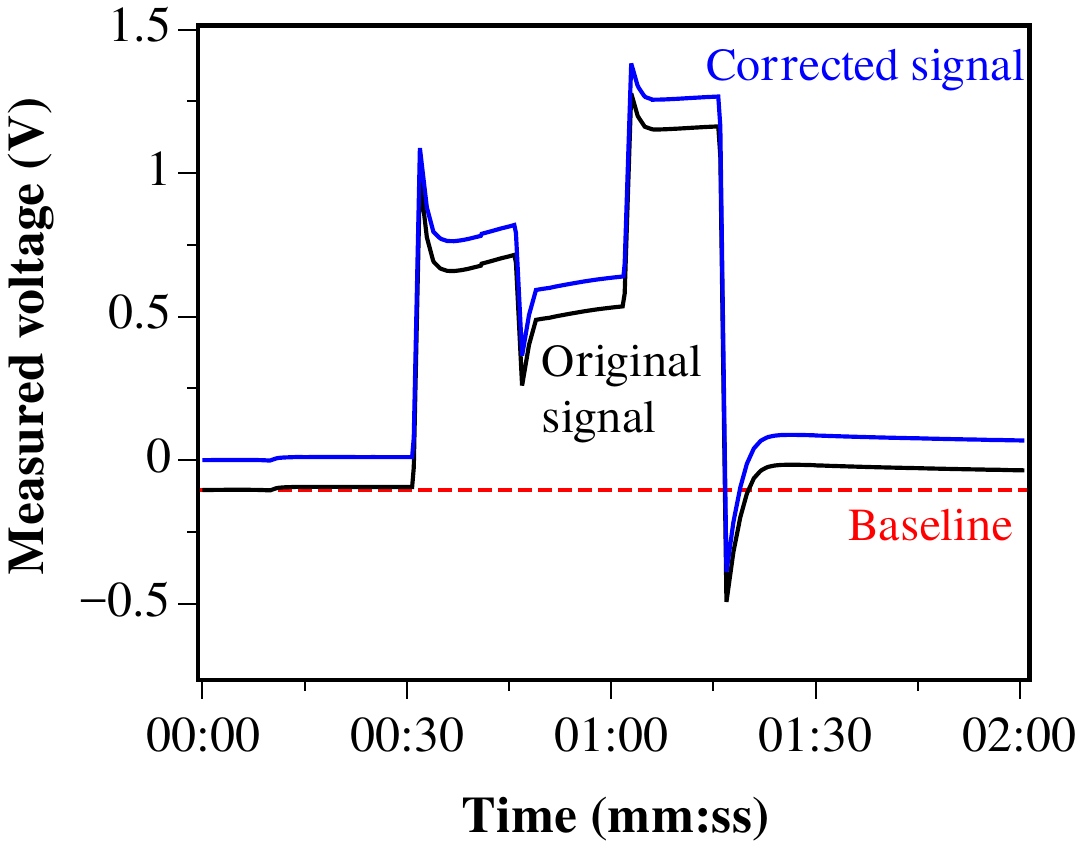}
    \caption{Measured output voltage and baseline correction of a ZnO and proteinoid mixture for a 2-bit string input.}
    \label{fig:2bit-voltage}
\end{figure}

\subsection{Logical circuits extraction}
The measured voltage output of the binary string $b = \lbrace 0,1\rbrace^2$ is depicted in Fig.~\ref{fig:2bit-voltage}. The measurements showed a constant DC bias, as can be seen in Fig.~\ref{fig:2bit-voltage}. The baseline was corrected by fitting a first-order polynomial using the procedure as described in~\cite{mazetBackgroundRemovalSpectra2005}.

\begin{figure}[!tbp]
     \centering
     \begin{subfigure}[c]{1\textwidth}
         \centering
         \caption{}
         \includegraphics[width=\textwidth, keepaspectratio]{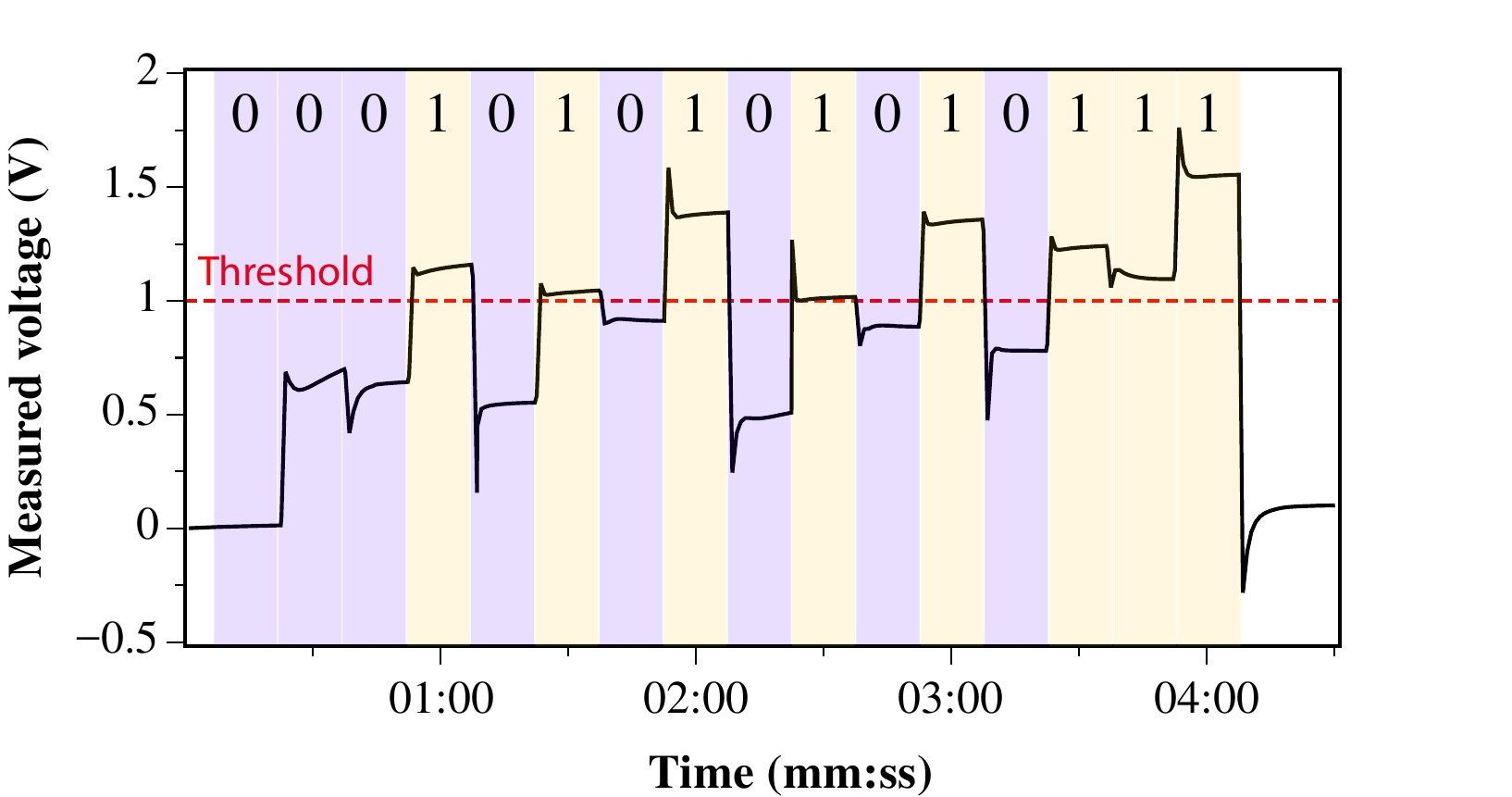}
         \label{fig:classification}
     \end{subfigure}
     \hfill
    \begin{subfigure}[c]{.45\linewidth}
        \centering
        \caption{}
        \begin{tabular}{@{}llllll@{}}
            \toprule
               & A & B & C & D & F \\ \midrule
            0  & 0 & 0 & 0 & 0 & 0 \\
            1  & 0 & 0 & 0 & 1 & 0 \\
            2  & 0 & 0 & 1 & 0 & 0 \\
            3  & 0 & 0 & 1 & 1 & 1 \\
            4  & 0 & 1 & 0 & 0 & 0 \\
            5  & 0 & 1 & 0 & 1 & 1 \\
            6  & 0 & 1 & 1 & 0 & 0 \\
            7  & 0 & 1 & 1 & 1 & 1 \\
            8  & 1 & 0 & 0 & 0 & 0 \\
            9  & 1 & 0 & 0 & 1 & 1 \\
            10 & 1 & 0 & 1 & 0 & 0 \\
            11 & 1 & 0 & 1 & 1 & 1 \\
            12 & 1 & 1 & 0 & 0 & 0 \\
            13 & 1 & 1 & 0 & 1 & 1 \\
            14 & 1 & 1 & 1 & 0 & 1 \\
            15 & 1 & 1 & 1 & 1 & 1 \\ \bottomrule
        \end{tabular}
        \label{tab:truth}
    \end{subfigure}
    \begin{subfigure}[c]{0.45\textwidth}
        \centering
        \caption{}
        \includegraphics[width=\textwidth, keepaspectratio]{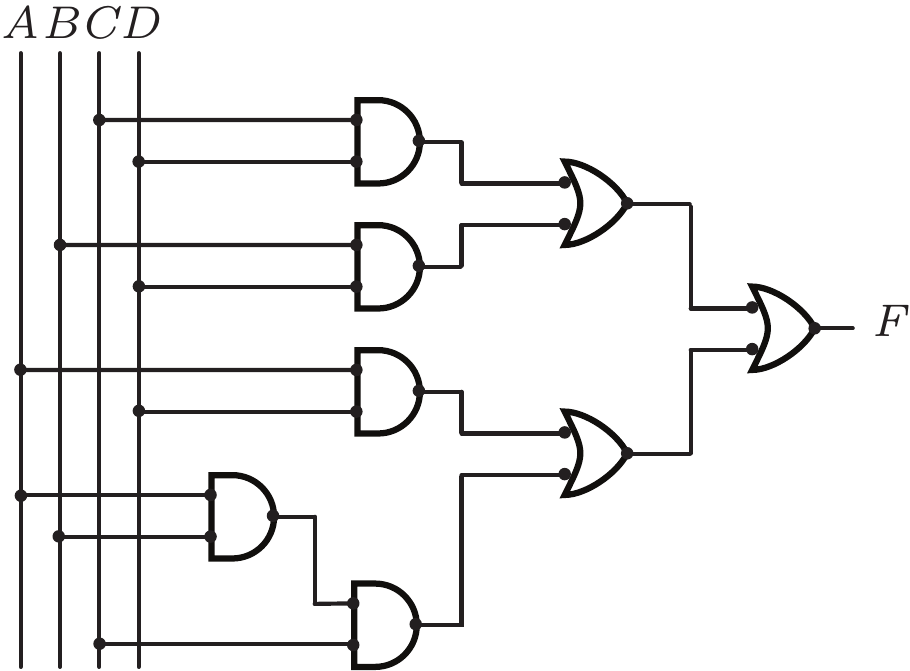}
        \label{fig:sop_circuit}
     \end{subfigure}
        \caption{(a) Pulse classification for a 4-bit string, (b) resulting truth table for classified pulses, and (c) extracted sum-of-products logic Boolean expression and realised logical circuit.}
        \label{fig:workflow}
\end{figure}

Logical circuits were extracted from the recorded data by classifying the measured electrical pulses according to a thresholding procedure. Each detected pulse was assigned a logical 1 (true) if its amplitude was greater than or equal to the selected threshold, or a logical 0 (false) if it was below the threshold. A total of 38 different thresholds were used, ranging from 5\% of the peak voltage to 100\% in 2.5\% increments. The classified binary pulse sequences were stored in truth tables, with each row representing a unique input-output combination observed in the experiment. The minimal sum-of-products (SOP) Boolean logic expression matching each truth table was then determined using the well-established Espresso logic minimisation algorithm~\cite{rudellMultipleValuedMinimizationPLA1987, braytonComparisonLogicMinimization1982}. From the computed minimal SOP expressions, the corresponding digital logic circuit was then realised. Fig.~\ref{fig:workflow} outlines the entire workflow, from pulse measurement to logic circuit extraction. Fig.~\ref{fig:classification} displays the classification procedure for a 4-bit string input, Fig.~\ref{fig:classification} shows the extracted truth table, and Fig.~\ref{fig:sop_circuit} shows the realised logic circuit.

\FloatBarrier
\section{ZnO}
Without repetition, a total of 2, 5, and 27 unique standard canonical SOP Boolean logical expressions were obtained for the experiments with 2-, 4-, and 8-bit inputs, respectively. Tables~\ref{tab:sop_2bit_ZnO}, \ref{tab:sop_4bit_ZnO}, and \ref{tab:sop_8bit_ZnO} show the four most frequent logical expressions identified for each input bit width. As anticipated, the logical complexity of the most common expressions increased with the number of input variables. For 2-bit inputs, the most frequent expression was a simple disjunctive (OR) term. However, for 4-bit inputs, the most common expressions were composed of four disjunctive terms. The most prevalent expressions for 8-bit inputs consisted of seven disjunctive terms and negated variables.
\begin{table}[!h]
    \centering
    \caption{Extracted sum-of-products (SOP) Boolean expression for a 2-bit string input with varying thresholds for ZnO colloidal suspension}
    \label{tab:sop_2bit_ZnO}
        \begin{tabular}{lr}
            \toprule
            SOP                            & Count   \\ \midrule
            $A\lor B$                      & 35      \\
            $A\land B$                     & 3       \\ \bottomrule
        \end{tabular}
\end{table}

\begin{table}[!h]
    \centering
    \caption{Extracted SOP Boolean expression for a 4-bit string input with varying thresholds for ZnO colloidal suspension}
    \label{tab:sop_4bit_ZnO}
    \begin{tabular}{lr}
        \toprule
        SOP                                                                            & Count  \\ \midrule
        $A \lor B \lor C \lor D$                                                       & 16     \\
        $(A \land B \land D \land \neg C) \lor (C \land D \land \neg  A \land \neg B)$ & 6      \\ 
        $(C \land D) \lor (A \land B \land D)$                                        & 4      \\ 
        $A \lor B \lor D$                                                              & 2      \\
        \bottomrule
    \end{tabular}
\end{table}

\begin{table}[!h]
    \centering
    \caption{Extracted SOP Boolean expression for a 8-bit string input with varying thresholds for ZnO colloidal suspension}
    \label{tab:sop_8bit_ZnO}
    \resizebox{1\textwidth}{!}{
        \begin{tabular}{lr}
            \toprule
            SOP                            & Count   \\ \midrule
            $\neg A \lor \neg B \lor \neg C \lor \neg D \lor \neg E \lor \neg F \lor \neg G$       & 4     \\
            $A \land C \land D \land E \land F \land G \land H \land \neg B$ & 2 \\ 
            $A \lor C \lor D \lor E \lor F \lor H \lor (B \land \neg G) \lor (G \land \neg B)$ & 1 \\ 
            $C \lor D \lor E \lor F \lor H \lor (A \land B) \lor (A \land \neg G) \lor (B \land \neg G) \lor (G \land \neg A \land \neg B)$ & 1\\
            \bottomrule
        \end{tabular}
    }
\end{table}

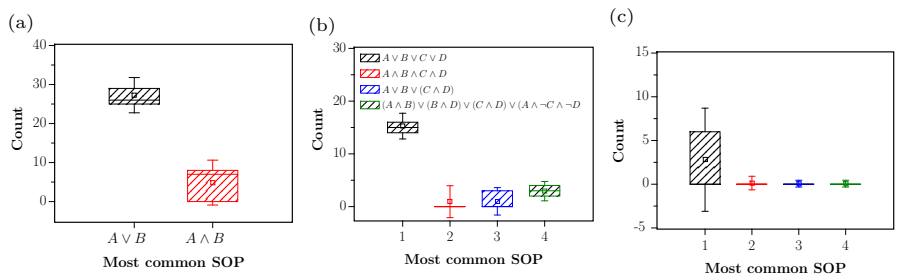
\begin{figure}[!tbp]
    \centering
    \begin{subfigure}[c]{0.32\textwidth}
        \centering
        \caption{}
        \resizebox{1\textwidth}{!}{
         \begin{tikzpicture}{0pt}{0pt}{481.8pt}{395.477pt}
	\color[rgb]{1,1,1}
	\fill(0pt,395.477pt) -- (481.8pt,395.477pt) -- (481.8pt,0pt) -- (0pt,0pt) -- (0pt,395.477pt);
\begin{scope}
	\clip(76.285pt,383.432pt) -- (476.781pt,383.432pt) -- (476.781pt,83.3112pt) -- (76.285pt,83.3112pt) -- (76.285pt,383.432pt);
	\definecolor{c}{rgb}{0,0,0}
	\fill [pattern color=c, pattern=north east lines](168.162pt,309.043pt) -- (252.744pt,309.043pt) -- (252.744pt,282.723pt) -- (168.162pt,282.723pt) -- (168.162pt,309.043pt);
	\color[rgb]{0,0,0}
	\draw[line width=1.5pt, line join=miter, line cap=rect](168.162pt,309.043pt) -- (252.744pt,309.043pt) -- (252.744pt,282.723pt) -- (168.162pt,282.723pt) -- (168.162pt,309.043pt);
	\color[rgb]{0,0,0}
	\draw[line width=1.5pt, line join=bevel, line cap=rect](210.453pt,327.358pt) -- (210.453pt,309.043pt);
	\draw[line width=1.5pt, line join=bevel, line cap=rect](210.453pt,267.918pt) -- (210.453pt,282.723pt);
	\draw[line width=1.5pt, line join=bevel, line cap=rect](201.995pt,267.918pt) -- (218.911pt,267.918pt);
	\draw[line width=1.5pt, line join=bevel, line cap=rect](201.995pt,327.358pt) -- (218.911pt,327.358pt);
	\draw[line width=1.5pt, line join=bevel, line cap=rect](168.162pt,289.303pt) -- (252.744pt,289.303pt);
	\draw[line width=1pt, line join=miter, line cap=rect](206.772pt,301.125pt) -- (213.799pt,301.125pt) -- (213.799pt,294.099pt) -- (206.772pt,294.099pt) -- (206.772pt,301.125pt);
	\definecolor{c}{rgb}{1,0,0}
	\fill [pattern color=c, pattern=north east lines](300.322pt,170.861pt) -- (384.905pt,170.861pt) -- (384.905pt,118.219pt) -- (300.322pt,118.219pt) -- (300.322pt,170.861pt);
	\color[rgb]{1,0,0}
	\draw[line width=1.5pt, line join=miter, line cap=rect](300.322pt,170.861pt) -- (384.905pt,170.861pt) -- (384.905pt,118.219pt) -- (300.322pt,118.219pt) -- (300.322pt,170.861pt);
	\draw[line width=1.5pt, line join=miter, line cap=rect](342.613pt,188.08pt) -- (342.613pt,170.861pt);
	\draw[line width=1.5pt, line join=miter, line cap=rect](342.613pt,112.406pt) -- (342.613pt,118.219pt);
	\draw[line width=1.5pt, line join=miter, line cap=rect](334.155pt,112.406pt) -- (351.072pt,112.406pt);
	\draw[line width=1.5pt, line join=miter, line cap=rect](334.155pt,188.08pt) -- (351.072pt,188.08pt);
	\draw[line width=1.5pt, line join=miter, line cap=rect](300.322pt,164.28pt) -- (384.905pt,164.28pt);
	\draw[line width=1pt, line join=miter, line cap=rect](339.267pt,153.574pt) -- (346.294pt,153.574pt) -- (346.294pt,146.547pt) -- (339.267pt,146.547pt) -- (339.267pt,153.574pt);
\end{scope}
\begin{scope}
	\color[rgb]{0,0,0}
	\draw[line width=2pt, line join=miter, line cap=rect](76.285pt,383.432pt) -- (476.781pt,383.432pt) -- (476.781pt,83.3112pt) -- (76.285pt,83.3112pt) -- (76.285pt,383.432pt);
	\color[rgb]{0,0,0}
	\pgftext[center, base, at={\pgfpoint{21.6277pt}{241.402pt}},rotate=90]{\fontsize{25}{0}\selectfont{\textbf{Count}}}
	\pgftext[center, base, at={\pgfpoint{57.2137pt}{112.734pt}}]{\fontsize{24}{0}\selectfont{0}}
	\pgftext[center, base, at={\pgfpoint{51.1912pt}{177.977pt}}]{\fontsize{24}{0}\selectfont{10}}
	\pgftext[center, base, at={\pgfpoint{51.1912pt}{244.225pt}}]{\fontsize{24}{0}\selectfont{20}}
	\pgftext[center, base, at={\pgfpoint{51.1912pt}{309.469pt}}]{\fontsize{24}{0}\selectfont{30}}
	\pgftext[center, base, at={\pgfpoint{51.1912pt}{375.716pt}}]{\fontsize{24}{0}\selectfont{40}}
	\draw[line width=1pt, line join=bevel, line cap=rect](76.285pt,151.12pt) -- (71.2662pt,151.12pt);
	\draw[line width=1pt, line join=bevel, line cap=rect](76.285pt,216.921pt) -- (71.2662pt,216.921pt);
	\draw[line width=1pt, line join=bevel, line cap=rect](76.285pt,282.723pt) -- (71.2662pt,282.723pt);
	\draw[line width=1pt, line join=bevel, line cap=rect](76.285pt,348.524pt) -- (71.2662pt,348.524pt);
	\draw[line width=1pt, line join=bevel, line cap=rect](76.285pt,118.219pt) -- (67.2512pt,118.219pt);
	\draw[line width=1pt, line join=bevel, line cap=rect](76.285pt,184.021pt) -- (67.2512pt,184.021pt);
	\draw[line width=1pt, line join=bevel, line cap=rect](76.285pt,249.822pt) -- (67.2512pt,249.822pt);
	\draw[line width=1pt, line join=bevel, line cap=rect](76.285pt,315.624pt) -- (67.2512pt,315.624pt);
	\draw[line width=1pt, line join=bevel, line cap=rect](76.285pt,381.425pt) -- (67.2512pt,381.425pt);
	\pgftext[center, base, at={\pgfpoint{277.027pt}{8.46914pt}}]{\fontsize{25}{0}\selectfont{\textbf{Most common SOP}}}
	\pgftext[center, base, at={\pgfpoint{197.739pt}{48.4937pt}}]{\fontsize{24}{0}\selectfont{$A\lor B$}}
	\pgftext[center, base, at={\pgfpoint{329.732pt}{48.4937pt}}]{\fontsize{24}{0}\selectfont{$A\land B$}}
	\draw[line width=1pt, line join=bevel, line cap=rect](210.453pt,83.3112pt) -- (210.453pt,74.2775pt);
	\draw[line width=1pt, line join=bevel, line cap=rect](342.613pt,83.3112pt) -- (342.613pt,74.2775pt);
\end{scope}
\end{tikzpicture}
         }
        \label{fig:BoxPlot_2bits_ZnO}
    \end{subfigure}
    \begin{subfigure}[c]{0.32\textwidth}
        \centering
        \caption{}
       \resizebox{1\textwidth}{!}{
         \begin{tikzpicture}{0pt}{0pt}{493.845pt}{394.474pt}
	\color[rgb]{1,1,1}
	\fill(0pt,394.474pt) -- (493.845pt,394.474pt) -- (493.845pt,0pt) -- (0pt,0pt) -- (0pt,394.474pt);
\begin{scope}
	\clip(76.285pt,382.429pt) -- (488.826pt,382.429pt) -- (488.826pt,82.3075pt) -- (76.285pt,82.3075pt) -- (76.285pt,382.429pt);
	\definecolor{c}{rgb}{0,0,0}
	\fill [pattern color=c, pattern=north east lines](133.852pt,252.868pt) -- (186.143pt,252.868pt) -- (186.143pt,234.646pt) -- (133.852pt,234.646pt) -- (133.852pt,252.868pt);
	\color[rgb]{0,0,0}
	\draw[line width=1.5pt, line join=miter, line cap=rect](133.852pt,252.868pt) -- (186.143pt,252.868pt) -- (186.143pt,234.646pt) -- (133.852pt,234.646pt) -- (133.852pt,252.868pt);
	\color[rgb]{0,0,0}
	\draw[line width=1.5pt, line join=bevel, line cap=rect](159.998pt,268.384pt) -- (159.998pt,252.868pt);
	\draw[line width=1.5pt, line join=bevel, line cap=rect](159.998pt,223.989pt) -- (159.998pt,234.646pt);
	\draw[line width=1.5pt, line join=bevel, line cap=rect](154.769pt,223.989pt) -- (165.227pt,223.989pt);
	\draw[line width=1.5pt, line join=bevel, line cap=rect](154.769pt,268.384pt) -- (165.227pt,268.384pt);
	\draw[line width=1.5pt, line join=bevel, line cap=rect](133.852pt,243.757pt) -- (186.143pt,243.757pt);
	\draw[line width=1pt, line join=miter, line cap=rect](156.585pt,249.934pt) -- (163.611pt,249.934pt) -- (163.611pt,242.907pt) -- (156.585pt,242.907pt) -- (156.585pt,249.934pt);
	\definecolor{c}{rgb}{1,0,0}
	\fill [pattern color=c, pattern=north east lines](215.557pt,107.092pt) -- (267.849pt,107.092pt) -- (267.849pt,107.092pt) -- (215.557pt,107.092pt) -- (215.557pt,107.092pt);
	\color[rgb]{1,0,0}
	\draw[line width=1.5pt, line join=miter, line cap=rect](215.557pt,107.092pt) -- (267.849pt,107.092pt) -- (267.849pt,107.092pt) -- (215.557pt,107.092pt) -- (215.557pt,107.092pt);
	\draw[line width=1.5pt, line join=miter, line cap=rect](241.703pt,143.156pt) -- (241.703pt,107.092pt);
	\draw[line width=1.5pt, line join=miter, line cap=rect](241.703pt,88.0363pt) -- (241.703pt,107.092pt);
	\draw[line width=1.5pt, line join=miter, line cap=rect](236.474pt,88.0363pt) -- (246.932pt,88.0363pt);
	\draw[line width=1.5pt, line join=miter, line cap=rect](236.474pt,143.156pt) -- (246.932pt,143.156pt);
	\draw[line width=1.5pt, line join=miter, line cap=rect](215.557pt,107.092pt) -- (267.849pt,107.092pt);
	\draw[line width=1pt, line join=miter, line cap=rect](237.889pt,119.446pt) -- (244.915pt,119.446pt) -- (244.915pt,112.42pt) -- (237.889pt,112.42pt) -- (237.889pt,119.446pt);
	\definecolor{c}{rgb}{0,0,1}
	\fill [pattern color=c, pattern=north east lines](297.263pt,134.425pt) -- (349.554pt,134.425pt) -- (349.554pt,107.092pt) -- (297.263pt,107.092pt) -- (297.263pt,134.425pt);
	\color[rgb]{0,0,1}
	\draw[line width=1.5pt, line join=miter, line cap=rect](297.263pt,134.425pt) -- (349.554pt,134.425pt) -- (349.554pt,107.092pt) -- (297.263pt,107.092pt) -- (297.263pt,134.425pt);
	\draw[line width=1.5pt, line join=miter, line cap=rect](323.408pt,139.874pt) -- (323.408pt,134.425pt);
	\draw[line width=1.5pt, line join=miter, line cap=rect](323.408pt,92.5324pt) -- (323.408pt,107.092pt);
	\draw[line width=1.5pt, line join=miter, line cap=rect](318.179pt,92.5324pt) -- (328.637pt,92.5324pt);
	\draw[line width=1.5pt, line join=miter, line cap=rect](318.179pt,139.874pt) -- (328.637pt,139.874pt);
	\draw[line width=1.5pt, line join=miter, line cap=rect](297.263pt,107.092pt) -- (349.554pt,107.092pt);
	\draw[line width=1pt, line join=miter, line cap=rect](320.196pt,119.446pt) -- (327.222pt,119.446pt) -- (327.222pt,112.42pt) -- (320.196pt,112.42pt) -- (320.196pt,119.446pt);
	\definecolor{c}{rgb}{0,0,0}
	\fill [pattern color=c, pattern=north east lines](378.968pt,143.536pt) -- (431.259pt,143.536pt) -- (431.259pt,125.314pt) -- (378.968pt,125.314pt) -- (378.968pt,143.536pt);
	\color[rgb]{0,0.501961,0}
	\draw[line width=1.5pt, line join=miter, line cap=rect](378.968pt,143.536pt) -- (431.259pt,143.536pt) -- (431.259pt,125.314pt) -- (378.968pt,125.314pt) -- (378.968pt,143.536pt);
	\draw[line width=1.5pt, line join=miter, line cap=rect](405.113pt,150.529pt) -- (405.113pt,143.536pt);
	\draw[line width=1.5pt, line join=miter, line cap=rect](405.113pt,117.107pt) -- (405.113pt,125.314pt);
	\draw[line width=1.5pt, line join=miter, line cap=rect](399.884pt,117.107pt) -- (410.343pt,117.107pt);
	\draw[line width=1.5pt, line join=miter, line cap=rect](399.884pt,150.529pt) -- (410.343pt,150.529pt);
	\draw[line width=1.5pt, line join=miter, line cap=rect](378.968pt,134.425pt) -- (431.259pt,134.425pt);
	\draw[line width=1pt, line join=miter, line cap=rect](401.5pt,137.514pt) -- (408.526pt,137.514pt) -- (408.526pt,130.487pt) -- (401.5pt,130.487pt) -- (401.5pt,137.514pt);
\end{scope}
\begin{scope}
	\color[rgb]{0,0,0}
	\draw[line width=2pt, line join=miter, line cap=rect](76.285pt,382.429pt) -- (488.826pt,382.429pt) -- (488.826pt,82.3075pt) -- (76.285pt,82.3075pt) -- (76.285pt,382.429pt);
	\color[rgb]{0,0,0}
	\pgftext[center, base, at={\pgfpoint{21.6277pt}{240.398pt}},rotate=90]{\fontsize{25}{0}\selectfont{\textbf{Count}}}
	\pgftext[center, base, at={\pgfpoint{57.2137pt}{101.692pt}}]{\fontsize{24}{0}\selectfont{0}}
	\pgftext[center, base, at={\pgfpoint{51.1912pt}{192.03pt}}]{\fontsize{24}{0}\selectfont{10}}
	\pgftext[center, base, at={\pgfpoint{51.1912pt}{283.371pt}}]{\fontsize{24}{0}\selectfont{20}}
	\pgftext[center, base, at={\pgfpoint{51.1912pt}{374.712pt}}]{\fontsize{24}{0}\selectfont{30}}
	\draw[line width=1pt, line join=bevel, line cap=rect](76.285pt,152.647pt) -- (71.2662pt,152.647pt);
	\draw[line width=1pt, line join=bevel, line cap=rect](76.285pt,243.757pt) -- (71.2662pt,243.757pt);
	\draw[line width=1pt, line join=bevel, line cap=rect](76.285pt,334.866pt) -- (71.2662pt,334.866pt);
	\draw[line width=1pt, line join=bevel, line cap=rect](76.285pt,107.092pt) -- (67.2512pt,107.092pt);
	\draw[line width=1pt, line join=bevel, line cap=rect](76.285pt,198.202pt) -- (67.2512pt,198.202pt);
	\draw[line width=1pt, line join=bevel, line cap=rect](76.285pt,289.312pt) -- (67.2512pt,289.312pt);
	\draw[line width=1pt, line join=bevel, line cap=rect](76.285pt,380.421pt) -- (67.2512pt,380.421pt);
	\pgftext[center, base, at={\pgfpoint{283.05pt}{8.46914pt}}]{\fontsize{25}{0}\selectfont{\textbf{Most common SOP}}}
	\pgftext[center, base, at={\pgfpoint{159.596pt}{47.4899pt}}]{\fontsize{24}{0}\selectfont{1}}
	\pgftext[center, base, at={\pgfpoint{241.904pt}{47.4899pt}}]{\fontsize{24}{0}\selectfont{2}}
	\pgftext[center, base, at={\pgfpoint{323.207pt}{47.4899pt}}]{\fontsize{24}{0}\selectfont{3}}
	\pgftext[center, base, at={\pgfpoint{405.515pt}{47.4899pt}}]{\fontsize{24}{0}\selectfont{4}}
	\draw[line width=1pt, line join=bevel, line cap=rect](159.998pt,82.3075pt) -- (159.998pt,73.2737pt);
	\draw[line width=1pt, line join=bevel, line cap=rect](241.703pt,82.3075pt) -- (241.703pt,73.2737pt);
	\draw[line width=1pt, line join=bevel, line cap=rect](323.408pt,82.3075pt) -- (323.408pt,73.2737pt);
	\draw[line width=1pt, line join=bevel, line cap=rect](405.113pt,82.3075pt) -- (405.113pt,73.2737pt);
	\definecolor{c}{rgb}{0,0,0}
	\fill [pattern color=c, pattern=north east lines](87.3262pt,370.384pt) -- (119.446pt,370.384pt) -- (119.446pt,357.335pt) -- (87.3262pt,357.335pt) -- (87.3262pt,370.384pt);
	\draw[line width=1.5pt, line join=miter, line cap=rect](87.3262pt,370.384pt) -- (119.446pt,370.384pt) -- (119.446pt,357.335pt) -- (87.3262pt,357.335pt) -- (87.3262pt,370.384pt);
	\pgftext[left, base, at={\pgfpoint{124.465pt}{358.072pt}}]{\fontsize{18}{0}\selectfont{$A \lor B \lor C \lor D$}}
	\definecolor{c}{rgb}{1,0,0}
	\fill [pattern color=c, pattern=north east lines](87.3262pt,343.282pt) -- (119.446pt,343.282pt) -- (119.446pt,330.234pt) -- (87.3262pt,330.234pt) -- (87.3262pt,343.282pt);
	\color[rgb]{1,0,0}
	\draw[line width=1.5pt, line join=miter, line cap=rect](87.3262pt,343.282pt) -- (119.446pt,343.282pt) -- (119.446pt,330.234pt) -- (87.3262pt,330.234pt) -- (87.3262pt,343.282pt);
	\color[rgb]{0,0,0}
	\pgftext[left, base, at={\pgfpoint{124.465pt}{330.971pt}}]{\fontsize{18}{0}\selectfont{$A \land B \land C \land D$}}
	\definecolor{c}{rgb}{0,0,1}
	\fill [pattern color=c, pattern=north east lines](87.3262pt,316.181pt) -- (119.446pt,316.181pt) -- (119.446pt,303.132pt) -- (87.3262pt,303.132pt) -- (87.3262pt,316.181pt);
	\color[rgb]{0,0,1}
	\draw[line width=1.5pt, line join=miter, line cap=rect](87.3262pt,316.181pt) -- (119.446pt,316.181pt) -- (119.446pt,303.132pt) -- (87.3262pt,303.132pt) -- (87.3262pt,316.181pt);
	\color[rgb]{0,0,0}
	\pgftext[left, base, at={\pgfpoint{124.465pt}{303.87pt}}]{\fontsize{18}{0}\selectfont{$A \lor B \lor (C \land D)$ }}
	\definecolor{c}{rgb}{0,0,0}
	\fill [pattern color=c, pattern=north east lines](87.3262pt,289.08pt) -- (119.446pt,289.08pt) -- (119.446pt,276.031pt) -- (87.3262pt,276.031pt) -- (87.3262pt,289.08pt);
	\color[rgb]{0,0.501961,0}
	\draw[line width=1.5pt, line join=miter, line cap=rect](87.3262pt,289.08pt) -- (119.446pt,289.08pt) -- (119.446pt,276.031pt) -- (87.3262pt,276.031pt) -- (87.3262pt,289.08pt);
	\color[rgb]{0,0,0}
	\pgftext[left, base, at={\pgfpoint{124.465pt}{276.768pt}}]{\fontsize{18}{0}\selectfont{$(A \land B) \lor (B \land D) \lor (C \land D) \lor (A \land \neg C \land \neg D$}}
\end{scope}
\end{tikzpicture}
         }
        \label{fig:BoxPlot_4bits_ZnO}
    \end{subfigure}
    \begin{subfigure}[c]{0.32\textwidth}
        \centering
        \caption{}
        \resizebox{1\textwidth}{!}{
         \begin{tikzpicture}{0pt}{0pt}{483.807pt}{419.567pt}
	\color[rgb]{1,1,1}
	\fill(0pt,419.567pt) -- (483.807pt,419.567pt) -- (483.807pt,0pt) -- (0pt,0pt) -- (0pt,419.567pt);
\begin{scope}
	\clip(78.2925pt,383.432pt) -- (478.789pt,383.432pt) -- (478.789pt,83.3112pt) -- (78.2925pt,83.3112pt) -- (78.2925pt,383.432pt);
	\definecolor{c}{rgb}{0,0,0}
	\fill [pattern color=c, pattern=north east lines](134.221pt,248.177pt) -- (184.971pt,248.177pt) -- (184.971pt,159.345pt) -- (134.221pt,159.345pt) -- (134.221pt,248.177pt);
	\color[rgb]{0,0,0}
	\draw[line width=1.5pt, line join=miter, line cap=rect](134.221pt,248.177pt) -- (184.971pt,248.177pt) -- (184.971pt,159.345pt) -- (134.221pt,159.345pt) -- (134.221pt,248.177pt);
	\color[rgb]{0,0,0}
	\draw[line width=1.5pt, line join=bevel, line cap=rect](159.596pt,288.112pt) -- (159.596pt,248.177pt);
	\draw[line width=1.5pt, line join=bevel, line cap=rect](159.596pt,113.488pt) -- (159.596pt,159.345pt);
	\draw[line width=1.5pt, line join=bevel, line cap=rect](154.521pt,113.488pt) -- (164.671pt,113.488pt);
	\draw[line width=1.5pt, line join=bevel, line cap=rect](154.521pt,288.112pt) -- (164.671pt,288.112pt);
	\draw[line width=1.5pt, line join=bevel, line cap=rect](134.221pt,159.345pt) -- (184.971pt,159.345pt);
	\draw[line width=1pt, line join=miter, line cap=rect](156.585pt,204.765pt) -- (163.611pt,204.765pt) -- (163.611pt,197.739pt) -- (156.585pt,197.739pt) -- (156.585pt,204.765pt);
	\definecolor{c}{rgb}{1,0,0}
	\fill [pattern color=c, pattern=north east lines](213.518pt,159.345pt) -- (264.267pt,159.345pt) -- (264.267pt,159.345pt) -- (213.518pt,159.345pt) -- (213.518pt,159.345pt);
	\color[rgb]{1,0,0}
	\draw[line width=1.5pt, line join=miter, line cap=rect](213.518pt,159.345pt) -- (264.267pt,159.345pt) -- (264.267pt,159.345pt) -- (213.518pt,159.345pt) -- (213.518pt,159.345pt);
	\draw[line width=1.5pt, line join=miter, line cap=rect](238.892pt,172.787pt) -- (238.892pt,159.345pt);
	\draw[line width=1.5pt, line join=miter, line cap=rect](238.892pt,149.851pt) -- (238.892pt,159.345pt);
	\draw[line width=1.5pt, line join=miter, line cap=rect](233.818pt,149.851pt) -- (243.967pt,149.851pt);
	\draw[line width=1.5pt, line join=miter, line cap=rect](233.818pt,172.787pt) -- (243.967pt,172.787pt);
	\draw[line width=1.5pt, line join=miter, line cap=rect](213.518pt,159.345pt) -- (264.267pt,159.345pt);
	\draw[line width=1pt, line join=miter, line cap=rect](235.881pt,164.615pt) -- (242.907pt,164.615pt) -- (242.907pt,157.589pt) -- (235.881pt,157.589pt) -- (235.881pt,164.615pt);
	\definecolor{c}{rgb}{0,0,1}
	\fill [pattern color=c, pattern=north east lines](292.814pt,159.345pt) -- (343.564pt,159.345pt) -- (343.564pt,159.345pt) -- (292.814pt,159.345pt) -- (292.814pt,159.345pt);
	\color[rgb]{0,0,0}
	\draw[line width=1.5pt, line join=miter, line cap=rect](292.814pt,159.345pt) -- (343.564pt,159.345pt) -- (343.564pt,159.345pt) -- (292.814pt,159.345pt) -- (292.814pt,159.345pt);
	\color[rgb]{0,0,1}
	\draw[line width=1.5pt, line join=miter, line cap=rect](318.189pt,166.066pt) -- (318.189pt,159.345pt);
	\draw[line width=1.5pt, line join=miter, line cap=rect](318.189pt,154.598pt) -- (318.189pt,159.345pt);
	\draw[line width=1.5pt, line join=miter, line cap=rect](313.114pt,154.598pt) -- (323.264pt,154.598pt);
	\draw[line width=1.5pt, line join=miter, line cap=rect](313.114pt,166.066pt) -- (323.264pt,166.066pt);
	\draw[line width=1.5pt, line join=miter, line cap=rect](292.814pt,159.345pt) -- (343.564pt,159.345pt);
	\draw[line width=1pt, line join=miter, line cap=rect](315.177pt,163.611pt) -- (322.204pt,163.611pt) -- (322.204pt,156.585pt) -- (315.177pt,156.585pt) -- (315.177pt,163.611pt);
	\definecolor{c}{rgb}{0,0.501961,0}
	\fill [pattern color=c, pattern=north east lines](372.11pt,159.345pt) -- (422.86pt,159.345pt) -- (422.86pt,159.345pt) -- (372.11pt,159.345pt) -- (372.11pt,159.345pt);
	\color[rgb]{0,0.501961,0}
	\draw[line width=1.5pt, line join=miter, line cap=rect](372.11pt,159.345pt) -- (422.86pt,159.345pt) -- (422.86pt,159.345pt) -- (372.11pt,159.345pt) -- (372.11pt,159.345pt);
	\draw[line width=1.5pt, line join=miter, line cap=rect](397.485pt,166.066pt) -- (397.485pt,159.345pt);
	\draw[line width=1.5pt, line join=miter, line cap=rect](397.485pt,154.598pt) -- (397.485pt,159.345pt);
	\draw[line width=1.5pt, line join=miter, line cap=rect](392.41pt,154.598pt) -- (402.56pt,154.598pt);
	\draw[line width=1.5pt, line join=miter, line cap=rect](392.41pt,166.066pt) -- (402.56pt,166.066pt);
	\draw[line width=1.5pt, line join=miter, line cap=rect](372.11pt,159.345pt) -- (422.86pt,159.345pt);
	\draw[line width=1pt, line join=miter, line cap=rect](394.474pt,163.611pt) -- (401.5pt,163.611pt) -- (401.5pt,156.585pt) -- (394.474pt,156.585pt) -- (394.474pt,163.611pt);
\end{scope}
\begin{scope}
	\color[rgb]{0,0,0}
	\draw[line width=2pt, line join=miter, line cap=rect](78.2925pt,383.432pt) -- (478.789pt,383.432pt) -- (478.789pt,83.3112pt) -- (78.2925pt,83.3112pt) -- (78.2925pt,383.432pt);
	\color[rgb]{1,1,1}
	\pgftext[center, base, at={\pgfpoint{281.042pt}{406.095pt}}]{\fontsize{15}{0}\selectfont{\textbf{Title}}}
	\color[rgb]{0,0,0}
	\pgftext[center, base, at={\pgfpoint{21.6277pt}{233.372pt}},rotate=90]{\fontsize{25}{0}\selectfont{\textbf{Count}}}
	\pgftext[center, base, at={\pgfpoint{49.1759pt}{79.6099pt}}]{\fontsize{24}{0}\selectfont{-5}}
	\pgftext[center, base, at={\pgfpoint{59.2212pt}{153.887pt}}]{\fontsize{24}{0}\selectfont{0}}
	\pgftext[center, base, at={\pgfpoint{59.2212pt}{227.161pt}}]{\fontsize{24}{0}\selectfont{5}}
	\pgftext[center, base, at={\pgfpoint{53.1987pt}{301.439pt}}]{\fontsize{24}{0}\selectfont{10}}
	\pgftext[center, base, at={\pgfpoint{53.1987pt}{375.716pt}}]{\fontsize{24}{0}\selectfont{15}}
	\draw[line width=1pt, line join=bevel, line cap=rect](78.2925pt,122.332pt) -- (73.2737pt,122.332pt);
	\draw[line width=1pt, line join=bevel, line cap=rect](78.2925pt,196.359pt) -- (73.2737pt,196.359pt);
	\draw[line width=1pt, line join=bevel, line cap=rect](78.2925pt,270.385pt) -- (73.2737pt,270.385pt);
	\draw[line width=1pt, line join=bevel, line cap=rect](78.2925pt,344.412pt) -- (73.2737pt,344.412pt);
	\draw[line width=1pt, line join=bevel, line cap=rect](78.2925pt,85.3187pt) -- (69.2587pt,85.3187pt);
	\draw[line width=1pt, line join=bevel, line cap=rect](78.2925pt,159.345pt) -- (69.2587pt,159.345pt);
	\draw[line width=1pt, line join=bevel, line cap=rect](78.2925pt,233.372pt) -- (69.2587pt,233.372pt);
	\draw[line width=1pt, line join=bevel, line cap=rect](78.2925pt,307.398pt) -- (69.2587pt,307.398pt);
	\draw[line width=1pt, line join=bevel, line cap=rect](78.2925pt,381.425pt) -- (69.2587pt,381.425pt);
	\pgftext[center, base, at={\pgfpoint{279.035pt}{8.46914pt}}]{\fontsize{25}{0}\selectfont{\textbf{Most common SOP}}}
	\pgftext[center, base, at={\pgfpoint{159.596pt}{48.4937pt}}]{\fontsize{24}{0}\selectfont{1}}
	\pgftext[center, base, at={\pgfpoint{238.892pt}{48.4937pt}}]{\fontsize{24}{0}\selectfont{2}}
	\pgftext[center, base, at={\pgfpoint{318.189pt}{48.4937pt}}]{\fontsize{24}{0}\selectfont{3}}
	\pgftext[center, base, at={\pgfpoint{397.485pt}{48.4937pt}}]{\fontsize{24}{0}\selectfont{4}}
	\draw[line width=1pt, line join=bevel, line cap=rect](159.596pt,83.3112pt) -- (159.596pt,74.2775pt);
	\draw[line width=1pt, line join=bevel, line cap=rect](238.892pt,83.3112pt) -- (238.892pt,74.2775pt);
	\draw[line width=1pt, line join=bevel, line cap=rect](318.189pt,83.3112pt) -- (318.189pt,74.2775pt);
	\draw[line width=1pt, line join=bevel, line cap=rect](397.485pt,83.3112pt) -- (397.485pt,74.2775pt);
\end{scope}
\end{tikzpicture}
         }
        \label{fig:BoxPlot_8bits_ZnO}
    \end{subfigure}
    \caption{Boxplots showing the distribution of the four most common sum-of-product expressions found in repeated experiments with (a) 2-bit, (b) 4-bit, and (c) 8-bit binary strings for ZnO colloidal suspension. Whiskers indicate 1.5 times the standard deviation.}
    \label{fig:boxplot_ZnO}
\end{figure}

\begin{figure}[!h]
    \centering
    \begin{subfigure}[c]{0.32\textwidth}
        \centering
        \caption{}
        \includegraphics[width=1\textwidth, keepaspectratio]{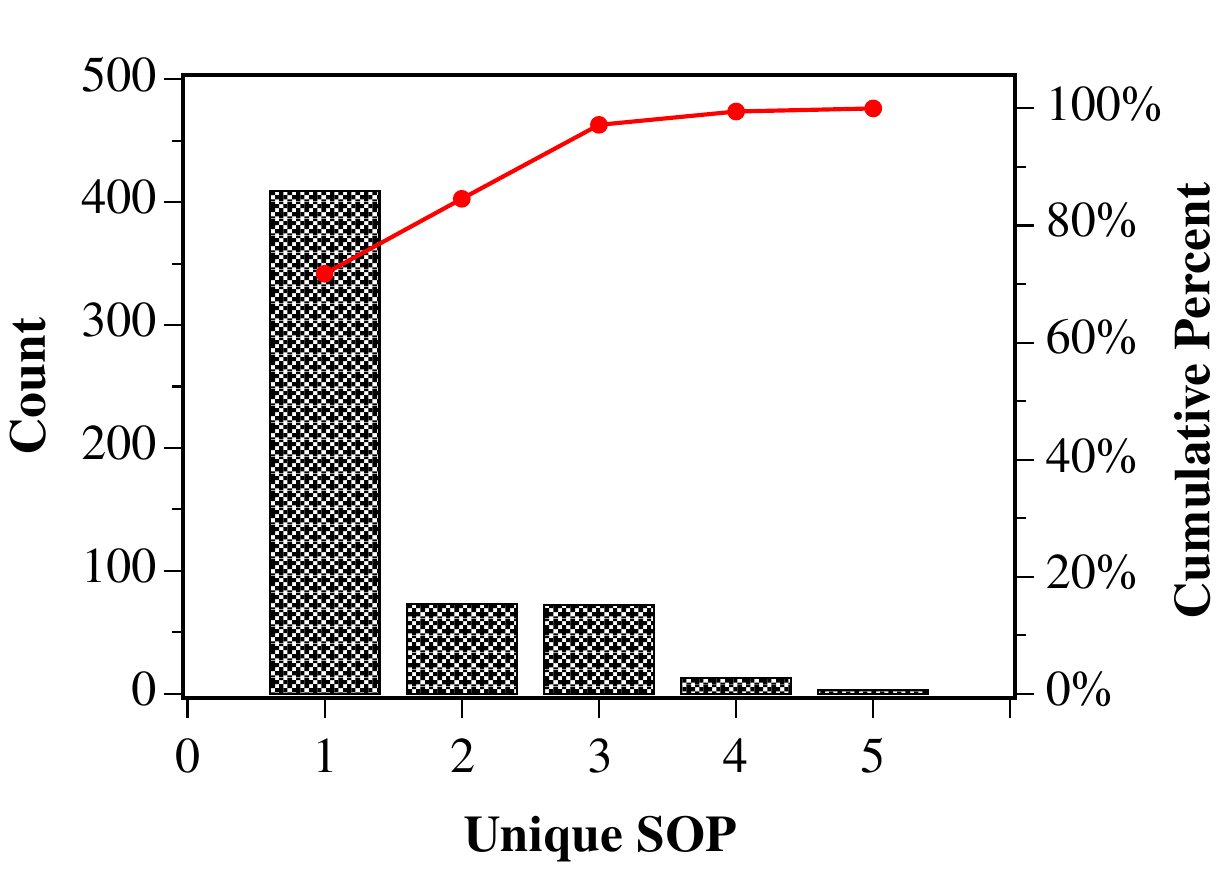}
        \label{fig:Pareto_2bits_ZnO}
    \end{subfigure}
    \begin{subfigure}[c]{0.32\textwidth}
        \centering
        \caption{}
        \includegraphics[width=1\textwidth, keepaspectratio]{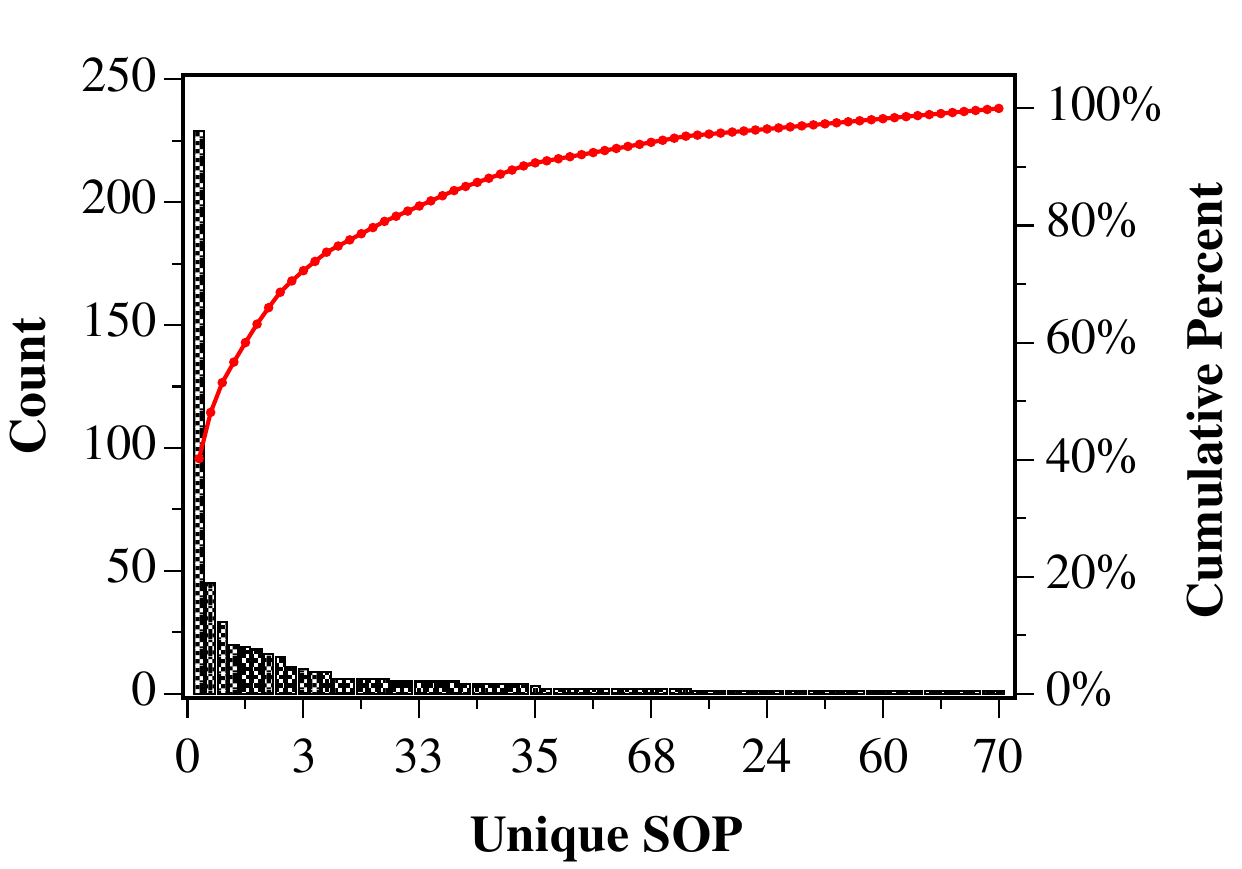}
        \label{fig:Pareto_4bits_ZnO}
    \end{subfigure}
    \begin{subfigure}[c]{0.32\textwidth}
        \centering
        \caption{}
        \includegraphics[width=1\textwidth, keepaspectratio]{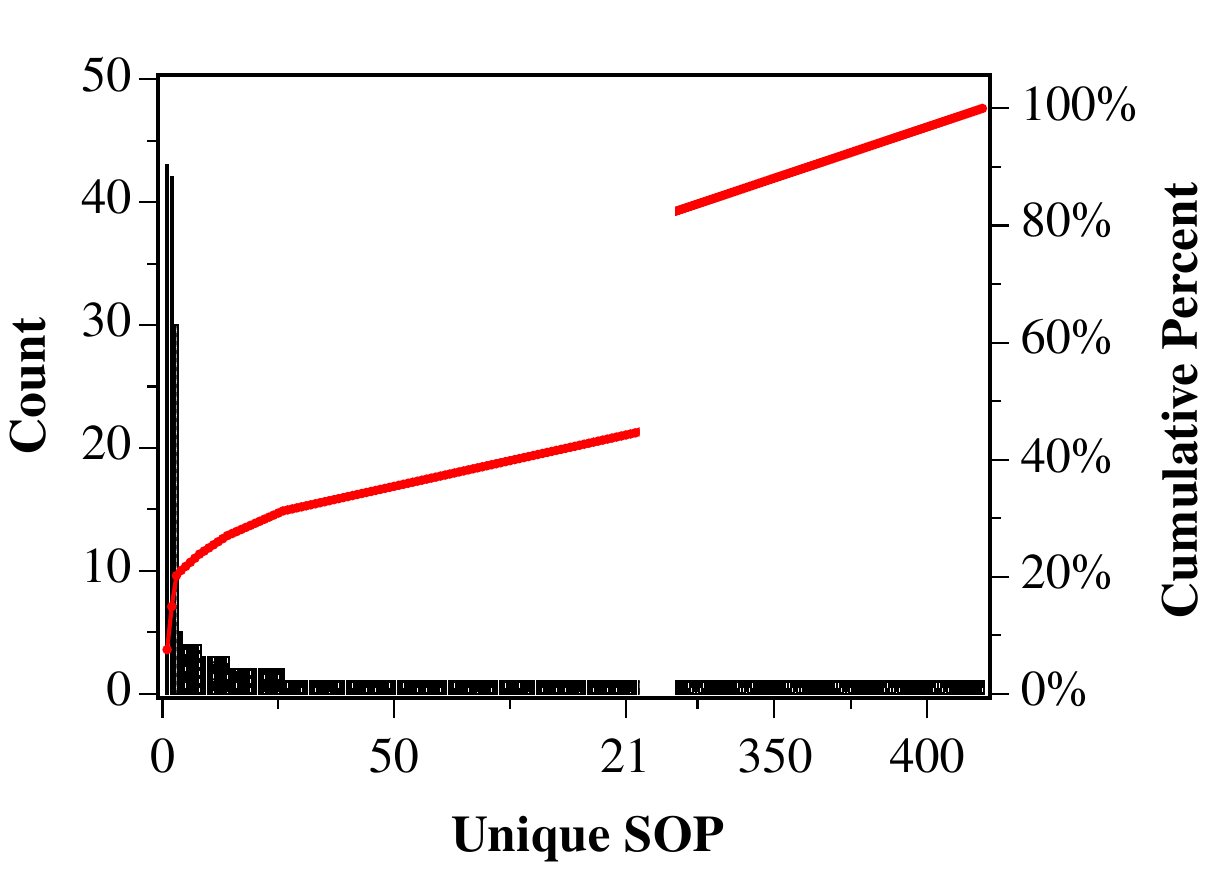}
        \label{fig:Pareto_8bits_ZnO}
    \end{subfigure}
    \caption{Pareto charts showing the frequency counts of unique sum-of-products (SOP) from repeated experiments for ZnO colloidal suspension. The red curve represents the cumulative percentage of the total count for (a) 2-bit, (b) 4-bit, and (c) 8-bit.}
    \label{fig:pareto_ZnO}
\end{figure}

The reproducibility of the extracted logic circuits was assessed by conducting 15 repetitions of the experiments for each binary input string. Prior to each repetition, the sample was manually shaken to disrupt any existing structures. Fig.~\ref{fig:boxplot_ZnO} displays boxplots summarising the results for 2-, 4-, and 8-bit, with the whiskers illustrating 1.5 times the standard deviation across the repeats. The consistency of the most common logic expressions establishes the ability to produce circuits that are reproducible for a given input sequence. Figure~\ref{fig:pareto_ZnO} shows the Pareto charts that depict the outcomes of the multiple trials conducted. These charts, a type of histogram that orders outcomes by frequency to visualise their distribution and relative dominance, show that the variability in the number of distinct SOP expressions increased as the number of input bits increased, reflecting the greater number of possible logic functions that could be realised. The recurring extraction of the most common logical expression underscores the robust and deterministic nature of the computation with the colloidal suspension. In addition, the increase in variability with input string size is in line with the exponential growth in potential Boolean functions.

\section{Proteinoid}

A total of 3, 9, and 17 unique standard logical expressions were obtained for the similar experiments using proteinoid and 2-, 4-, and 8-bit inputs. Tables~\ref{tab:sop_2bit_prot},~\ref{tab:sop_4bit_prot}, and~\ref{tab:sop_8bit_prot} show the four most common logical expressions identified for each input binary string. The logical complexity of the most common expressions increased with more input variables, although at different rates than in the previous ZnO nanoparticles experiments. For 2-bit inputs, the most common expression was a logical 0 term, whereas in the ZnO case it was a disjunctive term, showing a clear contrast between the materials. The most common expressions for 4-bit inputs showed a lower circuit complexity (lower number of extracted logic gates) when compared to those shown for the ZnO colloidal suspension. The most common expressions for 8 bits showed a large number of disjunctive terms, indicating a higher level of circuit complexity than that of the ZnO suspension. While the complexity increased as more inputs were added, both the expressions and their frequency highlighted the differences between the materials.

\begin{table}[!tbp]
\centering
\caption{Extracted sum-of-products (SOP) Boolean expression for a 2-bit string input with varying thresholds for proteinoid}
\label{tab:sop_2bit_prot}
\begin{tabular}{lr}
\toprule
SOP                            & Count   \\ \midrule
$0\quad(\mathrm{False})$                     & 19     \\
$A\lor B$                      & 16       \\ 
$A\land B$                      & 3       \\ \bottomrule
\end{tabular}
\end{table}

\begin{table}[!tbp]
\centering
\caption{Extracted SOP Boolean expression for a 4-bit string input with varying thresholds for proteinoid}
\label{tab:sop_4bit_prot}
\begin{tabular}{lr}
\toprule
SOP                            & Count   \\ \midrule
$A \lor B \lor C \lor D$  & 23     \\
$A \land B \land C \land D$ & 4 \\ 
$A \lor B \lor (C \land D)$ & 3 \\ 
$(A \land B) \lor (B \land D) \lor (C \land D) \lor (A \land \neg C \land \neg D) $ & 3\\
\bottomrule
\end{tabular}
\end{table}

\begin{table}[!tbp]
\centering
\caption{Extracted SOP Boolean expression for an 8-bit string input with varying thresholds for proteinoid}
\label{tab:sop_8bit_prot}
\resizebox{1\textwidth}{!}{
\begin{tabular}{lr}
\toprule
SOP                                                                                                                                                                                                                                                                                        & Count \\ \midrule
$\neg A \lor \neg B \lor \neg C \lor \neg D \lor \neg E \lor \neg F \lor \neg G \lor \neg H$                                                                                                                                                                                               & 19    \\
$(A \land \neg E) \lor (B \land \neg H) \lor (C \land \neg G) \lor (D \land \neg F) \lor (E \land \neg D) \lor (F \land \neg C) \lor (G \land \neg B) \lor (H \land \neg A)$                                                                                                               & 4     \\
$A \land B \land C \land D \land E \land F \land H \land \neg G$                                                                                                                                                                                                                           & 2     \\
\begin{tabular}[c]{@{}l@{}}$(C \land \neg B) \lor (C \land \neg D) \lor (D \land \neg E) \lor (E \land \neg G) \lor (F \land \neg H) \lor (G \land \neg F) \lor$ \\ $(H \land \neg E) \lor (A \land B \land \neg C) \lor (A \land H \land \neg B) \lor (B \land H \land \neg C$\end{tabular} & 1     \\ \bottomrule
\end{tabular}}
\end{table}

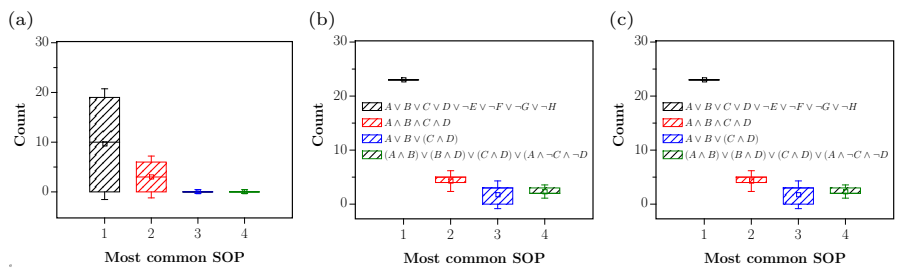
\begin{figure}[!tbp]
    \centering
    \begin{subfigure}[c]{0.32\textwidth}
        \centering
        \caption{}
       \resizebox{1\textwidth}{!}{
         e\begin{tikzpicture}{0pt}{0pt}{481.8pt}{395.477pt}
	\color[rgb]{1,1,1}
	\fill(0pt,395.477pt) -- (481.8pt,395.477pt) -- (481.8pt,0pt) -- (0pt,0pt) -- (0pt,395.477pt);
\begin{scope}
	\clip(76.285pt,383.432pt) -- (476.781pt,383.432pt) -- (476.781pt,83.3112pt) -- (76.285pt,83.3112pt) -- (76.285pt,383.432pt);
	\definecolor{c}{rgb}{0,0,0}
	\fill [pattern color=c, pattern=north east lines](132.214pt,288.363pt) -- (182.964pt,288.363pt) -- (182.964pt,127.62pt) -- (132.214pt,127.62pt) -- (132.214pt,288.363pt);
	\color[rgb]{0,0,0}
	\draw[line width=1.5pt, line join=miter, line cap=rect](132.214pt,288.363pt) -- (182.964pt,288.363pt) -- (182.964pt,127.62pt) -- (132.214pt,127.62pt) -- (132.214pt,288.363pt);
	\color[rgb]{0,0,0}
	\draw[line width=1.5pt, line join=miter, line cap=rect](157.589pt,303.049pt) -- (157.589pt,288.363pt);
	\draw[line width=1.5pt, line join=miter, line cap=rect](157.589pt,114.626pt) -- (157.589pt,127.62pt);
	\draw[line width=1.5pt, line join=miter, line cap=rect](152.514pt,114.626pt) -- (162.664pt,114.626pt);
	\draw[line width=1.5pt, line join=miter, line cap=rect](152.514pt,303.049pt) -- (162.664pt,303.049pt);
	\draw[line width=1.5pt, line join=miter, line cap=rect](132.214pt,212.221pt) -- (182.964pt,212.221pt);
	\draw[line width=1pt, line join=miter, line cap=rect](154.577pt,212.795pt) -- (161.604pt,212.795pt) -- (161.604pt,205.769pt) -- (154.577pt,205.769pt) -- (154.577pt,212.795pt);
	\definecolor{c}{rgb}{1,0,0}
	\fill [pattern color=c, pattern=north east lines](211.51pt,178.381pt) -- (262.26pt,178.381pt) -- (262.26pt,127.62pt) -- (211.51pt,127.62pt) -- (211.51pt,178.381pt);
	\color[rgb]{1,0,0}
	\draw[line width=1.5pt, line join=miter, line cap=rect](211.51pt,178.381pt) -- (262.26pt,178.381pt) -- (262.26pt,127.62pt) -- (211.51pt,127.62pt) -- (211.51pt,178.381pt);
	\draw[line width=1.5pt, line join=miter, line cap=rect](236.885pt,188.572pt) -- (236.885pt,178.381pt);
	\draw[line width=1.5pt, line join=miter, line cap=rect](236.885pt,117.429pt) -- (236.885pt,127.62pt);
	\draw[line width=1.5pt, line join=miter, line cap=rect](231.81pt,117.429pt) -- (241.96pt,117.429pt);
	\draw[line width=1.5pt, line join=miter, line cap=rect](231.81pt,188.572pt) -- (241.96pt,188.572pt);
	\draw[line width=1.5pt, line join=miter, line cap=rect](211.51pt,153pt) -- (262.26pt,153pt);
	\draw[line width=1pt, line join=miter, line cap=rect](233.874pt,156.585pt) -- (240.9pt,156.585pt) -- (240.9pt,149.559pt) -- (233.874pt,149.559pt) -- (233.874pt,156.585pt);
	\definecolor{c}{rgb}{0,0,1}
	\fill [pattern color=c, pattern=north east lines](290.806pt,127.62pt) -- (341.556pt,127.62pt) -- (341.556pt,127.62pt) -- (290.806pt,127.62pt) -- (290.806pt,127.62pt);
	\color[rgb]{0,0,0}
	\draw[line width=1.5pt, line join=miter, line cap=rect](290.806pt,127.62pt) -- (341.556pt,127.62pt) -- (341.556pt,127.62pt) -- (290.806pt,127.62pt) -- (290.806pt,127.62pt);
	\color[rgb]{0,0,1}
	\draw[line width=1.5pt, line join=miter, line cap=rect](316.181pt,131.46pt) -- (316.181pt,127.62pt);
	\draw[line width=1.5pt, line join=miter, line cap=rect](316.181pt,124.907pt) -- (316.181pt,127.62pt);
	\draw[line width=1.5pt, line join=miter, line cap=rect](311.106pt,124.907pt) -- (321.256pt,124.907pt);
	\draw[line width=1.5pt, line join=miter, line cap=rect](311.106pt,131.46pt) -- (321.256pt,131.46pt);
	\draw[line width=1.5pt, line join=miter, line cap=rect](290.806pt,127.62pt) -- (341.556pt,127.62pt);
	\draw[line width=1pt, line join=miter, line cap=rect](313.17pt,131.491pt) -- (320.196pt,131.491pt) -- (320.196pt,124.465pt) -- (313.17pt,124.465pt) -- (313.17pt,131.491pt);
	\definecolor{c}{rgb}{0,0.501961,0}
	\fill [pattern color=c, pattern=north east lines](370.103pt,127.62pt) -- (420.852pt,127.62pt) -- (420.852pt,127.62pt) -- (370.103pt,127.62pt) -- (370.103pt,127.62pt);
	\color[rgb]{0,0.501961,0}
	\draw[line width=1.5pt, line join=miter, line cap=rect](370.103pt,127.62pt) -- (420.852pt,127.62pt) -- (420.852pt,127.62pt) -- (370.103pt,127.62pt) -- (370.103pt,127.62pt);
	\draw[line width=1.5pt, line join=miter, line cap=rect](395.477pt,131.46pt) -- (395.477pt,127.62pt);
	\draw[line width=1.5pt, line join=miter, line cap=rect](395.477pt,124.907pt) -- (395.477pt,127.62pt);
	\draw[line width=1.5pt, line join=miter, line cap=rect](390.403pt,124.907pt) -- (400.552pt,124.907pt);
	\draw[line width=1.5pt, line join=miter, line cap=rect](390.403pt,131.46pt) -- (400.552pt,131.46pt);
	\draw[line width=1.5pt, line join=miter, line cap=rect](370.103pt,127.62pt) -- (420.852pt,127.62pt);
	\draw[line width=1pt, line join=miter, line cap=rect](392.466pt,131.491pt) -- (399.492pt,131.491pt) -- (399.492pt,124.465pt) -- (392.466pt,124.465pt) -- (392.466pt,131.491pt);
\end{scope}
\begin{scope}
	\color[rgb]{0,0,0}
	\draw[line width=2pt, line join=miter, line cap=rect](76.285pt,383.432pt) -- (476.781pt,383.432pt) -- (476.781pt,83.3112pt) -- (76.285pt,83.3112pt) -- (76.285pt,383.432pt);
	\color[rgb]{0,0,0}
	\pgftext[center, base, at={\pgfpoint{21.6277pt}{241.402pt}},rotate=90]{\fontsize{25}{0}\selectfont{\textbf{Count}}}
	\pgftext[center, base, at={\pgfpoint{57.2137pt}{121.767pt}}]{\fontsize{24}{0}\selectfont{0}}
	\pgftext[center, base, at={\pgfpoint{51.1912pt}{206.082pt}}]{\fontsize{24}{0}\selectfont{10}}
	\pgftext[center, base, at={\pgfpoint{51.1912pt}{291.401pt}}]{\fontsize{24}{0}\selectfont{20}}
	\pgftext[center, base, at={\pgfpoint{51.1912pt}{375.716pt}}]{\fontsize{24}{0}\selectfont{30}}
	\draw[line width=1pt, line join=bevel, line cap=rect](76.285pt,169.92pt) -- (71.2662pt,169.92pt);
	\draw[line width=1pt, line join=bevel, line cap=rect](76.285pt,254.522pt) -- (71.2662pt,254.522pt);
	\draw[line width=1pt, line join=bevel, line cap=rect](76.285pt,339.124pt) -- (71.2662pt,339.124pt);
	\draw[line width=1pt, line join=bevel, line cap=rect](76.285pt,127.62pt) -- (67.2512pt,127.62pt);
	\draw[line width=1pt, line join=bevel, line cap=rect](76.285pt,212.221pt) -- (67.2512pt,212.221pt);
	\draw[line width=1pt, line join=bevel, line cap=rect](76.285pt,296.823pt) -- (67.2512pt,296.823pt);
	\draw[line width=1pt, line join=bevel, line cap=rect](76.285pt,381.425pt) -- (67.2512pt,381.425pt);
	\pgftext[center, base, at={\pgfpoint{277.027pt}{8.46914pt}}]{\fontsize{25}{0}\selectfont{\textbf{Most common SOP}}}
	\pgftext[center, base, at={\pgfpoint{157.589pt}{48.4937pt}}]{\fontsize{24}{0}\selectfont{1}}
	\pgftext[center, base, at={\pgfpoint{236.885pt}{48.4937pt}}]{\fontsize{24}{0}\selectfont{2}}
	\pgftext[center, base, at={\pgfpoint{316.181pt}{48.4937pt}}]{\fontsize{24}{0}\selectfont{3}}
	\pgftext[center, base, at={\pgfpoint{395.477pt}{48.4937pt}}]{\fontsize{24}{0}\selectfont{4}}
	\draw[line width=1pt, line join=bevel, line cap=rect](157.589pt,83.3112pt) -- (157.589pt,74.2775pt);
	\draw[line width=1pt, line join=bevel, line cap=rect](236.885pt,83.3112pt) -- (236.885pt,74.2775pt);
	\draw[line width=1pt, line join=bevel, line cap=rect](316.181pt,83.3112pt) -- (316.181pt,74.2775pt);
	\draw[line width=1pt, line join=bevel, line cap=rect](395.477pt,83.3112pt) -- (395.477pt,74.2775pt);
\end{scope}
\end{tikzpicture}
         }
        \label{fig:BoxPlot_2bits_proteinoid}
    \end{subfigure}
    \begin{subfigure}[c]{0.32\textwidth}
        \centering
        \caption{}
       \resizebox{1\textwidth}{!}{
         \begin{tikzpicture}{0pt}{0pt}{481.8pt}{394.474pt}
	\color[rgb]{1,1,1}
	\fill(0pt,394.474pt) -- (481.8pt,394.474pt) -- (481.8pt,0pt) -- (0pt,0pt) -- (0pt,394.474pt);
\begin{scope}
	\clip(76.285pt,382.429pt) -- (476.781pt,382.429pt) -- (476.781pt,82.3075pt) -- (76.285pt,82.3075pt) -- (76.285pt,382.429pt);
	\definecolor{c}{rgb}{0,0,0}
	\fill [pattern color=c, pattern=north east lines](132.214pt,316.645pt) -- (182.964pt,316.645pt) -- (182.964pt,316.645pt) -- (132.214pt,316.645pt) -- (132.214pt,316.645pt);
	\color[rgb]{0,0,0}
	\draw[line width=1.5pt, line join=miter, line cap=rect](132.214pt,316.645pt) -- (182.964pt,316.645pt) -- (182.964pt,316.645pt) -- (132.214pt,316.645pt) -- (132.214pt,316.645pt);
	\color[rgb]{0,0,0}
	\draw[line width=1.5pt, line join=miter, line cap=rect](152.514pt,316.645pt) -- (162.664pt,316.645pt);
	\draw[line width=1.5pt, line join=miter, line cap=rect](152.514pt,316.645pt) -- (162.664pt,316.645pt);
	\draw[line width=1.5pt, line join=miter, line cap=rect](132.214pt,316.645pt) -- (182.964pt,316.645pt);
	\draw[line width=1pt, line join=miter, line cap=rect](154.577pt,320.196pt) -- (161.604pt,320.196pt) -- (161.604pt,313.17pt) -- (154.577pt,313.17pt) -- (154.577pt,320.196pt);
	\definecolor{c}{rgb}{1,0,0}
	\fill [pattern color=c, pattern=north east lines](211.51pt,152.647pt) -- (262.26pt,152.647pt) -- (262.26pt,143.536pt) -- (211.51pt,143.536pt) -- (211.51pt,152.647pt);
	\color[rgb]{1,0,0}
	\draw[line width=1.5pt, line join=miter, line cap=rect](211.51pt,152.647pt) -- (262.26pt,152.647pt) -- (262.26pt,143.536pt) -- (211.51pt,143.536pt) -- (211.51pt,152.647pt);
	\draw[line width=1.5pt, line join=miter, line cap=rect](236.885pt,163.457pt) -- (236.885pt,152.647pt);
	\draw[line width=1.5pt, line join=miter, line cap=rect](236.885pt,128.474pt) -- (236.885pt,143.536pt);
	\draw[line width=1.5pt, line join=miter, line cap=rect](231.81pt,128.474pt) -- (241.96pt,128.474pt);
	\draw[line width=1.5pt, line join=miter, line cap=rect](231.81pt,163.457pt) -- (241.96pt,163.457pt);
	\draw[line width=1.5pt, line join=miter, line cap=rect](211.51pt,152.647pt) -- (262.26pt,152.647pt);
	\draw[line width=1pt, line join=miter, line cap=rect](233.874pt,149.559pt) -- (240.9pt,149.559pt) -- (240.9pt,142.532pt) -- (233.874pt,142.532pt) -- (233.874pt,149.559pt);
	\definecolor{c}{rgb}{0,0,1}
	\fill [pattern color=c, pattern=north east lines](290.806pt,134.425pt) -- (341.556pt,134.425pt) -- (341.556pt,107.092pt) -- (290.806pt,107.092pt) -- (290.806pt,134.425pt);
	\color[rgb]{0,0,1}
	\draw[line width=1.5pt, line join=miter, line cap=rect](290.806pt,134.425pt) -- (341.556pt,134.425pt) -- (341.556pt,107.092pt) -- (290.806pt,107.092pt) -- (290.806pt,134.425pt);
	\draw[line width=1.5pt, line join=miter, line cap=rect](316.181pt,146.253pt) -- (316.181pt,134.425pt);
	\draw[line width=1.5pt, line join=miter, line cap=rect](316.181pt,99.5163pt) -- (316.181pt,107.092pt);
	\draw[line width=1.5pt, line join=miter, line cap=rect](311.106pt,99.5163pt) -- (321.256pt,99.5163pt);
	\draw[line width=1.5pt, line join=miter, line cap=rect](311.106pt,146.253pt) -- (321.256pt,146.253pt);
	\draw[line width=1.5pt, line join=miter, line cap=rect](290.806pt,134.425pt) -- (341.556pt,134.425pt);
	\draw[line width=1pt, line join=miter, line cap=rect](313.17pt,126.472pt) -- (320.196pt,126.472pt) -- (320.196pt,119.446pt) -- (313.17pt,119.446pt) -- (313.17pt,126.472pt);
	\definecolor{c}{rgb}{0,0,0}
	\fill [pattern color=c, pattern=north east lines](370.103pt,134.425pt) -- (420.852pt,134.425pt) -- (420.852pt,125.314pt) -- (370.103pt,125.314pt) -- (370.103pt,134.425pt);
	\color[rgb]{0,0.501961,0}
	\draw[line width=1.5pt, line join=miter, line cap=rect](370.103pt,134.425pt) -- (420.852pt,134.425pt) -- (420.852pt,125.314pt) -- (370.103pt,125.314pt) -- (370.103pt,134.425pt);
	\draw[line width=1.5pt, line join=miter, line cap=rect](395.477pt,139.51pt) -- (395.477pt,134.425pt);
	\draw[line width=1.5pt, line join=miter, line cap=rect](395.477pt,117.193pt) -- (395.477pt,125.314pt);
	\draw[line width=1.5pt, line join=miter, line cap=rect](390.403pt,117.193pt) -- (400.552pt,117.193pt);
	\draw[line width=1.5pt, line join=miter, line cap=rect](390.403pt,139.51pt) -- (400.552pt,139.51pt);
	\draw[line width=1.5pt, line join=miter, line cap=rect](370.103pt,125.314pt) -- (420.852pt,125.314pt);
	\draw[line width=1pt, line join=miter, line cap=rect](392.466pt,131.491pt) -- (399.492pt,131.491pt) -- (399.492pt,124.465pt) -- (392.466pt,124.465pt) -- (392.466pt,131.491pt);
\end{scope}
\begin{scope}
	\color[rgb]{0,0,0}
	\draw[line width=2pt, line join=miter, line cap=rect](76.285pt,382.429pt) -- (476.781pt,382.429pt) -- (476.781pt,82.3075pt) -- (76.285pt,82.3075pt) -- (76.285pt,382.429pt);
	\color[rgb]{0,0,0}
	\pgftext[center, base, at={\pgfpoint{21.6277pt}{240.398pt}},rotate=90]{\fontsize{25}{0}\selectfont{\textbf{Count}}}
	\pgftext[center, base, at={\pgfpoint{57.2137pt}{101.692pt}}]{\fontsize{24}{0}\selectfont{0}}
	\pgftext[center, base, at={\pgfpoint{51.1912pt}{192.03pt}}]{\fontsize{24}{0}\selectfont{10}}
	\pgftext[center, base, at={\pgfpoint{51.1912pt}{283.371pt}}]{\fontsize{24}{0}\selectfont{20}}
	\pgftext[center, base, at={\pgfpoint{51.1912pt}{374.712pt}}]{\fontsize{24}{0}\selectfont{30}}
	\draw[line width=1pt, line join=bevel, line cap=rect](76.285pt,152.647pt) -- (71.2662pt,152.647pt);
	\draw[line width=1pt, line join=bevel, line cap=rect](76.285pt,243.757pt) -- (71.2662pt,243.757pt);
	\draw[line width=1pt, line join=bevel, line cap=rect](76.285pt,334.866pt) -- (71.2662pt,334.866pt);
	\draw[line width=1pt, line join=bevel, line cap=rect](76.285pt,107.092pt) -- (67.2512pt,107.092pt);
	\draw[line width=1pt, line join=bevel, line cap=rect](76.285pt,198.202pt) -- (67.2512pt,198.202pt);
	\draw[line width=1pt, line join=bevel, line cap=rect](76.285pt,289.312pt) -- (67.2512pt,289.312pt);
	\draw[line width=1pt, line join=bevel, line cap=rect](76.285pt,380.421pt) -- (67.2512pt,380.421pt);
	\pgftext[center, base, at={\pgfpoint{277.027pt}{8.46914pt}}]{\fontsize{25}{0}\selectfont{\textbf{Most common SOP}}}
	\pgftext[center, base, at={\pgfpoint{157.589pt}{47.4899pt}}]{\fontsize{24}{0}\selectfont{1}}
	\pgftext[center, base, at={\pgfpoint{236.885pt}{47.4899pt}}]{\fontsize{24}{0}\selectfont{2}}
	\pgftext[center, base, at={\pgfpoint{316.181pt}{47.4899pt}}]{\fontsize{24}{0}\selectfont{3}}
	\pgftext[center, base, at={\pgfpoint{395.477pt}{47.4899pt}}]{\fontsize{24}{0}\selectfont{4}}
	\draw[line width=1pt, line join=bevel, line cap=rect](157.589pt,82.3075pt) -- (157.589pt,73.2737pt);
	\draw[line width=1pt, line join=bevel, line cap=rect](236.885pt,82.3075pt) -- (236.885pt,73.2737pt);
	\draw[line width=1pt, line join=bevel, line cap=rect](316.181pt,82.3075pt) -- (316.181pt,73.2737pt);
	\draw[line width=1pt, line join=bevel, line cap=rect](395.477pt,82.3075pt) -- (395.477pt,73.2737pt);
	\definecolor{c}{rgb}{0,0,0}
	\fill [pattern color=c, pattern=north east lines](88.33pt,278.039pt) -- (120.45pt,278.039pt) -- (120.45pt,264.99pt) -- (88.33pt,264.99pt) -- (88.33pt,278.039pt);
	\draw[line width=1.5pt, line join=miter, line cap=rect](88.33pt,278.039pt) -- (120.45pt,278.039pt) -- (120.45pt,264.99pt) -- (88.33pt,264.99pt) -- (88.33pt,278.039pt);
	\pgftext[left, base, at={\pgfpoint{125.469pt}{265.727pt}}]{\fontsize{18}{0}\selectfont{$A \lor B \lor C \lor D \lor \neg E \lor \neg F \lor \neg G \lor \neg H$}}
	\definecolor{c}{rgb}{1,0,0}
	\fill [pattern color=c, pattern=north east lines](88.33pt,250.937pt) -- (120.45pt,250.937pt) -- (120.45pt,237.889pt) -- (88.33pt,237.889pt) -- (88.33pt,250.937pt);
	\color[rgb]{1,0,0}
	\draw[line width=1.5pt, line join=miter, line cap=rect](88.33pt,250.937pt) -- (120.45pt,250.937pt) -- (120.45pt,237.889pt) -- (88.33pt,237.889pt) -- (88.33pt,250.937pt);
	\color[rgb]{0,0,0}
	\pgftext[left, base, at={\pgfpoint{125.469pt}{238.626pt}}]{\fontsize{18}{0}\selectfont{$A \land B \land C \land D$}}
	\definecolor{c}{rgb}{0,0,1}
	\fill [pattern color=c, pattern=north east lines](88.33pt,223.836pt) -- (120.45pt,223.836pt) -- (120.45pt,210.787pt) -- (88.33pt,210.787pt) -- (88.33pt,223.836pt);
	\color[rgb]{0,0,1}
	\draw[line width=1.5pt, line join=miter, line cap=rect](88.33pt,223.836pt) -- (120.45pt,223.836pt) -- (120.45pt,210.787pt) -- (88.33pt,210.787pt) -- (88.33pt,223.836pt);
	\color[rgb]{0,0,0}
	\pgftext[left, base, at={\pgfpoint{125.469pt}{211.525pt}}]{\fontsize{18}{0}\selectfont{$A \lor B \lor (C \land D)$}}
	\definecolor{c}{rgb}{0,0,0}
	\fill [pattern color=c, pattern=north east lines](88.33pt,196.735pt) -- (120.45pt,196.735pt) -- (120.45pt,183.686pt) -- (88.33pt,183.686pt) -- (88.33pt,196.735pt);
	\color[rgb]{0,0.501961,0}
	\draw[line width=1.5pt, line join=miter, line cap=rect](88.33pt,196.735pt) -- (120.45pt,196.735pt) -- (120.45pt,183.686pt) -- (88.33pt,183.686pt) -- (88.33pt,196.735pt);
	\color[rgb]{0,0,0}
	\pgftext[left, base, at={\pgfpoint{125.469pt}{184.423pt}}]{\fontsize{18}{0}\selectfont{$(A \land B) \lor (B \land D) \lor (C \land D) \lor (A \land \neg C \land \neg D$}}
\end{scope}
\end{tikzpicture}
         }
        \label{fig:BoxPlot_4bits_proteinoid}
    \end{subfigure}
    \begin{subfigure}[c]{0.32\textwidth}
        \centering
        \caption{}
        \resizebox{1\textwidth}{!}{
         \begin{tikzpicture}{0pt}{0pt}{481.8pt}{394.474pt}
	\color[rgb]{1,1,1}
	\fill(0pt,394.474pt) -- (481.8pt,394.474pt) -- (481.8pt,0pt) -- (0pt,0pt) -- (0pt,394.474pt);
\begin{scope}
	\clip(76.285pt,382.429pt) -- (476.781pt,382.429pt) -- (476.781pt,82.3075pt) -- (76.285pt,82.3075pt) -- (76.285pt,382.429pt);
	\definecolor{c}{rgb}{0,0,0}
	\fill [pattern color=c, pattern=north east lines](132.214pt,316.645pt) -- (182.964pt,316.645pt) -- (182.964pt,316.645pt) -- (132.214pt,316.645pt) -- (132.214pt,316.645pt);
	\color[rgb]{0,0,0}
	\draw[line width=1.5pt, line join=miter, line cap=rect](132.214pt,316.645pt) -- (182.964pt,316.645pt) -- (182.964pt,316.645pt) -- (132.214pt,316.645pt) -- (132.214pt,316.645pt);
	\color[rgb]{0,0,0}
	\draw[line width=1.5pt, line join=miter, line cap=rect](152.514pt,316.645pt) -- (162.664pt,316.645pt);
	\draw[line width=1.5pt, line join=miter, line cap=rect](152.514pt,316.645pt) -- (162.664pt,316.645pt);
	\draw[line width=1.5pt, line join=miter, line cap=rect](132.214pt,316.645pt) -- (182.964pt,316.645pt);
	\draw[line width=1pt, line join=miter, line cap=rect](154.577pt,320.196pt) -- (161.604pt,320.196pt) -- (161.604pt,313.17pt) -- (154.577pt,313.17pt) -- (154.577pt,320.196pt);
	\definecolor{c}{rgb}{1,0,0}
	\fill [pattern color=c, pattern=north east lines](211.51pt,152.647pt) -- (262.26pt,152.647pt) -- (262.26pt,143.536pt) -- (211.51pt,143.536pt) -- (211.51pt,152.647pt);
	\color[rgb]{1,0,0}
	\draw[line width=1.5pt, line join=miter, line cap=rect](211.51pt,152.647pt) -- (262.26pt,152.647pt) -- (262.26pt,143.536pt) -- (211.51pt,143.536pt) -- (211.51pt,152.647pt);
	\draw[line width=1.5pt, line join=miter, line cap=rect](236.885pt,163.457pt) -- (236.885pt,152.647pt);
	\draw[line width=1.5pt, line join=miter, line cap=rect](236.885pt,128.474pt) -- (236.885pt,143.536pt);
	\draw[line width=1.5pt, line join=miter, line cap=rect](231.81pt,128.474pt) -- (241.96pt,128.474pt);
	\draw[line width=1.5pt, line join=miter, line cap=rect](231.81pt,163.457pt) -- (241.96pt,163.457pt);
	\draw[line width=1.5pt, line join=miter, line cap=rect](211.51pt,152.647pt) -- (262.26pt,152.647pt);
	\draw[line width=1pt, line join=miter, line cap=rect](233.874pt,149.559pt) -- (240.9pt,149.559pt) -- (240.9pt,142.532pt) -- (233.874pt,142.532pt) -- (233.874pt,149.559pt);
	\definecolor{c}{rgb}{0,0,1}
	\fill [pattern color=c, pattern=north east lines](290.806pt,134.425pt) -- (341.556pt,134.425pt) -- (341.556pt,107.092pt) -- (290.806pt,107.092pt) -- (290.806pt,134.425pt);
	\color[rgb]{0,0,1}
	\draw[line width=1.5pt, line join=miter, line cap=rect](290.806pt,134.425pt) -- (341.556pt,134.425pt) -- (341.556pt,107.092pt) -- (290.806pt,107.092pt) -- (290.806pt,134.425pt);
	\draw[line width=1.5pt, line join=miter, line cap=rect](316.181pt,146.253pt) -- (316.181pt,134.425pt);
	\draw[line width=1.5pt, line join=miter, line cap=rect](316.181pt,99.5163pt) -- (316.181pt,107.092pt);
	\draw[line width=1.5pt, line join=miter, line cap=rect](311.106pt,99.5163pt) -- (321.256pt,99.5163pt);
	\draw[line width=1.5pt, line join=miter, line cap=rect](311.106pt,146.253pt) -- (321.256pt,146.253pt);
	\draw[line width=1.5pt, line join=miter, line cap=rect](290.806pt,134.425pt) -- (341.556pt,134.425pt);
	\draw[line width=1pt, line join=miter, line cap=rect](313.17pt,126.472pt) -- (320.196pt,126.472pt) -- (320.196pt,119.446pt) -- (313.17pt,119.446pt) -- (313.17pt,126.472pt);
	\definecolor{c}{rgb}{0,0,0}
	\fill [pattern color=c, pattern=north east lines](370.103pt,134.425pt) -- (420.852pt,134.425pt) -- (420.852pt,125.314pt) -- (370.103pt,125.314pt) -- (370.103pt,134.425pt);
	\color[rgb]{0,0.501961,0}
	\draw[line width=1.5pt, line join=miter, line cap=rect](370.103pt,134.425pt) -- (420.852pt,134.425pt) -- (420.852pt,125.314pt) -- (370.103pt,125.314pt) -- (370.103pt,134.425pt);
	\draw[line width=1.5pt, line join=miter, line cap=rect](395.477pt,139.51pt) -- (395.477pt,134.425pt);
	\draw[line width=1.5pt, line join=miter, line cap=rect](395.477pt,117.193pt) -- (395.477pt,125.314pt);
	\draw[line width=1.5pt, line join=miter, line cap=rect](390.403pt,117.193pt) -- (400.552pt,117.193pt);
	\draw[line width=1.5pt, line join=miter, line cap=rect](390.403pt,139.51pt) -- (400.552pt,139.51pt);
	\draw[line width=1.5pt, line join=miter, line cap=rect](370.103pt,125.314pt) -- (420.852pt,125.314pt);
	\draw[line width=1pt, line join=miter, line cap=rect](392.466pt,131.491pt) -- (399.492pt,131.491pt) -- (399.492pt,124.465pt) -- (392.466pt,124.465pt) -- (392.466pt,131.491pt);
\end{scope}
\begin{scope}
	\color[rgb]{0,0,0}
	\draw[line width=2pt, line join=miter, line cap=rect](76.285pt,382.429pt) -- (476.781pt,382.429pt) -- (476.781pt,82.3075pt) -- (76.285pt,82.3075pt) -- (76.285pt,382.429pt);
	\color[rgb]{0,0,0}
	\pgftext[center, base, at={\pgfpoint{21.6277pt}{240.398pt}},rotate=90]{\fontsize{25}{0}\selectfont{\textbf{Count}}}
	\pgftext[center, base, at={\pgfpoint{57.2137pt}{101.692pt}}]{\fontsize{24}{0}\selectfont{0}}
	\pgftext[center, base, at={\pgfpoint{51.1912pt}{192.03pt}}]{\fontsize{24}{0}\selectfont{10}}
	\pgftext[center, base, at={\pgfpoint{51.1912pt}{283.371pt}}]{\fontsize{24}{0}\selectfont{20}}
	\pgftext[center, base, at={\pgfpoint{51.1912pt}{374.712pt}}]{\fontsize{24}{0}\selectfont{30}}
	\draw[line width=1pt, line join=bevel, line cap=rect](76.285pt,152.647pt) -- (71.2662pt,152.647pt);
	\draw[line width=1pt, line join=bevel, line cap=rect](76.285pt,243.757pt) -- (71.2662pt,243.757pt);
	\draw[line width=1pt, line join=bevel, line cap=rect](76.285pt,334.866pt) -- (71.2662pt,334.866pt);
	\draw[line width=1pt, line join=bevel, line cap=rect](76.285pt,107.092pt) -- (67.2512pt,107.092pt);
	\draw[line width=1pt, line join=bevel, line cap=rect](76.285pt,198.202pt) -- (67.2512pt,198.202pt);
	\draw[line width=1pt, line join=bevel, line cap=rect](76.285pt,289.312pt) -- (67.2512pt,289.312pt);
	\draw[line width=1pt, line join=bevel, line cap=rect](76.285pt,380.421pt) -- (67.2512pt,380.421pt);
	\pgftext[center, base, at={\pgfpoint{277.027pt}{8.46914pt}}]{\fontsize{25}{0}\selectfont{\textbf{Most common SOP}}}
	\pgftext[center, base, at={\pgfpoint{157.589pt}{47.4899pt}}]{\fontsize{24}{0}\selectfont{1}}
	\pgftext[center, base, at={\pgfpoint{236.885pt}{47.4899pt}}]{\fontsize{24}{0}\selectfont{2}}
	\pgftext[center, base, at={\pgfpoint{316.181pt}{47.4899pt}}]{\fontsize{24}{0}\selectfont{3}}
	\pgftext[center, base, at={\pgfpoint{395.477pt}{47.4899pt}}]{\fontsize{24}{0}\selectfont{4}}
	\draw[line width=1pt, line join=bevel, line cap=rect](157.589pt,82.3075pt) -- (157.589pt,73.2737pt);
	\draw[line width=1pt, line join=bevel, line cap=rect](236.885pt,82.3075pt) -- (236.885pt,73.2737pt);
	\draw[line width=1pt, line join=bevel, line cap=rect](316.181pt,82.3075pt) -- (316.181pt,73.2737pt);
	\draw[line width=1pt, line join=bevel, line cap=rect](395.477pt,82.3075pt) -- (395.477pt,73.2737pt);
	\definecolor{c}{rgb}{0,0,0}
	\fill [pattern color=c, pattern=north east lines](88.33pt,278.039pt) -- (120.45pt,278.039pt) -- (120.45pt,264.99pt) -- (88.33pt,264.99pt) -- (88.33pt,278.039pt);
	\draw[line width=1.5pt, line join=miter, line cap=rect](88.33pt,278.039pt) -- (120.45pt,278.039pt) -- (120.45pt,264.99pt) -- (88.33pt,264.99pt) -- (88.33pt,278.039pt);
	\pgftext[left, base, at={\pgfpoint{125.469pt}{265.727pt}}]{\fontsize{18}{0}\selectfont{$A \lor B \lor C \lor D \lor \neg E \lor \neg F \lor \neg G \lor \neg H$}}
	\definecolor{c}{rgb}{1,0,0}
	\fill [pattern color=c, pattern=north east lines](88.33pt,250.937pt) -- (120.45pt,250.937pt) -- (120.45pt,237.889pt) -- (88.33pt,237.889pt) -- (88.33pt,250.937pt);
	\color[rgb]{1,0,0}
	\draw[line width=1.5pt, line join=miter, line cap=rect](88.33pt,250.937pt) -- (120.45pt,250.937pt) -- (120.45pt,237.889pt) -- (88.33pt,237.889pt) -- (88.33pt,250.937pt);
	\color[rgb]{0,0,0}
	\pgftext[left, base, at={\pgfpoint{125.469pt}{238.626pt}}]{\fontsize{18}{0}\selectfont{$A \land B \land C \land D$}}
	\definecolor{c}{rgb}{0,0,1}
	\fill [pattern color=c, pattern=north east lines](88.33pt,223.836pt) -- (120.45pt,223.836pt) -- (120.45pt,210.787pt) -- (88.33pt,210.787pt) -- (88.33pt,223.836pt);
	\color[rgb]{0,0,1}
	\draw[line width=1.5pt, line join=miter, line cap=rect](88.33pt,223.836pt) -- (120.45pt,223.836pt) -- (120.45pt,210.787pt) -- (88.33pt,210.787pt) -- (88.33pt,223.836pt);
	\color[rgb]{0,0,0}
	\pgftext[left, base, at={\pgfpoint{125.469pt}{211.525pt}}]{\fontsize{18}{0}\selectfont{$A \lor B \lor (C \land D)$}}
	\definecolor{c}{rgb}{0,0,0}
	\fill [pattern color=c, pattern=north east lines](88.33pt,196.735pt) -- (120.45pt,196.735pt) -- (120.45pt,183.686pt) -- (88.33pt,183.686pt) -- (88.33pt,196.735pt);
	\color[rgb]{0,0.501961,0}
	\draw[line width=1.5pt, line join=miter, line cap=rect](88.33pt,196.735pt) -- (120.45pt,196.735pt) -- (120.45pt,183.686pt) -- (88.33pt,183.686pt) -- (88.33pt,196.735pt);
	\color[rgb]{0,0,0}
	\pgftext[left, base, at={\pgfpoint{125.469pt}{184.423pt}}]{\fontsize{18}{0}\selectfont{$(A \land B) \lor (B \land D) \lor (C \land D) \lor (A \land \neg C \land \neg D$}}
\end{scope}
\end{tikzpicture}
         }
        \label{fig:BoxPlot_8bits_proteinoid}
    \end{subfigure}
    \caption{Boxplots showing the distribution of the four most common SOP expressions found in repeated experiments with (a) 2-bit, (b) 4-bit, and (c) 8-bit binary strings for proteinoid. Whiskers indicate 1.5 times the standard deviation.}
    \label{fig:boxplot_prot}
\end{figure}

\begin{figure}[!tbp]
    \centering
    \begin{subfigure}[c]{0.32\textwidth}
        \centering
        \caption{}
        \includegraphics[width=1\textwidth, keepaspectratio]{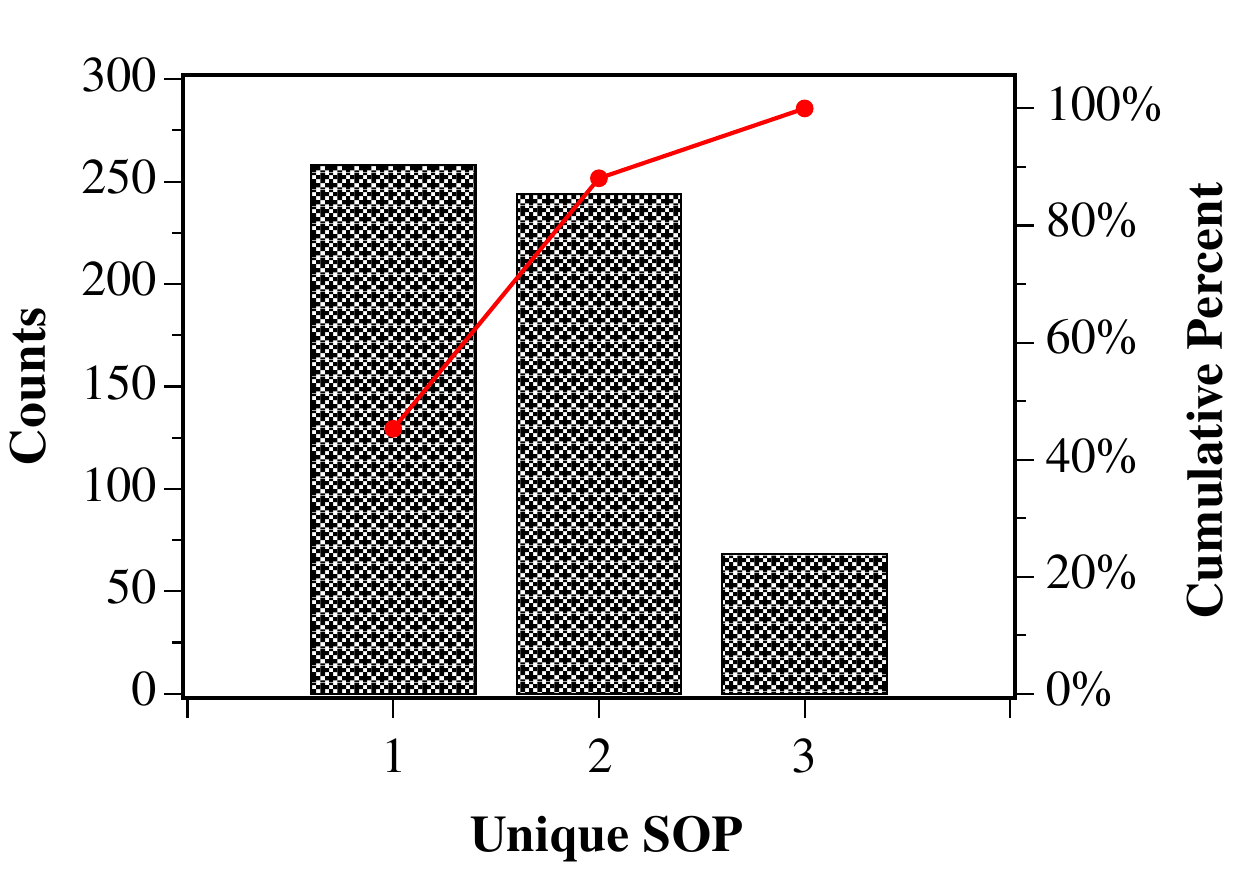}
        \label{fig:Pareto_2bits_proteinoid}
    \end{subfigure}
    \begin{subfigure}[c]{0.32\textwidth}
        \centering
        \caption{}
        \includegraphics[width=1\textwidth, keepaspectratio]{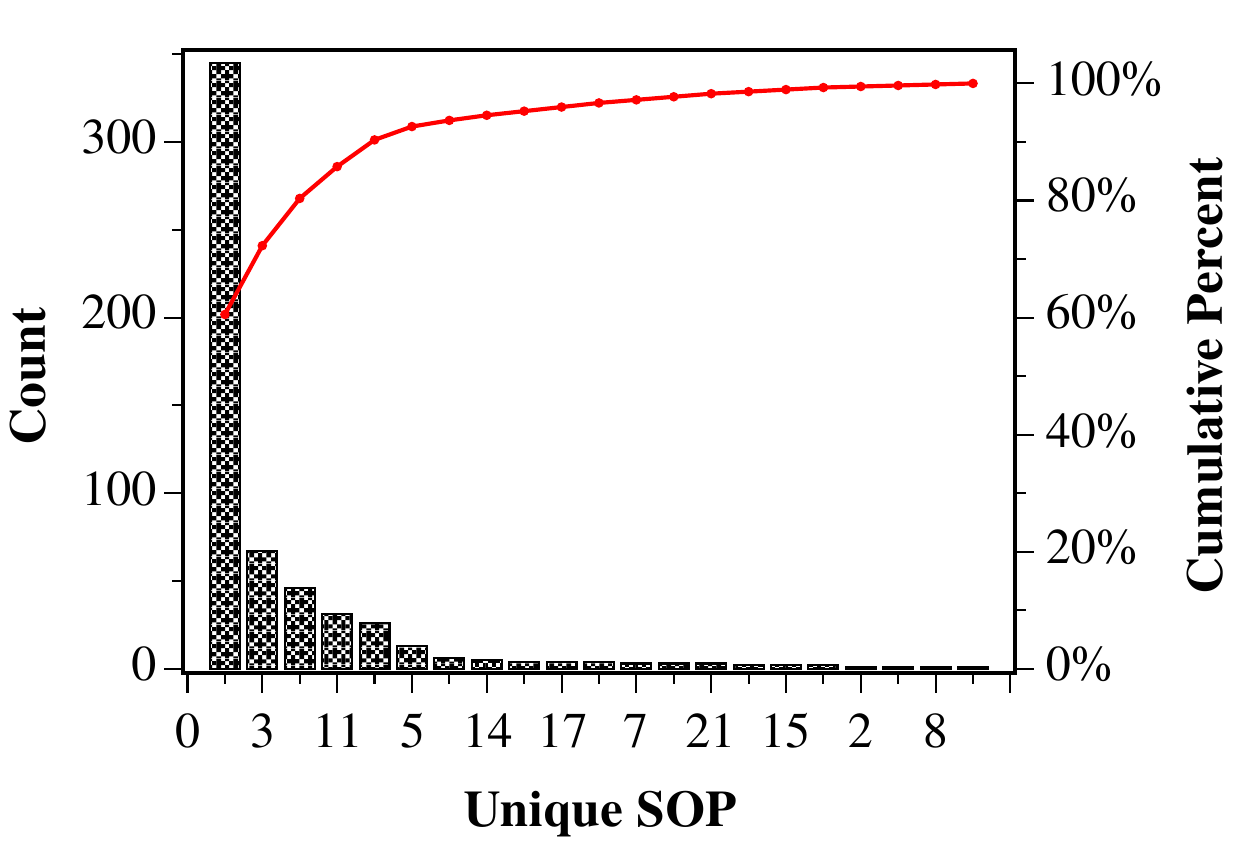}
        \label{fig:Pareto_4bits_proteinoid}
    \end{subfigure}
    \begin{subfigure}[c]{0.32\textwidth}
        \centering
        \caption{}
        \includegraphics[width=1\textwidth, keepaspectratio]{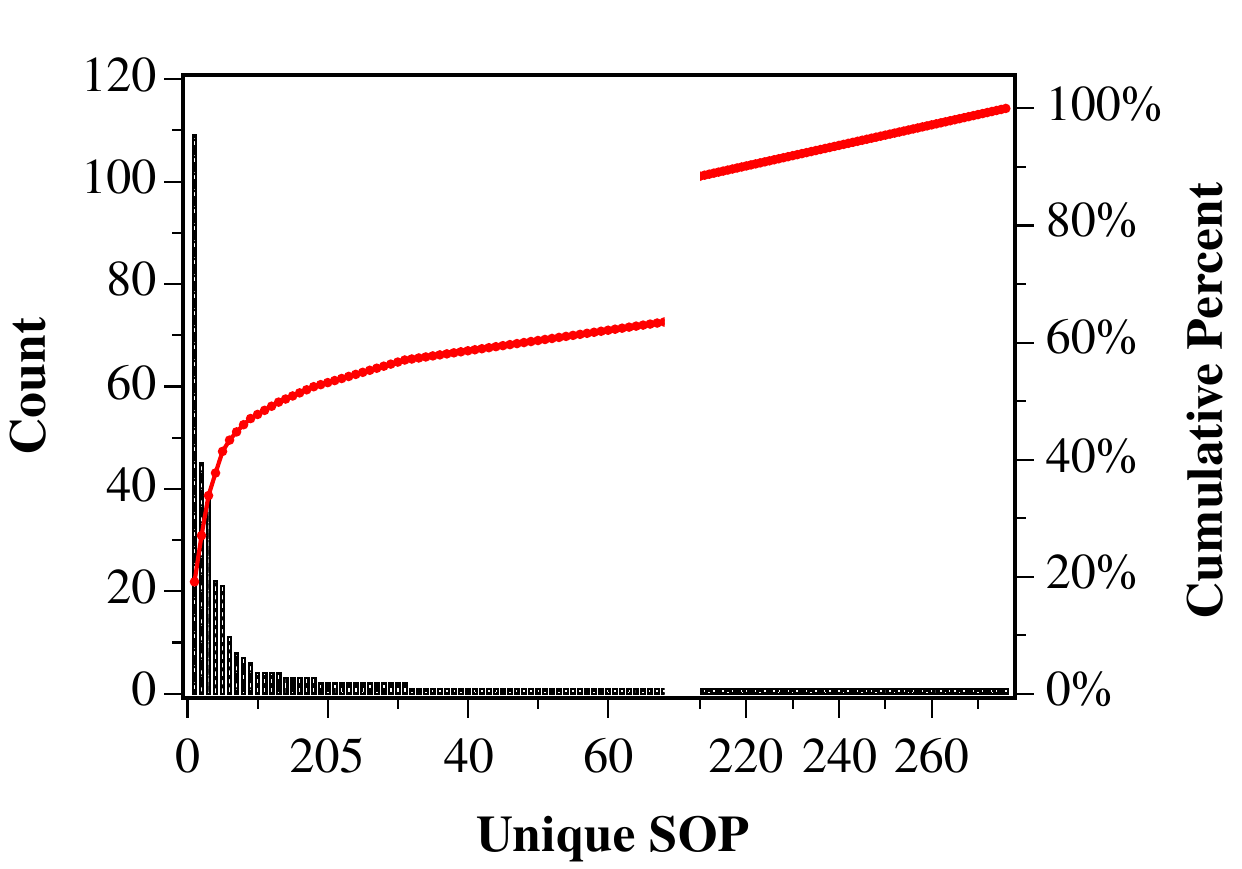}
        \label{fig:Pareto_8bits_proteinoid}
    \end{subfigure}

    \caption{Pareto charts showing the frequency counts of unique sum-of-products from repeated experiments for proteinoid. The red curve represents the cumulative percentage of the total count for (a) 2-bit, (b) 4-bit, and (c) 8-bit.}
    \label{fig:pareto_proteinoid}
\end{figure}

The study reproducibility of the extracted logic circuits using proteinoid was assessed across 15 repetitions of experiments for each binary input string. This was conducted in a similar manner to the prior ZnO colloidal suspension experiments, wherein the samples were shaken before each repetition to disrupt any existing structures. Boxplots in Fig.~\ref{fig:boxplot_prot} illustrate the outcomes for 2-, 4-, and 8-bit inputs, with whiskers denoting 1.5 times the standard deviation across the repeats. The consistent occurrence of the most frequent logical expressions demonstrates the ability to create logical circuits for a particular input order in a deterministic manner, although it is less robust compared to ZnO. Pareto charts in Fig.~\ref{fig:pareto_proteinoid} illustrate the results from various tests. Conversely, these histograms indicate that proteinoids yield fewer extracted SOPs in comparison to the ZnO colloidal suspension.

\section{ZnO + proteinoids}
A total of 4, 6, and 24 unique standard canonical sum-of-products (SOP) Boolean logical expressions were obtained for the 2-, 4-, 8-bit input experiments with a mixture of ZnO nanoparticles and proteinoid, respectively. The four most frequent logical expressions identified for each input binary string are displayed in Tables~\ref{tab:sop_2bit}, ~\ref{tab:sop_4bit} and ~\ref{tab:sop_8bit}. Again, the logical complexity of the frequent expressions escalated with greater input variables, however, in a manner distinct from either the isolated ZnO colloidal suspension or proteinoid systems. For 2-bit inputs, the dominant expression was a simple conjunctive term, contrasting the disjunctive forms seen previously for ZnO suspension and logical 0 for proteinoid. The most common 4-bit input expressions comprised 3 disjunctive and negated clauses, while the sample with just ZnO yielded 4 disjunctive and 8 conjunctive terms. The most frequent 8-bit input expressions consisted of 8 disjunctive and negated variables, exhibiting an intermediate complexity. The responses of this mixture highlight the uniqueness of the responses of each material.
\begin{table}[!ht]
\centering
\caption{Extracted sum-of-products Boolean expression for a 2-bit string input with varying thresholds}
\label{tab:sop_2bit}
\begin{tabular}{lr}
\toprule
SOP                            & Count   \\ \midrule
$A\land B$                     & 21     \\
$A\lor B$                      & 14     \\ 
$B$                            & 2      \\
$A$                            & 1       \\ \bottomrule
\end{tabular}
\end{table}
\begin{table}[!ht]
\centering
\caption{Extracted sum-of-products Boolean expression for a 2-bit string input with varying thresholds}
\label{tab:sop_4bit}
\begin{tabular}{@{}lr@{}}
\toprule
SOP                                                                                                                                                 & Count \\ \midrule
$\neg A \lor \neg B \lor \neg C$                                                                                                                            & 15   \\
$(A \land \neg B) \lor (B \land \neg C) \lor (D \land \neg A)$     & 7    \\
$A \land B \land D \land \neg C$ & 5    \\
$(A \land B \land \neg C) \lor (A \land D \land \neg B) \lor (B \land D \land \neg C)$                                                                          & 4    \\ \bottomrule
\end{tabular}
\end{table}

\begin{table}[!ht]
\centering
\caption{Extracted sum-of-products Boolean expression for a 2 bit string input with varying thresholds}
\label{tab:sop_8bit}
\resizebox{1\textwidth}{!}{
\begin{tabular}{@{}lr@{}}
\toprule
SOP                                                                                                                                                                                       & Count \\ \midrule
$ A \lor  B \lor C \lor  D \lor  E \lor  F \lor  G \lor  H$& 6\\
$A \land B \land C \land D \land E \land F \land G \land \neg H$& 3\\
$(A \land \neg D) \lor (B \land \neg G) \lor (C \land \neg F) \lor (D \land \neg E) \lor (E \land \neg C) \lor (F \land \neg B) \lor (G \land \neg A) \lor (G \land \neg H)$& 2\\
$(A \lor C \lor D \lor E \lor (B \land F) \lor (B \land G) \lor (B \land H) \lor (F \land G) \lor (F \land H) \lor (G \land H)$& 1\\ \bottomrule
\end{tabular}}
\end{table}

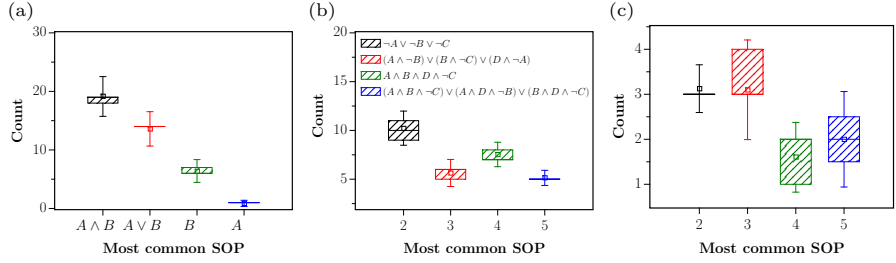
\begin{figure}[!tb]
    \centering
    \begin{subfigure}[c]{0.32\textwidth}
        \centering
        \caption{}
        \resizebox{1\textwidth}{!}{
         \begin{tikzpicture}{0pt}{0pt}{481.8pt}{394.474pt}
	\color[rgb]{1,1,1}
	\fill(0pt,394.474pt) -- (481.8pt,394.474pt) -- (481.8pt,0pt) -- (0pt,0pt) -- (0pt,394.474pt);
\begin{scope}
	\clip(76.285pt,382.429pt) -- (476.781pt,382.429pt) -- (476.781pt,82.3075pt) -- (76.285pt,82.3075pt) -- (76.285pt,382.429pt);
	\definecolor{c}{rgb}{0,0,0}
	\fill [pattern color=c, pattern=north east lines](132.214pt,271.849pt) -- (182.964pt,271.849pt) -- (182.964pt,261.979pt) -- (132.214pt,261.979pt) -- (132.214pt,271.849pt);
	\color[rgb]{0,0,0}
	\draw[line width=1.5pt, line join=miter, line cap=rect](132.214pt,271.849pt) -- (182.964pt,271.849pt) -- (182.964pt,261.979pt) -- (132.214pt,261.979pt) -- (132.214pt,271.849pt);
	\color[rgb]{0,0,0}
	\draw[line width=1.5pt, line join=bevel, line cap=rect](157.589pt,306.678pt) -- (157.589pt,271.849pt);
	\draw[line width=1.5pt, line join=bevel, line cap=rect](157.589pt,239.652pt) -- (157.589pt,261.979pt);
	\draw[line width=1.5pt, line join=bevel, line cap=rect](152.514pt,239.652pt) -- (162.664pt,239.652pt);
	\draw[line width=1.5pt, line join=bevel, line cap=rect](152.514pt,306.678pt) -- (162.664pt,306.678pt);
	\draw[line width=1.5pt, line join=bevel, line cap=rect](132.214pt,271.849pt) -- (182.964pt,271.849pt);
	\draw[line width=1pt, line join=miter, line cap=rect](154.577pt,277.035pt) -- (161.604pt,277.035pt) -- (161.604pt,270.009pt) -- (154.577pt,270.009pt) -- (154.577pt,277.035pt);
	\definecolor{c}{rgb}{1,0,0}
	\fill [pattern color=c, pattern=north east lines](211.51pt,222.498pt) -- (262.26pt,222.498pt) -- (262.26pt,222.498pt) -- (211.51pt,222.498pt) -- (211.51pt,222.498pt);
	\color[rgb]{1,0,0}
	\draw[line width=1.5pt, line join=miter, line cap=rect](211.51pt,222.498pt) -- (262.26pt,222.498pt) -- (262.26pt,222.498pt) -- (211.51pt,222.498pt) -- (211.51pt,222.498pt);
	\draw[line width=1.5pt, line join=miter, line cap=rect](236.885pt,247.519pt) -- (236.885pt,222.498pt);
	\draw[line width=1.5pt, line join=miter, line cap=rect](236.885pt,189.581pt) -- (236.885pt,222.498pt);
	\draw[line width=1.5pt, line join=miter, line cap=rect](231.81pt,189.581pt) -- (241.96pt,189.581pt);
	\draw[line width=1.5pt, line join=miter, line cap=rect](231.81pt,247.519pt) -- (241.96pt,247.519pt);
	\draw[line width=1.5pt, line join=miter, line cap=rect](211.51pt,222.498pt) -- (262.26pt,222.498pt);
	\draw[line width=1pt, line join=miter, line cap=rect](233.874pt,221.829pt) -- (240.9pt,221.829pt) -- (240.9pt,214.802pt) -- (233.874pt,214.802pt) -- (233.874pt,221.829pt);
	\definecolor{c}{rgb}{0,0.501961,0}
	\fill [pattern color=c, pattern=north east lines](290.806pt,153.406pt) -- (341.556pt,153.406pt) -- (341.556pt,143.536pt) -- (290.806pt,143.536pt) -- (290.806pt,153.406pt);
	\color[rgb]{0,0.501961,0}
	\draw[line width=1.5pt, line join=miter, line cap=rect](290.806pt,153.406pt) -- (341.556pt,153.406pt) -- (341.556pt,143.536pt) -- (290.806pt,143.536pt) -- (290.806pt,153.406pt);
	\draw[line width=1.5pt, line join=miter, line cap=rect](316.181pt,166.707pt) -- (316.181pt,153.406pt);
	\draw[line width=1.5pt, line join=miter, line cap=rect](316.181pt,128.262pt) -- (316.181pt,143.536pt);
	\draw[line width=1.5pt, line join=miter, line cap=rect](311.106pt,128.262pt) -- (321.256pt,128.262pt);
	\draw[line width=1.5pt, line join=miter, line cap=rect](311.106pt,166.707pt) -- (321.256pt,166.707pt);
	\draw[line width=1.5pt, line join=miter, line cap=rect](290.806pt,143.536pt) -- (341.556pt,143.536pt);
	\draw[line width=1pt, line join=miter, line cap=rect](313.17pt,150.563pt) -- (320.196pt,150.563pt) -- (320.196pt,143.536pt) -- (313.17pt,143.536pt) -- (313.17pt,150.563pt);
	\definecolor{c}{rgb}{0,0,1}
	\fill [pattern color=c, pattern=north east lines](370.103pt,94.1852pt) -- (420.852pt,94.1852pt) -- (420.852pt,94.1852pt) -- (370.103pt,94.1852pt) -- (370.103pt,94.1852pt);
	\color[rgb]{0,0,1}
	\draw[line width=1.5pt, line join=miter, line cap=rect](370.103pt,94.1852pt) -- (420.852pt,94.1852pt) -- (420.852pt,94.1852pt) -- (370.103pt,94.1852pt) -- (370.103pt,94.1852pt);
	\draw[line width=1.5pt, line join=miter, line cap=rect](395.477pt,98.0787pt) -- (395.477pt,94.1852pt);
	\draw[line width=1.5pt, line join=miter, line cap=rect](395.477pt,87.6597pt) -- (395.477pt,94.1852pt);
	\draw[line width=1.5pt, line join=miter, line cap=rect](390.403pt,87.6597pt) -- (400.552pt,87.6597pt);
	\draw[line width=1.5pt, line join=miter, line cap=rect](390.403pt,98.0787pt) -- (400.552pt,98.0787pt);
	\draw[line width=1.5pt, line join=miter, line cap=rect](370.103pt,94.1852pt) -- (420.852pt,94.1852pt);
	\draw[line width=1pt, line join=miter, line cap=rect](392.466pt,96.36pt) -- (399.492pt,96.36pt) -- (399.492pt,89.3337pt) -- (392.466pt,89.3337pt) -- (392.466pt,96.36pt);
\end{scope}
\begin{scope}
	\color[rgb]{0,0,0}
	\draw[line width=2pt, line join=miter, line cap=rect](76.285pt,382.429pt) -- (476.781pt,382.429pt) -- (476.781pt,82.3075pt) -- (76.285pt,82.3075pt) -- (76.285pt,382.429pt);
	\color[rgb]{0,0,0}
	\pgftext[center, base, at={\pgfpoint{21.6277pt}{232.368pt}},rotate=90]{\fontsize{25}{0}\selectfont{\textbf{Count}}}
	\pgftext[center, base, at={\pgfpoint{57.2137pt}{78.6062pt}}]{\fontsize{24}{0}\selectfont{0}}
	\pgftext[center, base, at={\pgfpoint{51.1912pt}{176.974pt}}]{\fontsize{24}{0}\selectfont{10}}
	\pgftext[center, base, at={\pgfpoint{51.1912pt}{276.345pt}}]{\fontsize{24}{0}\selectfont{20}}
	\pgftext[center, base, at={\pgfpoint{51.1912pt}{374.712pt}}]{\fontsize{24}{0}\selectfont{30}}
	\draw[line width=1pt, line join=bevel, line cap=rect](76.285pt,133.666pt) -- (71.2662pt,133.666pt);
	\draw[line width=1pt, line join=bevel, line cap=rect](76.285pt,232.368pt) -- (71.2662pt,232.368pt);
	\draw[line width=1pt, line join=bevel, line cap=rect](76.285pt,331.07pt) -- (71.2662pt,331.07pt);
	\draw[line width=1pt, line join=bevel, line cap=rect](76.285pt,84.315pt) -- (67.2512pt,84.315pt);
	\draw[line width=1pt, line join=bevel, line cap=rect](76.285pt,183.017pt) -- (67.2512pt,183.017pt);
	\draw[line width=1pt, line join=bevel, line cap=rect](76.285pt,281.719pt) -- (67.2512pt,281.719pt);
	\draw[line width=1pt, line join=bevel, line cap=rect](76.285pt,380.421pt) -- (67.2512pt,380.421pt);
	\pgftext[center, base, at={\pgfpoint{277.027pt}{8.46914pt}}]{\fontsize{25}{0}\selectfont{\textbf{Most common SOP}}}
	\pgftext[center, base, at={\pgfpoint{145.042pt}{47.4899pt}}]{\fontsize{24}{0}\selectfont{$A\land B$}}
	\pgftext[center, base, at={\pgfpoint{223.836pt}{47.4899pt}}]{\fontsize{24}{0}\selectfont{$A\lor B$}}
	\pgftext[center, base, at={\pgfpoint{303.132pt}{47.4899pt}}]{\fontsize{24}{0}\selectfont{$B$}}
	\pgftext[center, base, at={\pgfpoint{382.931pt}{47.4899pt}}]{\fontsize{24}{0}\selectfont{$A$}}
	\draw[line width=1pt, line join=bevel, line cap=rect](157.589pt,82.3075pt) -- (157.589pt,73.2737pt);
	\draw[line width=1pt, line join=bevel, line cap=rect](236.885pt,82.3075pt) -- (236.885pt,73.2737pt);
	\draw[line width=1pt, line join=bevel, line cap=rect](316.181pt,82.3075pt) -- (316.181pt,73.2737pt);
	\draw[line width=1pt, line join=bevel, line cap=rect](395.477pt,82.3075pt) -- (395.477pt,73.2737pt);
\end{scope}
\end{tikzpicture}
         }
        \label{fig:BoxPlot2bit}
    \end{subfigure}
    \begin{subfigure}[c]{0.32\textwidth}
        \centering
        \caption{}
        \resizebox{1\textwidth}{!}{
         \begin{tikzpicture}{0pt}{0pt}{481.8pt}{394.474pt}
	\color[rgb]{1,1,1}
	\fill(0pt,394.474pt) -- (481.8pt,394.474pt) -- (481.8pt,0pt) -- (0pt,0pt) -- (0pt,394.474pt);
\begin{scope}
	\clip(76.285pt,382.429pt) -- (476.781pt,382.429pt) -- (476.781pt,82.3075pt) -- (76.285pt,82.3075pt) -- (76.285pt,382.429pt);
	\definecolor{c}{rgb}{0,0,0}
	\fill [pattern color=c, pattern=north east lines](132.214pt,232.368pt) -- (182.964pt,232.368pt) -- (182.964pt,199.467pt) -- (132.214pt,199.467pt) -- (132.214pt,232.368pt);
	\color[rgb]{0,0,0}
	\draw[line width=1.5pt, line join=miter, line cap=rect](132.214pt,232.368pt) -- (182.964pt,232.368pt) -- (182.964pt,199.467pt) -- (132.214pt,199.467pt) -- (132.214pt,232.368pt);
	\color[rgb]{0,0,0}
	\draw[line width=1.5pt, line join=bevel, line cap=rect](157.589pt,248.48pt) -- (157.589pt,232.368pt);
	\draw[line width=1.5pt, line join=bevel, line cap=rect](157.589pt,190.949pt) -- (157.589pt,199.467pt);
	\draw[line width=1.5pt, line join=bevel, line cap=rect](152.514pt,190.949pt) -- (162.664pt,190.949pt);
	\draw[line width=1.5pt, line join=bevel, line cap=rect](152.514pt,248.48pt) -- (162.664pt,248.48pt);
	\draw[line width=1.5pt, line join=miter, line cap=rect](132.214pt,215.918pt) -- (182.964pt,215.918pt);
	\draw[line width=1pt, line join=miter, line cap=rect](154.577pt,222.832pt) -- (161.604pt,222.832pt) -- (161.604pt,215.806pt) -- (154.577pt,215.806pt) -- (154.577pt,222.832pt);
	\definecolor{c}{rgb}{1,0,0}
	\fill [pattern color=c, pattern=north east lines](211.51pt,150.116pt) -- (262.26pt,150.116pt) -- (262.26pt,133.666pt) -- (211.51pt,133.666pt) -- (211.51pt,150.116pt);
	\color[rgb]{1,0,0}
	\draw[line width=1.5pt, line join=miter, line cap=rect](211.51pt,150.116pt) -- (262.26pt,150.116pt) -- (262.26pt,133.666pt) -- (211.51pt,133.666pt) -- (211.51pt,150.116pt);
	\draw[line width=1.5pt, line join=miter, line cap=rect](236.885pt,166.945pt) -- (236.885pt,150.116pt);
	\draw[line width=1.5pt, line join=miter, line cap=rect](236.885pt,121.324pt) -- (236.885pt,133.666pt);
	\draw[line width=1.5pt, line join=miter, line cap=rect](231.81pt,121.324pt) -- (241.96pt,121.324pt);
	\draw[line width=1.5pt, line join=miter, line cap=rect](231.81pt,166.945pt) -- (241.96pt,166.945pt);
	\draw[line width=1.5pt, line join=miter, line cap=rect](211.51pt,150.116pt) -- (262.26pt,150.116pt);
	\draw[line width=1pt, line join=miter, line cap=rect](233.874pt,147.551pt) -- (240.9pt,147.551pt) -- (240.9pt,140.525pt) -- (233.874pt,140.525pt) -- (233.874pt,147.551pt);
	\definecolor{c}{rgb}{0,0.501961,0}
	\fill [pattern color=c, pattern=north east lines](290.806pt,183.017pt) -- (341.556pt,183.017pt) -- (341.556pt,166.567pt) -- (290.806pt,166.567pt) -- (290.806pt,183.017pt);
	\color[rgb]{0,0.501961,0}
	\draw[line width=1.5pt, line join=miter, line cap=rect](290.806pt,183.017pt) -- (341.556pt,183.017pt) -- (341.556pt,166.567pt) -- (290.806pt,166.567pt) -- (290.806pt,183.017pt);
	\draw[line width=1.5pt, line join=miter, line cap=rect](316.181pt,195.915pt) -- (316.181pt,183.017pt);
	\draw[line width=1.5pt, line join=miter, line cap=rect](316.181pt,154.766pt) -- (316.181pt,166.567pt);
	\draw[line width=1.5pt, line join=miter, line cap=rect](311.106pt,154.766pt) -- (321.256pt,154.766pt);
	\draw[line width=1.5pt, line join=miter, line cap=rect](311.106pt,195.915pt) -- (321.256pt,195.915pt);
	\draw[line width=1.5pt, line join=miter, line cap=rect](290.806pt,166.567pt) -- (341.556pt,166.567pt);
	\draw[line width=1pt, line join=miter, line cap=rect](313.17pt,178.667pt) -- (320.196pt,178.667pt) -- (320.196pt,171.641pt) -- (313.17pt,171.641pt) -- (313.17pt,178.667pt);
	\definecolor{c}{rgb}{0,0,1}
	\fill [pattern color=c, pattern=north east lines](370.103pt,133.666pt) -- (420.852pt,133.666pt) -- (420.852pt,133.666pt) -- (370.103pt,133.666pt) -- (370.103pt,133.666pt);
	\color[rgb]{0,0,1}
	\draw[line width=1.5pt, line join=miter, line cap=rect](370.103pt,133.666pt) -- (420.852pt,133.666pt) -- (420.852pt,133.666pt) -- (370.103pt,133.666pt) -- (370.103pt,133.666pt);
	\draw[line width=1.5pt, line join=miter, line cap=rect](395.477pt,148.602pt) -- (395.477pt,133.666pt);
	\draw[line width=1.5pt, line join=miter, line cap=rect](395.477pt,123.117pt) -- (395.477pt,133.666pt);
	\draw[line width=1.5pt, line join=miter, line cap=rect](390.403pt,123.117pt) -- (400.552pt,123.117pt);
	\draw[line width=1.5pt, line join=miter, line cap=rect](390.403pt,148.602pt) -- (400.552pt,148.602pt);
	\draw[line width=1.5pt, line join=miter, line cap=rect](370.103pt,133.666pt) -- (420.852pt,133.666pt);
	\draw[line width=1pt, line join=miter, line cap=rect](392.466pt,139.521pt) -- (399.492pt,139.521pt) -- (399.492pt,132.495pt) -- (392.466pt,132.495pt) -- (392.466pt,139.521pt);
\end{scope}
\begin{scope}
	\color[rgb]{0,0,0}
	\draw[line width=2pt, line join=miter, line cap=rect](76.285pt,382.429pt) -- (476.781pt,382.429pt) -- (476.781pt,82.3075pt) -- (76.285pt,82.3075pt) -- (76.285pt,382.429pt);
	\color[rgb]{0,0,0}
	\pgftext[center, base, at={\pgfpoint{21.6277pt}{240.398pt}},rotate=90]{\fontsize{25}{0}\selectfont{\textbf{Count}}}
	\pgftext[center, base, at={\pgfpoint{57.2137pt}{127.79pt}}]{\fontsize{24}{0}\selectfont{5}}
	\pgftext[center, base, at={\pgfpoint{51.1912pt}{210.097pt}}]{\fontsize{24}{0}\selectfont{10}}
	\pgftext[center, base, at={\pgfpoint{51.1912pt}{292.405pt}}]{\fontsize{24}{0}\selectfont{15}}
	\pgftext[center, base, at={\pgfpoint{51.1912pt}{374.712pt}}]{\fontsize{24}{0}\selectfont{20}}
	\draw[line width=1pt, line join=bevel, line cap=rect](76.285pt,92.5401pt) -- (71.2662pt,92.5401pt);
	\draw[line width=1pt, line join=bevel, line cap=rect](76.285pt,174.792pt) -- (71.2662pt,174.792pt);
	\draw[line width=1pt, line join=bevel, line cap=rect](76.285pt,257.044pt) -- (71.2662pt,257.044pt);
	\draw[line width=1pt, line join=bevel, line cap=rect](76.285pt,339.295pt) -- (71.2662pt,339.295pt);
	\draw[line width=1pt, line join=bevel, line cap=rect](76.285pt,133.666pt) -- (67.2512pt,133.666pt);
	\draw[line width=1pt, line join=bevel, line cap=rect](76.285pt,215.918pt) -- (67.2512pt,215.918pt);
	\draw[line width=1pt, line join=bevel, line cap=rect](76.285pt,298.17pt) -- (67.2512pt,298.17pt);
	\draw[line width=1pt, line join=bevel, line cap=rect](76.285pt,380.421pt) -- (67.2512pt,380.421pt);
	\pgftext[center, base, at={\pgfpoint{277.027pt}{8.46914pt}}]{\fontsize{25}{0}\selectfont{\textbf{Most common SOP}}}
	\pgftext[center, base, at={\pgfpoint{157.589pt}{47.4899pt}}]{\fontsize{24}{0}\selectfont{2}}
	\pgftext[center, base, at={\pgfpoint{236.885pt}{47.4899pt}}]{\fontsize{24}{0}\selectfont{3}}
	\pgftext[center, base, at={\pgfpoint{316.181pt}{47.4899pt}}]{\fontsize{24}{0}\selectfont{4}}
	\pgftext[center, base, at={\pgfpoint{395.477pt}{47.4899pt}}]{\fontsize{24}{0}\selectfont{5}}
	\draw[line width=1pt, line join=bevel, line cap=rect](157.589pt,82.3075pt) -- (157.589pt,73.2737pt);
	\draw[line width=1pt, line join=bevel, line cap=rect](236.885pt,82.3075pt) -- (236.885pt,73.2737pt);
	\draw[line width=1pt, line join=bevel, line cap=rect](316.181pt,82.3075pt) -- (316.181pt,73.2737pt);
	\draw[line width=1pt, line join=bevel, line cap=rect](395.477pt,82.3075pt) -- (395.477pt,73.2737pt);
	\definecolor{c}{rgb}{0,0,0}
	\fill [pattern color=c, pattern=north east lines](86.3225pt,369.38pt) -- (118.442pt,369.38pt) -- (118.442pt,356.331pt) -- (86.3225pt,356.331pt) -- (86.3225pt,369.38pt);
	\draw[line width=1.5pt, line join=miter, line cap=rect](86.3225pt,369.38pt) -- (118.442pt,369.38pt) -- (118.442pt,356.331pt) -- (86.3225pt,356.331pt) -- (86.3225pt,369.38pt);
	\pgftext[left, base, at={\pgfpoint{123.461pt}{357.068pt}}]{\fontsize{18}{0}\selectfont{$\neg A \lor \neg B \lor \neg C$     }}
	\definecolor{c}{rgb}{1,0,0}
	\fill [pattern color=c, pattern=north east lines](86.3225pt,342.279pt) -- (118.442pt,342.279pt) -- (118.442pt,329.23pt) -- (86.3225pt,329.23pt) -- (86.3225pt,342.279pt);
	\color[rgb]{1,0,0}
	\draw[line width=1.5pt, line join=miter, line cap=rect](86.3225pt,342.279pt) -- (118.442pt,342.279pt) -- (118.442pt,329.23pt) -- (86.3225pt,329.23pt) -- (86.3225pt,342.279pt);
	\color[rgb]{0,0,0}
	\pgftext[left, base, at={\pgfpoint{123.461pt}{329.967pt}}]{\fontsize{18}{0}\selectfont{$(A \land \neg B) \lor (B \land \neg C) \lor (D \land \neg A)$}}
	\definecolor{c}{rgb}{0,0.501961,0}
	\fill [pattern color=c, pattern=north east lines](86.3225pt,315.177pt) -- (118.442pt,315.177pt) -- (118.442pt,302.129pt) -- (86.3225pt,302.129pt) -- (86.3225pt,315.177pt);
	\color[rgb]{0,0.501961,0}
	\draw[line width=1.5pt, line join=miter, line cap=rect](86.3225pt,315.177pt) -- (118.442pt,315.177pt) -- (118.442pt,302.129pt) -- (86.3225pt,302.129pt) -- (86.3225pt,315.177pt);
	\color[rgb]{0,0,0}
	\pgftext[left, base, at={\pgfpoint{123.461pt}{302.866pt}}]{\fontsize{18}{0}\selectfont{$A \land B \land D \land \neg C$}}
	\definecolor{c}{rgb}{0,0,1}
	\fill [pattern color=c, pattern=north east lines](86.3225pt,288.076pt) -- (118.442pt,288.076pt) -- (118.442pt,275.027pt) -- (86.3225pt,275.027pt) -- (86.3225pt,288.076pt);
	\color[rgb]{0,0,1}
	\draw[line width=1.5pt, line join=miter, line cap=rect](86.3225pt,288.076pt) -- (118.442pt,288.076pt) -- (118.442pt,275.027pt) -- (86.3225pt,275.027pt) -- (86.3225pt,288.076pt);
	\color[rgb]{0,0,0}
	\pgftext[left, base, at={\pgfpoint{123.461pt}{275.765pt}}]{\fontsize{18}{0}\selectfont{$(A \land B \land \neg C) \lor (A \land D \land \neg B) \lor (B \land D \land \neg C)$}}
\end{scope}
\end{tikzpicture}
         }
        \label{fig:BoxPlot4bit}
    \end{subfigure}
    \begin{subfigure}[c]{0.32\textwidth}
        \centering
        \caption{}
        \resizebox{1\textwidth}{!}{
         \begin{tikzpicture}{0pt}{0pt}{469.755pt}{382.429pt}
	\color[rgb]{1,1,1}
	\fill(0pt,382.429pt) -- (469.755pt,382.429pt) -- (469.755pt,0pt) -- (0pt,0pt) -- (0pt,382.429pt);
\begin{scope}
	\clip(64.24pt,382.429pt) -- (464.736pt,382.429pt) -- (464.736pt,82.3075pt) -- (64.24pt,82.3075pt) -- (64.24pt,382.429pt);
	\definecolor{c}{rgb}{0,0,0}
	\fill [pattern color=c, pattern=north east lines](120.169pt,269.381pt) -- (170.919pt,269.381pt) -- (170.919pt,269.381pt) -- (120.169pt,269.381pt) -- (120.169pt,269.381pt);
	\color[rgb]{0,0,0}
	\draw[line width=1.5pt, line join=miter, line cap=rect](120.169pt,269.381pt) -- (170.919pt,269.381pt) -- (170.919pt,269.381pt) -- (120.169pt,269.381pt) -- (120.169pt,269.381pt);
	\color[rgb]{0,0,0}
	\draw[line width=1.5pt, line join=bevel, line cap=rect](145.544pt,317.893pt) -- (145.544pt,269.381pt);
	\draw[line width=1.5pt, line join=bevel, line cap=rect](145.544pt,239.376pt) -- (145.544pt,269.381pt);
	\draw[line width=1.5pt, line join=bevel, line cap=rect](140.469pt,239.376pt) -- (150.619pt,239.376pt);
	\draw[line width=1.5pt, line join=bevel, line cap=rect](140.469pt,317.893pt) -- (150.619pt,317.893pt);
	\draw[line width=1.5pt, line join=bevel, line cap=rect](120.169pt,269.381pt) -- (170.919pt,269.381pt);
	\draw[line width=1pt, line join=miter, line cap=rect](142.532pt,282.054pt) -- (149.559pt,282.054pt) -- (149.559pt,275.027pt) -- (142.532pt,275.027pt) -- (142.532pt,282.054pt);
	\definecolor{c}{rgb}{1,0,0}
	\fill [pattern color=c, pattern=north east lines](199.465pt,343.408pt) -- (250.215pt,343.408pt) -- (250.215pt,269.381pt) -- (199.465pt,269.381pt) -- (199.465pt,343.408pt);
	\color[rgb]{1,0,0}
	\draw[line width=1.5pt, line join=miter, line cap=rect](199.465pt,343.408pt) -- (250.215pt,343.408pt) -- (250.215pt,269.381pt) -- (199.465pt,269.381pt) -- (199.465pt,343.408pt);
	\draw[line width=1.5pt, line join=miter, line cap=rect](224.84pt,358.716pt) -- (224.84pt,343.408pt);
	\draw[line width=1.5pt, line join=miter, line cap=rect](224.84pt,194.852pt) -- (224.84pt,269.381pt);
	\draw[line width=1.5pt, line join=miter, line cap=rect](219.765pt,194.852pt) -- (229.915pt,194.852pt);
	\draw[line width=1.5pt, line join=miter, line cap=rect](219.765pt,358.716pt) -- (229.915pt,358.716pt);
	\draw[line width=1.5pt, line join=miter, line cap=rect](199.465pt,269.381pt) -- (250.215pt,269.381pt);
	\draw[line width=1pt, line join=miter, line cap=rect](221.829pt,280.046pt) -- (228.855pt,280.046pt) -- (228.855pt,273.02pt) -- (221.829pt,273.02pt) -- (221.829pt,280.046pt);
	\definecolor{c}{rgb}{0,0.501961,0}
	\fill [pattern color=c, pattern=north east lines](278.761pt,195.355pt) -- (329.511pt,195.355pt) -- (329.511pt,121.328pt) -- (278.761pt,121.328pt) -- (278.761pt,195.355pt);
	\color[rgb]{0,0.501961,0}
	\draw[line width=1.5pt, line join=miter, line cap=rect](278.761pt,195.355pt) -- (329.511pt,195.355pt) -- (329.511pt,121.328pt) -- (278.761pt,121.328pt) -- (278.761pt,195.355pt);
	\draw[line width=1.5pt, line join=miter, line cap=rect](304.136pt,223.085pt) -- (304.136pt,195.355pt);
	\draw[line width=1.5pt, line join=miter, line cap=rect](304.136pt,108.403pt) -- (304.136pt,121.328pt);
	\draw[line width=1.5pt, line join=miter, line cap=rect](299.061pt,108.403pt) -- (309.211pt,108.403pt);
	\draw[line width=1.5pt, line join=miter, line cap=rect](299.061pt,223.085pt) -- (309.211pt,223.085pt);
	\draw[line width=1.5pt, line join=miter, line cap=rect](278.761pt,195.355pt) -- (329.511pt,195.355pt);
	\draw[line width=1pt, line join=miter, line cap=rect](301.125pt,169.634pt) -- (308.151pt,169.634pt) -- (308.151pt,162.607pt) -- (301.125pt,162.607pt) -- (301.125pt,169.634pt);
	\definecolor{c}{rgb}{0,0,1}
	\fill [pattern color=c, pattern=north east lines](358.058pt,232.368pt) -- (408.807pt,232.368pt) -- (408.807pt,158.342pt) -- (358.058pt,158.342pt) -- (358.058pt,232.368pt);
	\color[rgb]{0,0,1}
	\draw[line width=1.5pt, line join=miter, line cap=rect](358.058pt,232.368pt) -- (408.807pt,232.368pt) -- (408.807pt,158.342pt) -- (358.058pt,158.342pt) -- (358.058pt,232.368pt);
	\draw[line width=1.5pt, line join=miter, line cap=rect](383.432pt,273.872pt) -- (383.432pt,232.368pt);
	\draw[line width=1.5pt, line join=miter, line cap=rect](383.432pt,116.838pt) -- (383.432pt,158.342pt);
	\draw[line width=1.5pt, line join=miter, line cap=rect](378.358pt,116.838pt) -- (388.507pt,116.838pt);
	\draw[line width=1.5pt, line join=miter, line cap=rect](378.358pt,273.872pt) -- (388.507pt,273.872pt);
	\draw[line width=1.5pt, line join=miter, line cap=rect](358.058pt,195.355pt) -- (408.807pt,195.355pt);
	\draw[line width=1pt, line join=miter, line cap=rect](380.421pt,198.742pt) -- (387.447pt,198.742pt) -- (387.447pt,191.716pt) -- (380.421pt,191.716pt) -- (380.421pt,198.742pt);
\end{scope}
\begin{scope}
	\color[rgb]{0,0,0}
	\draw[line width=2pt, line join=miter, line cap=rect](64.24pt,382.429pt) -- (464.736pt,382.429pt) -- (464.736pt,82.3075pt) -- (64.24pt,82.3075pt) -- (64.24pt,382.429pt);
	\color[rgb]{0,0,0}
	\pgftext[center, base, at={\pgfpoint{21.6277pt}{233.372pt}},rotate=90]{\fontsize{25}{0}\selectfont{\textbf{Count}}}
	\pgftext[center, base, at={\pgfpoint{45.1687pt}{115.745pt}}]{\fontsize{24}{0}\selectfont{1}}
	\pgftext[center, base, at={\pgfpoint{45.1687pt}{190.022pt}}]{\fontsize{24}{0}\selectfont{2}}
	\pgftext[center, base, at={\pgfpoint{45.1687pt}{263.296pt}}]{\fontsize{24}{0}\selectfont{3}}
	\pgftext[center, base, at={\pgfpoint{45.1687pt}{337.574pt}}]{\fontsize{24}{0}\selectfont{4}}
	\draw[line width=1pt, line join=bevel, line cap=rect](64.24pt,158.342pt) -- (59.2212pt,158.342pt);
	\draw[line width=1pt, line join=bevel, line cap=rect](64.24pt,232.368pt) -- (59.2212pt,232.368pt);
	\draw[line width=1pt, line join=bevel, line cap=rect](64.24pt,306.395pt) -- (59.2212pt,306.395pt);
	\draw[line width=1pt, line join=bevel, line cap=rect](64.24pt,380.421pt) -- (59.2212pt,380.421pt);
	\draw[line width=1pt, line join=bevel, line cap=rect](64.24pt,121.328pt) -- (55.2062pt,121.328pt);
	\draw[line width=1pt, line join=bevel, line cap=rect](64.24pt,195.355pt) -- (55.2062pt,195.355pt);
	\draw[line width=1pt, line join=bevel, line cap=rect](64.24pt,269.381pt) -- (55.2062pt,269.381pt);
	\draw[line width=1pt, line join=bevel, line cap=rect](64.24pt,343.408pt) -- (55.2062pt,343.408pt);
	\pgftext[center, base, at={\pgfpoint{264.982pt}{8.46914pt}}]{\fontsize{25}{0}\selectfont{\textbf{Most common SOP}}}
	\pgftext[center, base, at={\pgfpoint{145.544pt}{47.4899pt}}]{\fontsize{24}{0}\selectfont{2}}
	\pgftext[center, base, at={\pgfpoint{224.84pt}{47.4899pt}}]{\fontsize{24}{0}\selectfont{3}}
	\pgftext[center, base, at={\pgfpoint{304.136pt}{47.4899pt}}]{\fontsize{24}{0}\selectfont{4}}
	\pgftext[center, base, at={\pgfpoint{383.432pt}{47.4899pt}}]{\fontsize{24}{0}\selectfont{5}}
	\draw[line width=1pt, line join=bevel, line cap=rect](145.544pt,82.3075pt) -- (145.544pt,73.2737pt);
	\draw[line width=1pt, line join=bevel, line cap=rect](224.84pt,82.3075pt) -- (224.84pt,73.2737pt);
	\draw[line width=1pt, line join=bevel, line cap=rect](304.136pt,82.3075pt) -- (304.136pt,73.2737pt);
	\draw[line width=1pt, line join=bevel, line cap=rect](383.432pt,82.3075pt) -- (383.432pt,73.2737pt);
\end{scope}
\end{tikzpicture}
         }
        \label{fig:BoxPlot8bit}
    \end{subfigure}
    \caption{Boxplots showing the distribution of the four most common sum-of-product expressions found in repeated experiments with (a) 2-bit, (b) 4-bit, and (c) 8-bit binary strings. Whiskers indicate 1.5 times the standard deviation.}
    \label{fig:boxplot}
\end{figure}

\begin{figure}
    \centering
    \begin{subfigure}[c]{0.32\textwidth}
        \centering
        \caption{}
        \includegraphics[width=1\textwidth, keepaspectratio]{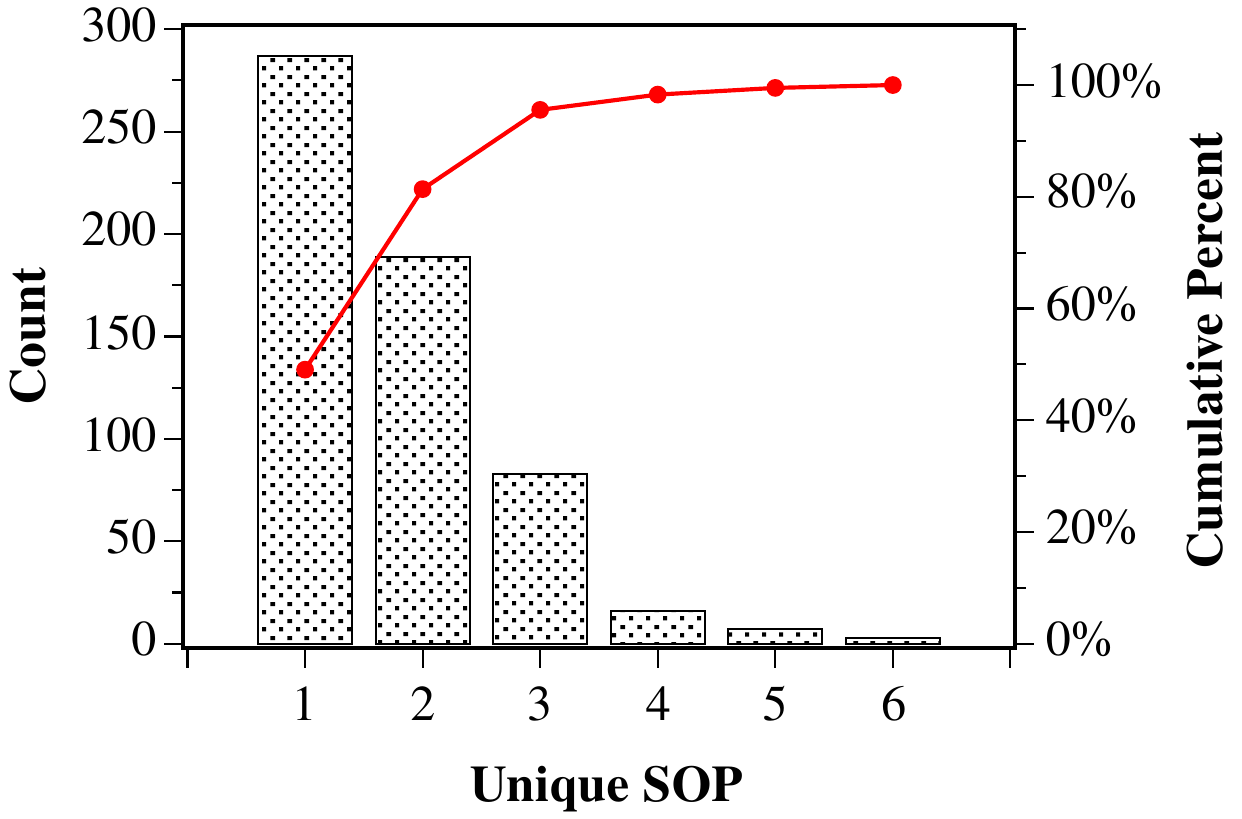}
        \label{fig:Pareto2bit}
    \end{subfigure}
    \begin{subfigure}[c]{0.32\textwidth}
        \centering
        \caption{}
        \includegraphics[width=1\textwidth, keepaspectratio]{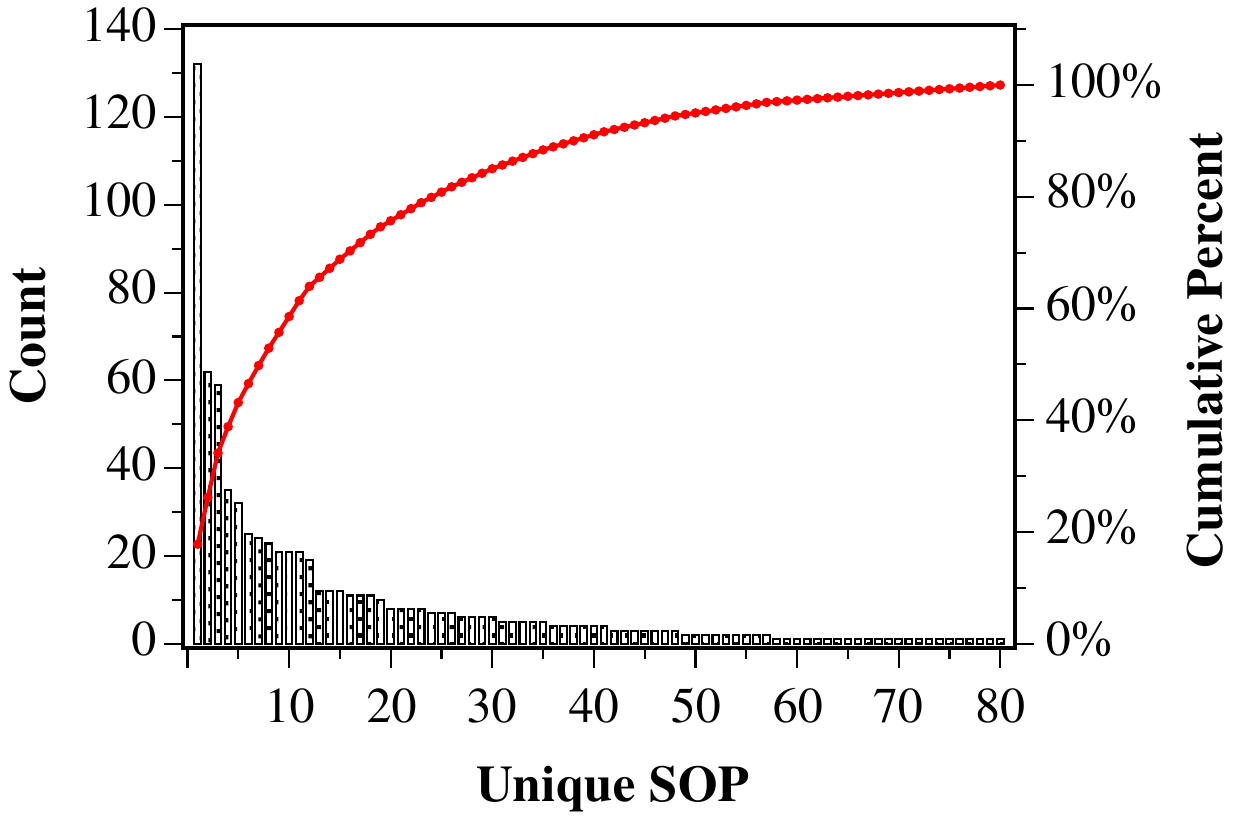}
        \label{fig:Pareto4bit}
    \end{subfigure}
    \begin{subfigure}[c]{0.32\textwidth}
        \centering
        \caption{}
        \includegraphics[width=1\textwidth, keepaspectratio]{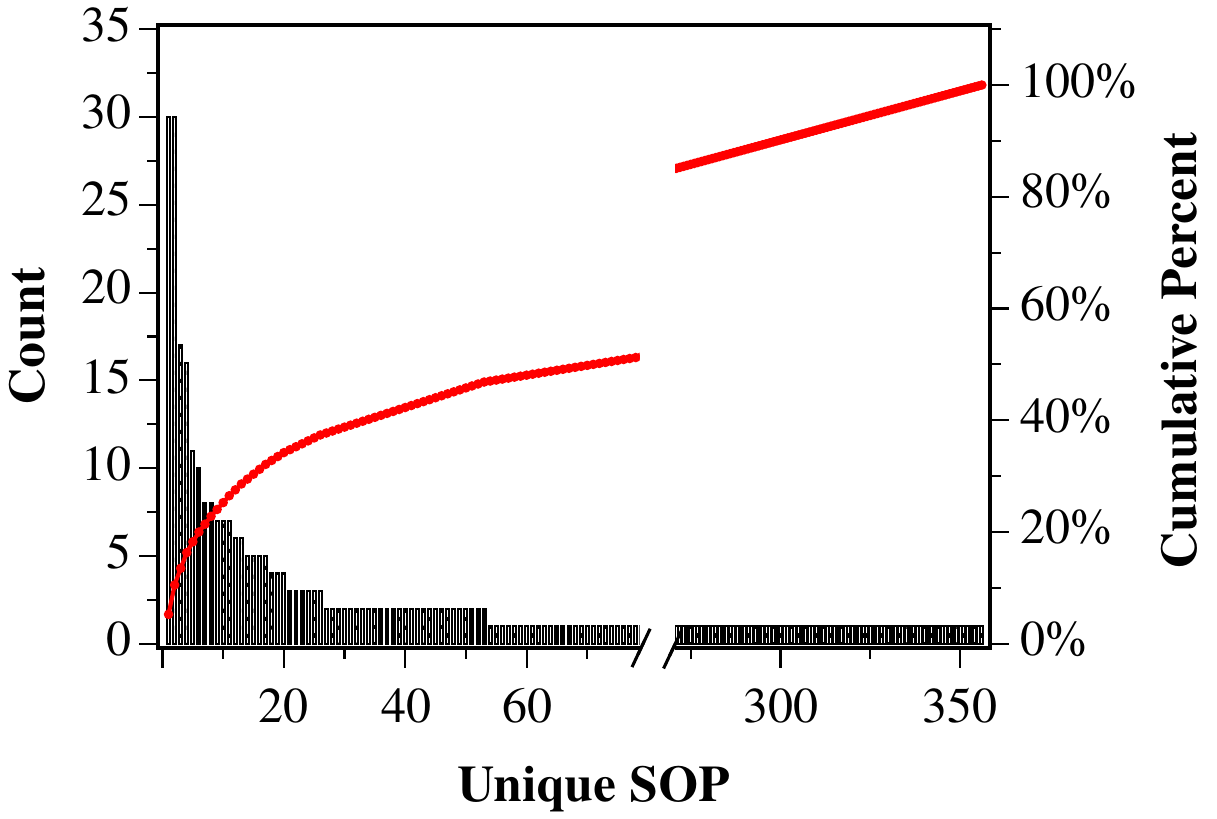}
        \label{fig:Pareto8bit}
    \end{subfigure}
    \caption{Pareto charts showing the frequency counts of unique SOP from repeated experiments. The red curve represents the cumulative percentage of the total count for (a) 2-bit, (b) 4-bit, and (c) 8-bit binary strings.}
    \label{fig:pareto}
\end{figure}

\FloatBarrier
Using the same methodology as before, the reproducibility of the extracted logic circuits using the ZnO + protenoids mixture was tested over 15 repetitions for each binary input string, with shaking before each repetition. Fig.~\ref{fig:boxplot} shows the boxplots for 2-, 4-, and 8-bit inputs, including whiskers indicating 1.5 times the standard deviation across repetitions. While more varied than the individual systems, the recurrence of common logical expressions shows how the circuits are deterministically derived from the mixture for a given binary string. The Pareto charts in Fig.~\ref{fig:pareto} indicate that the mixture displayed an intermediate behaviour, with a number of unique SOP occurring between its separate components, with its number again increasing exponentially with more input bits.
\FloatBarrier
\section{Complexity}
\begin{figure}[!ht]
    \centering
    \includegraphics[width=.75\textwidth, keepaspectratio]{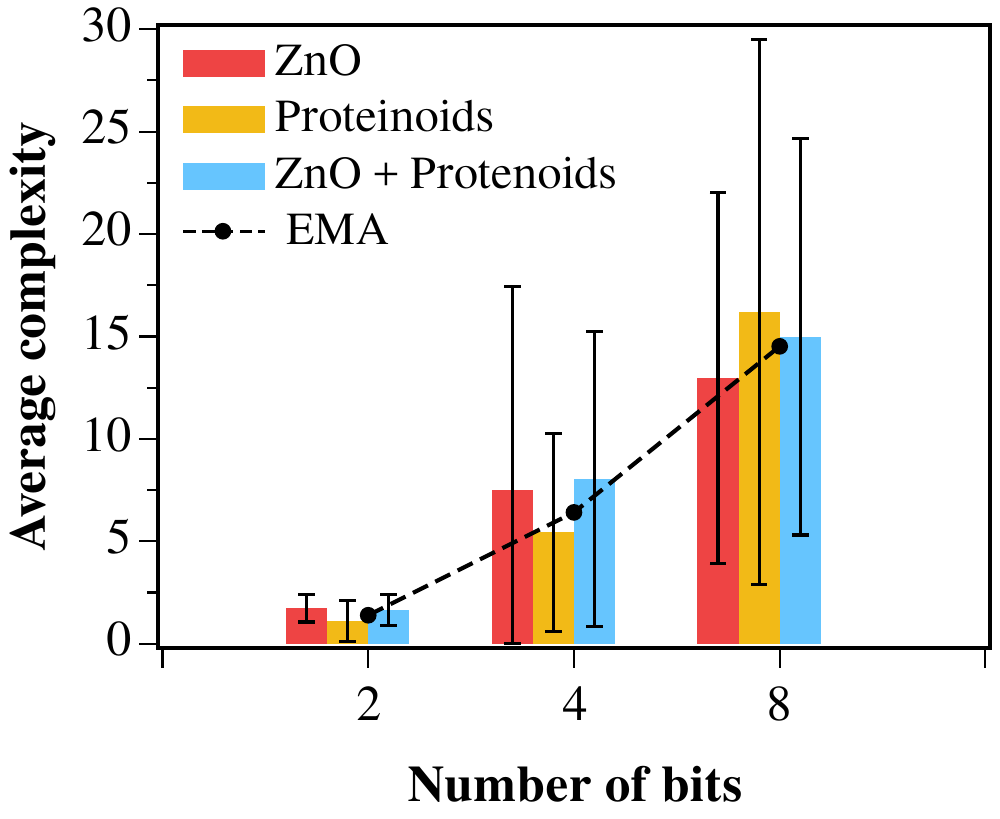}
    \caption{Complexity of the obtained logical expressions for ZnO, proteinoids, and mixture of ZnO and proteinoid for 2-, 4-, and 8-bit.}
    \label{fig:complexity}
\end{figure}

The complexity of the resulting logical expressions was quantified by determining the number of clauses in their conjunctive normal form representation. Fig.~\ref{fig:complexity} shows the average complexity values and standard deviations for logical expressions modelling the behaviour of ZnO nanoparticles, proteinoid microspheres and mixtures thereof for 2-, 4- and 8-bit strings. Interestingly, the complexity trends observed in the mixed systems appear to result from a weighted combination of the complexities of the individual components. By considering the average complexity as a material property, it is possible to construct a reasonable effective medium approximation for the mixture
\begin{equation}
    f \frac{\epsilon_1 - \epsilon_{\text{eff}}}{\epsilon_1 + 2\epsilon_{\text{eff}}} + (1-f) \frac{\epsilon_2 - \epsilon_{\text{eff}}}{\epsilon_2 + 2\epsilon_{\text{eff}}} = 0,
\end{equation}
where $f$ is the fraction of the first component of the mixture, $\epsilon_1$ is the property of the first component of the mixture, $\epsilon_2$ is the property of the second component of the mixture, and $\epsilon_{\text{eff}}$ is the effective component of the mixture. The predicted complexity values from this simple model, shown by the black circles in Fig.~\ref{fig:complexity}, are in good agreement with the measured complexities. This implies that the hybrid system's complexity is a result of the distinct contributions made by the zinc oxide nanoparticles and proteinoid microspheres.

\section{Conclusions}
In this work, the capabilities of colloidal suspensions of ZnO, protenoids and their mixture have been investigated using a bare-bones experimental setup. We demonstrate the feasibility of studying and using these materials for computational purposes. In particular, we verify unique responses and complexities for each material, with the combined system exhibiting averaged rather than enhanced complexity. While ZnO yields more complex logic expressions, protenoids enabled a more robust response. This research highlights colloidal suspensions as promising materials for unconventional computing, while illustrating the non-trivial relationship between device complexity and material behaviour. Further study may reveal ways to tune and optimise the computational repertoires of multicomponent colloidal logic.
\section*{Acknowledgement}
PM and AA are supported by EPSRC Grant EP/W010887/1 ``Computing with proteinoids''. The authors are grateful to David Paton for helping with SEM imaging. NRH, RF, AA, and AC received support from the European Innovation Council and the SMEs Executive Agency (EISMEA) under grant agreement No. 964388.

\FloatBarrier
\section*{Data availability statement}
Datasets that support the findings of this study are available from the corresponding authors upon reasonable request.
\emergencystretch 10em
\printbibliography

\end{document}